\documentclass[useAMS,usenatbib]{mn2e}

\usepackage{latexsym}
\usepackage{afterpage}
\usepackage{here}
\usepackage{longtable}
\usepackage[dvips]{graphicx}
\usepackage{natbib}
\usepackage{psfrag}
\usepackage{rotating}
\usepackage{amssymb}

\def\degsq{$\deg^2$ }
\def\Hzero{$H_0$}
\def\OM{$\Omega_M$}
\def\OL{$\Omega_{\Lambda}$}

\def\SfiveG{$S_{\rm 5.0~GHz}$}
\def\StwoG{$S_{\rm 2.7~GHz}$}
\def\SoneG{$S_{\rm 1.4~GHz}$}
\def\SfourM{$S_{\rm 408~MHz}$}
\def\SthreeM{$S_{\rm 365~MHz}$}
\def\SoneM{$S_{\rm 178~MHz}$}

\def\AJ{{\it Astron. J.}}
\def\ApJ{{\it Astroph. J.}}
\def\ApJS{{\it Astroph. J. Sup.}}
\def\MNRAS{{\it Mon. Not. R. Astr. Soc.}}
\def\AA{{\it Astron. Astrophys.}}
\def\AAS{{\it Astron. Astrophys. Sup.}}
\def\ASPC{{\it Astr. Soc. Pac. Conf. Ser.}}
\def\PASP{{\it Publi. Astr. Soc. Pac.}}

\title[The CoNFIG Sample]{The Combined NVSS-FIRST Galaxies (CoNFIG)
  Sample - I. Sample Definition, Classification and Evolution} 

\author[Melanie A. Gendre and J. V. Wall]{M. A. Gendre$^{1}$\thanks{E-mail:
mgendre@phas.ubc.ca; jvw@phas.ubc.ca} and J. V.
Wall$^{1}$\\
$^{1}$Department of Physics and Astronomy, The University of British Colombia, 6224
Agricultural Rd, Vancouver, BC, V6T 1Z1, Canada}

\begin{document}

\date{Accepted . Received ; in original form }

\pagerange{\pageref{firstpage}--\pageref{lastpage}} \pubyear{2008}

\maketitle

\label{firstpage}

%Abstract
\begin{abstract} 
The CoNFIG (Combined NVSS-FIRST Galaxies) sample is a new sample of 274
bright radio sources at 1.4~GHz. It was defined by selecting all sources
with \SoneG$\ge$1.3~Jy from the NRAO-VLA Sky Survey (NVSS) in the North
field of the Faint Images of the Radio Sky at Twenty-cm (FIRST)
survey. New radio observations obtained with the VLA for 31 of the sources
are presented. The sample has complete FRI/FRII morphology
identification; optical identifications and redshifts are available for  
80\% and 89\% of the sample respectively, yielding a mean redshift
of $\sim$0.71. One of the goals of this survey is to get better
definitions of luminosity distributions and source counts of FRI/FRII
sources separately, in order to determine the evolution of the
luminosity function for each type of source. We present a preliminary
analysis, showing that these data are an important step towards
examining various evolutionary schemes for these objects and to
confirm or correct the dual population unified scheme for radio
AGN. Improving our understanding of radio galaxy evolution will give
better insight into the role of AGN feedback in galaxy formation.  
\end{abstract}

%Keywords

%%%%%%%%%%%%%%%%%%%%%%%% INTRO %%%%%%%%%%%%%%%%%%%%%%%%%%%%%%%%%%%%%%%%%%%%%%%%%%%%%%%%%%%
\section{Introduction}

%Historical Background
\cite{Long66} determined that powerful radio sources undergo strong
differential evolution, the first indication of cosmic
downsizing. Since then our understanding of the space density of AGN
as a function of cosmic epoch has steadily continued to advance.\\   
\indent
With the development of evolutionary models for radio sources came the
idea of a dual-population model. The initial version of this dichotomy
\citep{Long66} was in terms of low and high luminosities. But with
high-frequency surveys, and the large number of flat/inverted
spectrum sources revealed in them, an alternative classification
emerged, based exclusively on the source spectra: sources with a
spectral index $\alpha \le -0.5$ (where $S^{\alpha}_{\nu} \propto
\nu^{\alpha}$),  corresponding to optically thin synchrotron
radiation, were classified as steep-spectrum, whereas sources with
lower spectral indices ($\alpha \ge -0.5$) inevitably showed features
of synchrotron self-absorption and were classified as
flat-spectrum. Initial indications suggested that these two
populations underwent different evolution
\citep{Schmidt76,Masson77}. However, \cite{Dunlop90} studied the radio
luminosity function (density of sources with a given luminosity per
unit of co-moving volume) of these two classes of radio sources, and
came to the conclusion that both populations were undergoing a similar
evolution, implying that they might not actually be distinct.\\  
%Definition of the FR classes 
\indent
The Fanaroff-Riley (FR) classification \citep{FR74}, originally a
sub-classification for steep-spectrum objects, was employed to provide
a further categorization of radio sources. This classification
divides radio sources into two classes of double-lobed
sources based on the appearance of their jets. The FRI objects have the
highest brightness along the jets and core, reside in moderately rich
cluster environments \citep{Hill91} and include sources with
irregular structure \citep{Parma92}. In contrast, FRII sources show
hot spots in the lobes and more collimated jets, are found in more
isolated environments and display stronger emission lines
\citep{Rawlings89,Baum89}. \cite{FR74} found these two classes to be
divided in radio power, with a break luminosity $P_{\rm 178~MHz} \sim
10^{25}\, {\rm W   Hz^{-1}sr^{-1}}$, with FRII sources lying above
this limit. Subsequently \cite{Owen94} showed that the break was a
function of both radio and optical luminosity.\\
%Unified Models 
\indent
During the 1980s the 'unification' hypothesis emerged to describe how
viewing aspect could relate RQSOs, (Radio Quasi-Stellar Objects of either
flat or steep-spectrum)  to FRII radio galaxies
\citep[e.g.][]{Peacock87,Scheuer87,Barthel89}. However, the scheme did not include
lower-luminosity AGNs such as FRI galaxies and BL Lac objects. The
unifying connection between these was introduced by
\cite{Marcha95}. The unified model of AGN proposed
by \cite{Wall97} and \cite{Jack99} assumes that the cosmic evolution
of radio loud AGN is based on a division of the radio sources into a
low-luminosity ($P_{\rm 178~MHz}<10^{25}\, {\rm W Hz^{-1}sr^{-1}}$)
component corresponding to FRIs, and a high-luminosity component
corresponding to FRIIs. In this scheme, the various forms of AGN
observed (FRI and FRII extended double sources, flat- and
steep-spectrum RQSOs and BL~Lac objects) result from the
orientation of the extended parent objects with respect to the
observer's line-of-sight. Indeed, because
the double-sided ejection of synchrotron blobs in AGN is at
relativistic speed, the orientation of the ejection axis to the
line-of-sight becomes crucial: sources viewed side-on appear as
double radio galaxies (FRI or FRII) and sources viewed along the jets
appear as RQSOs (beamed counterparts of FRII sources) or BL~Lac objects
(beamed counterparts of FRI sources). The relativistically-boosted jet
emission in the beamed counterparts of the extended sources dominates
the extended emission, making the overall radio emission appear
compact down to VLBI scales.\\
\indent
Initially, in modelling the space density of radio AGN, it was assumed
for simplicity that the low-luminosity radio galaxies including FRI
sources showed no cosmic evolution \citep{WPL80,Jack99}, the strong
cosmic evolution confined only to the higher luminosities and the FRII
galaxies. With the advent of large-scale redshift surveys for nearby
galaxies, many authors, including \cite{Brown01},
\cite{Snellen01}, \cite{Willott01}, \cite{Sadler07} and
\cite{Rigby07}, found significant evolution for low power sources --
but mild evolution in comparison with that of the high-luminosity
sources. \cite{Rigby07} argued that if both FRIs and FRIIs have
similar evolution, the dual-population scheme could be reduced to a
single-population model.\\

%Issues of dual-population model
The FRI/FRII dual-population scheme has encountered several
problems. One of these concerns the correspondence between FRI galaxies
and BL~Lac objects. \cite{Urry95} noted that some BL~Lac objects have
non-FRI-like morphologies and that the density of FRI sources is too
low to account for the entire BL~Lac population, a concern also
raised by \cite{Wall97}. Looking at
BL Lac objects from another point of view, \cite{March96}
demonstrate that only about one third of low-luminosity core
dominated radio sources - which are supposedly the beamed counterpart
of FRI sources - are conventional BL~Lac objects. Most of the remaining
sources have optical classification such as Seyfert objects or
elliptical galaxies.\\ 
\indent
A related issue concerns the existence of FRI RQSOs. Until
recently, these objects were thought not to exist, leading to the
hypothesis that FRI and FRII central engines were of
different nature \citep{Baum95} and that the torus opening in FRI
sources was too small to observe a quasar nucleus
\citep{Falcke95}. However, the discovery of an FRI QSO,
E1821+643, by \cite{Blun01} overthrew those assumptions. More recent
VLA observations \citep{Heywood07} uncovered another 4 sources of this
type.\\ 
\indent
Finally, if sources with different FR classes undergo different
evolution, this might imply that their fundamental characteristics,
such as the black hole spin or jet composition, are different
too. However, the existence of hybrid sources, which display both FRI
and FRII morphological characteristics \citep{Capetti95}, then becomes
puzzling. Based on observations of hybrid sources, \cite{Kaiser97}
argued that the FR dichotomy is based purely on the interaction
between the jets and the environmental medium, and not on intrinsic
properties of the central engine.  This view is also supported by
\cite{Gopal00} and \cite{Gawr06}. However, \cite{Wang92} suggested that
some AGN engines could be capable of ejecting jets of unequal power,
resolving the problem of hybrid sources. \cite{Gopal00}
found no evidence for such a process in their sample.\\  

%Other Schemes 
\indent
Other schemes have been suggested to resolve these
difficulties. \cite{Kaiser07} explained the FR dichotomy by
postulating that all sources start as FRII objects and used an analytical
model in which the evolution of the radio sources is governed by
energy losses from both radiating relativistic electrons in the lobes
and turbulence in the jets.\\
\indent
\cite{Willott01} used an approach to a dual-population
unified scheme based on the luminosity of sources instead of
morphology. Optical spectra of FRII sources are heterogeneous and they
can display both strong  and weak low-excitation emission lines
\citep{Laing94}. Therefore, radio sources can also be grouped based on
their emission lines, with one population composed of
low-luminosity sources having weak emission lines (containing both FRI
and FRII objects), and the other composed of high-luminosity sources
with strong emission lines (containing only FRII objects). With this
model, \cite{Willott01} concluded that high luminosity objects undergo
a stronger evolution with epoch than low luminosity sources, as found
by by e.g. \cite{Long66}, \cite{WPL80} and \cite{Urry95}, and that the
radio luminosity function has the form of a broken power-law, similar
to the conclusions of \cite{Dunlop90}. We note that, since
conventional accretion-disk systems are expected to be strong X-ray
sources, luminosity-based dual-population models are also used in
modelling the luminosity function of X-ray selected RQSOs
\citep{Hardcastle06}.\\ 

%Why so important to understand radio galaxies?
Defining the relation between the different radio sub-populations
together with their cosmic evolution is becoming fundamental to our
understanding of galaxy formation.\\
The current paradigm for galaxy formation follows hierarchical
build-up in a Cold Dark Matter (CDM) universe. Nevertheless, serious
difficulties arise from this model in its simplest form, as discussed
by \cite{Bower06}. It implies that current epoch galaxies must be the
largest and bluest and have the highest star forming rate of all
galaxies. Observations show that they are red, old galaxies, whereas
the bulk of star formation is observed at earlier epochs. This is
known as downsizing, first described by \cite{Cowie96}. AGN negative
feedback could be the key to understanding this phenomenon. The
ignition of the nucleus in a star forming galaxy could eject the gas
into the inter-galactic medium, thus reducing or even stopping star
formation, breaking the hierarchical buildup
\citep{Silk98,Granato01,Quilis01}. Note that there is also some
positive AGN feedback \citep{vanB04,Klamer04} in which  the pressure
from the jets compresses the inter-stellar medium and induces star
formation. The balance between these processes remains to be
understood; establishing the cosmic behaviour of the radio AGN is
important in revealing the precise role of the feedback mechanisms.\\

%Intro to CoNFIG
Although FRI and FRII sources show different evolution, they also lie
in different luminosity ranges. There is therefore a possibility that
both types may show similar evolution for overlapping luminosities
(i.e. high-luminosity FRIs and low-luminosity FRIIs).
In order to sort out the FR dichotomy and its details, accurate models
of the evolution of each population are needed. This implies compiling
accurate statistics, such as luminosity distributions and source
counts, for both types separately. This is the goal of the
CoNFIG (Combined NVSS-FIRST Galaxies) sample presented here, a new
sample of bright radio sources with complete morphological
identification.\\ 
\indent
The CoNFIG sample is defined as all
sources with \SoneG$\ge$1.3~Jy from the NRAO-VLA Sky Survey 
\citep[NVSS,][]{Condon98} catalogue within the north region of the Faint Images
of Radio Sources at Twenty-cm survey \citep[FIRST,][]{White97}, a 1.5~sr 
region defined roughly by $-8^{\circ} \le {\rm Dec}\le 64^{\circ}$
and 7 hr $\le {\rm RA}\le$ 17 hr. The flux density limit of 1.3~Jy
was chosen so that the number of sources in the sample was of
statistical significance while allowing us to identify the morphology for
each source individually. Optical identifications were obtained from
the  SuperCOSMOS Sky Survey \citep[SSS,][]{Hambly01} and redshift
information extracted using the SIMBAD  database. With the accompanying VLA
observations described here, the sample has complete morphology
information, a median flux density of \SoneG$\sim$1.96~Jy and optical
identifications and redshift information for $\sim$80\% and $\sim$89\%
of the sources respectively.\\ 

%Structure Paper
The structure of this paper is as follows. The details of the
construction of the CoNFIG sample are explained in \S\ref{sample}
while \S\ref{morpho} describes how the morphologies were
determined. Optical identifications and redshift information are
discussed in \S\ref{IDandz} and \S\ref{lumdist} outlines the
computation of the morphology-dependent luminosity
distributions. Finally, \S\ref{extSC} describes and discusses 
the FRI/FRII source counts.\\ 
\indent
Throughout this paper, we assume a standard $\Lambda$CDM
cosmology with \Hzero=70 km s$^{-1}$ Mpc$^{-1}$, \OM=0.3 and \OL=0.7.

%%%%%%%%%%%%%%%%%%%%%%%%% DEFINITION %%%%%%%%%%%%%%%%%%%%%%%%%%%%%%%%%%%%%%%%%%%%%%%%%%%%%%%%%%
\section{The CoNFIG Sample}\label{sample}

\subsection{Radio Surveys}\label{survey}

\subsubsection{NVSS}

The NRAO-VLA Sky Survey (NVSS) is a 1.4~GHz continuum survey covering the
entire sky north of $-40^{\circ}$ declination (corresponding to an
area of 10.3 sr). The completeness limit is about 2.5~mJy with an rms
brightness fluctuation of about 0.45~mJy/beam. The survey has yielded
over 1.8 million sources, implying a surface density of $\sim$50
sources per square degree. It was carried out with the Very Large
Array (VLA) in D and DnC configuration (the D configuration being
the most compact VLA configuration with a maximum antenna separation
of $\sim$1 km), providing an angular resolution of about
45 arcsec FWHM \citep{Condon98}.\\
\indent  
Since the median angular size of faint extragalactic sources at these flux
density levels is $\lesssim$10 arcsec \citep{Condon98}, most sources
are unresolved, and the NVSS flux density measurements are quite
accurate. However, the large beam size does not allow one to determine precise
structure of sources or to determine positions accurate enough to
establish optical counterparts.

\subsubsection{FIRST}

The Faint Images of the Radio Sky at Twenty-cm (FIRST) is another 1.4~GHz
continuum survey, covering an area of $\sim$9030 \degsq over the North
Galactic Pole. The completeness limit is about 1~mJy with a typical rms
of 0.15~mJy. The survey yielded $\sim$811,000 sources, implying a
surface density of $\sim$90 sources per square degree. It was carried
out with the VLA in B configuration (the B
configuration having a maximum antenna separation of $\sim$10 km), which
provides an angular resolution of about 5 arcsec FWHM
\citep{White97}. This survey complements the NVSS survey well,
providing a beam size small enough to resolve the structure of most
nearby extended radio sources.

\subsection{Sample Definition}\label{samdef}

The CoNFIG sample is defined as all sources with \SoneG$\ge$1.3~Jy from
the NVSS catalogue within the north region of the FIRST survey (1.5 sr
region defined roughly by $-8^{\circ} \le  {\rm Dec}\le 64^{\circ}$
and 7 hr $\le {\rm RA}\le$ 17 hr), resulting in a sample of 261
objects.\\
\indent
Very large sources resolved in NVSS within this initial sample, such
as 3CRR sources \citep{Laing83}, need to be considered. In
these cases, two or more NVSS sources with \SoneG$\ge$1.3~Jy are
actually components of a much larger resolved source. After collecting the
resolved source components together, the refined sample contains 248
sources.\\
\indent
Multi-component sources, where each component has \SoneG$<$1.3~Jy but
with a total flux density \SoneG$\ge$1.3~Jy, also need to be
considered. For this purpose, NVSS sources with
0.1~Jy$\le$\SoneG$\le$1.3~Jy were selected and, if any other source in
the catalogue was located within 4 arcmin of the listed sources,
they were set aside as candidate extended sources. The final decision
on whether or not the sources are actually components of a resolved
source was made by visual inspection of the NVSS contour plot. In this
manner, 32 more sources were added to the sample.\\
\indent
Finally, after looking at the NVSS and FIRST contour plots and optical
information of all the sources, 7 sources\footnote{Note that 3 of these sources (CoNFIG-015, 076 and
225) were deleted later in the sample construction process, based on
the 8~GHz VLA observations described in \S\ref{proposal}.} were deleted, either
because the source is a globular cluster, or because the source
actually consists of independent sources, each with
\SoneG$<$1.3~Jy.\\ 
\indent
The final sample, referred to as the CoNFIG Sample, consists of
274 sources. Details of the sources can be found in Appendix~\ref{details}.

%%%%%%%%%%%%%%%%%%%%%%%%% MORPH  %%%%%%%%%%%%%%%%%%%%%%%%%%%%%%%%%%%%%%%%%%%%%%%%%%%%%%%%%%

\section{Morphology}\label{morpho}

\subsection{Initial Classification}\label{FirstID}

\begin{figure}
  \begin{minipage}{8.0cm}
    \centerline{
      \includegraphics[angle=270,scale=0.2]{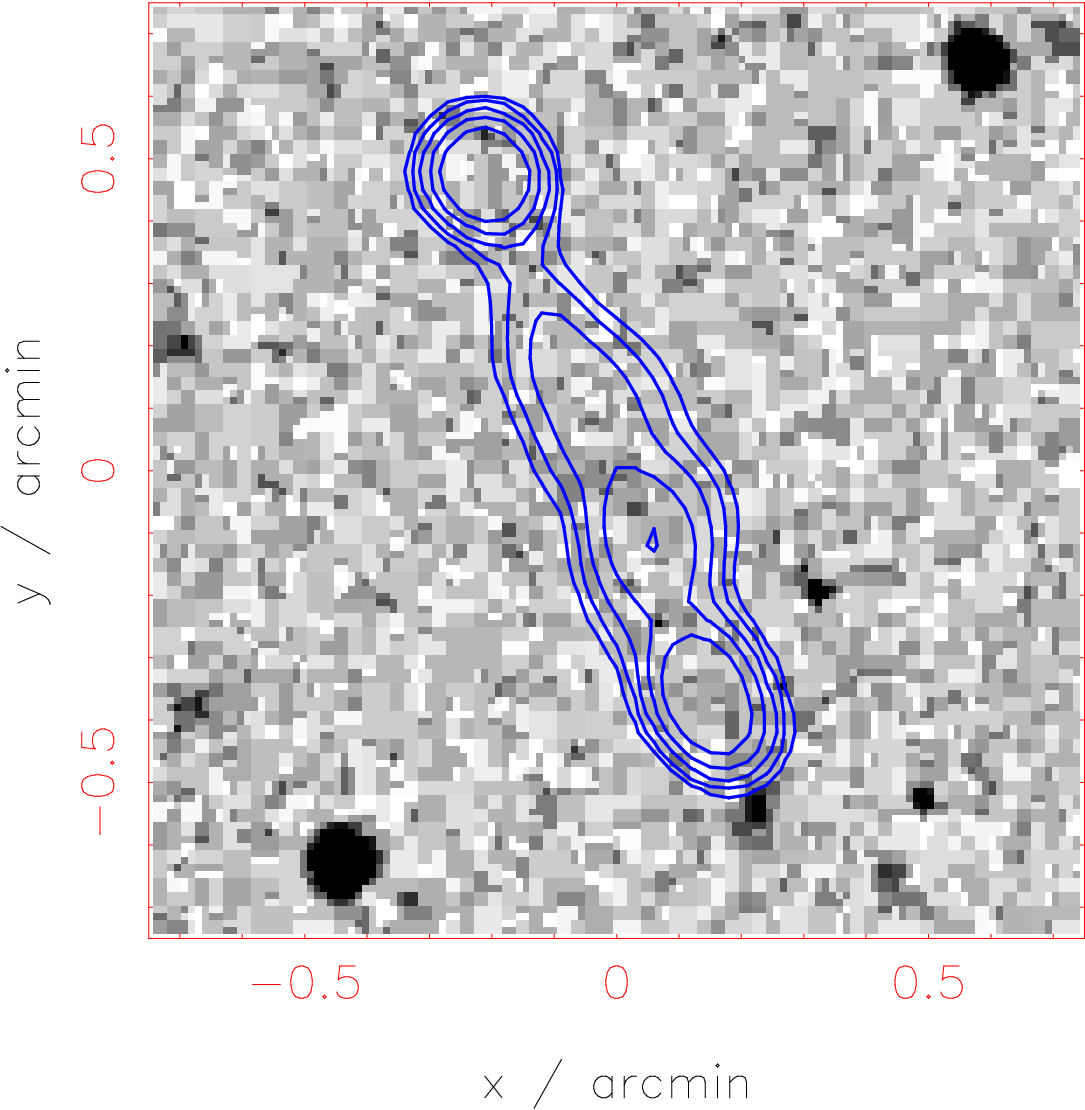}}
    \caption[Typical contour plot of FRII
      source]{\label{contour1}Typical FIRST contour plot of FRI
      source (here, 3C~284) against SSS grey-scale optical
      background. The hot spots are located along the
      jets and more towards the core of the source.\\}
  \end{minipage}
  \vfill
  \vfill
  \begin{minipage}{8.0cm}
    \centerline{
      \includegraphics[angle=270,scale=0.2]{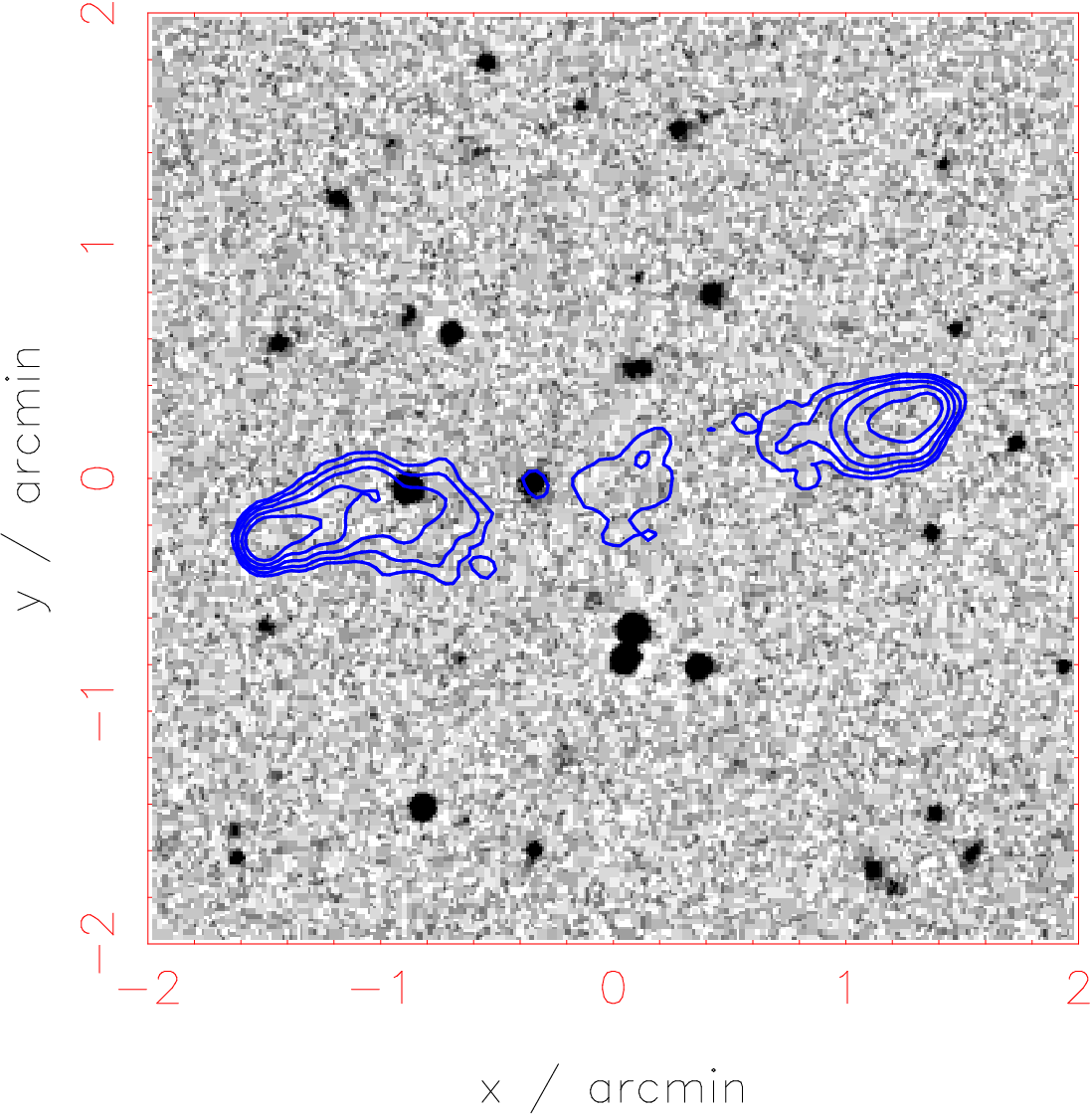}}
    \caption[Typical contour plot of FRI
      source]{\label{contour2}Typical FIRST contour plot of FRII
      source (here, 3C~223) against SSS grey-scale optical
      background. The hot spots are located at the edge of the
      lobes.\\}
  \end{minipage}
\end{figure}

\begin{figure*}
    \caption[]{\label{contsamp}Contour plots from NVSS (outermost contours)
      and FIRST (innermost contours) of sources in the CoNFIG
      sample. The complete set of images can be found in
      Appendix~\ref{ContPlot}.}
    \begin{minipage}{16cm}
      \begin{minipage}{5cm}
	\tiny
	\mbox{}
	\centerline{\includegraphics[scale=0.2,angle=270]{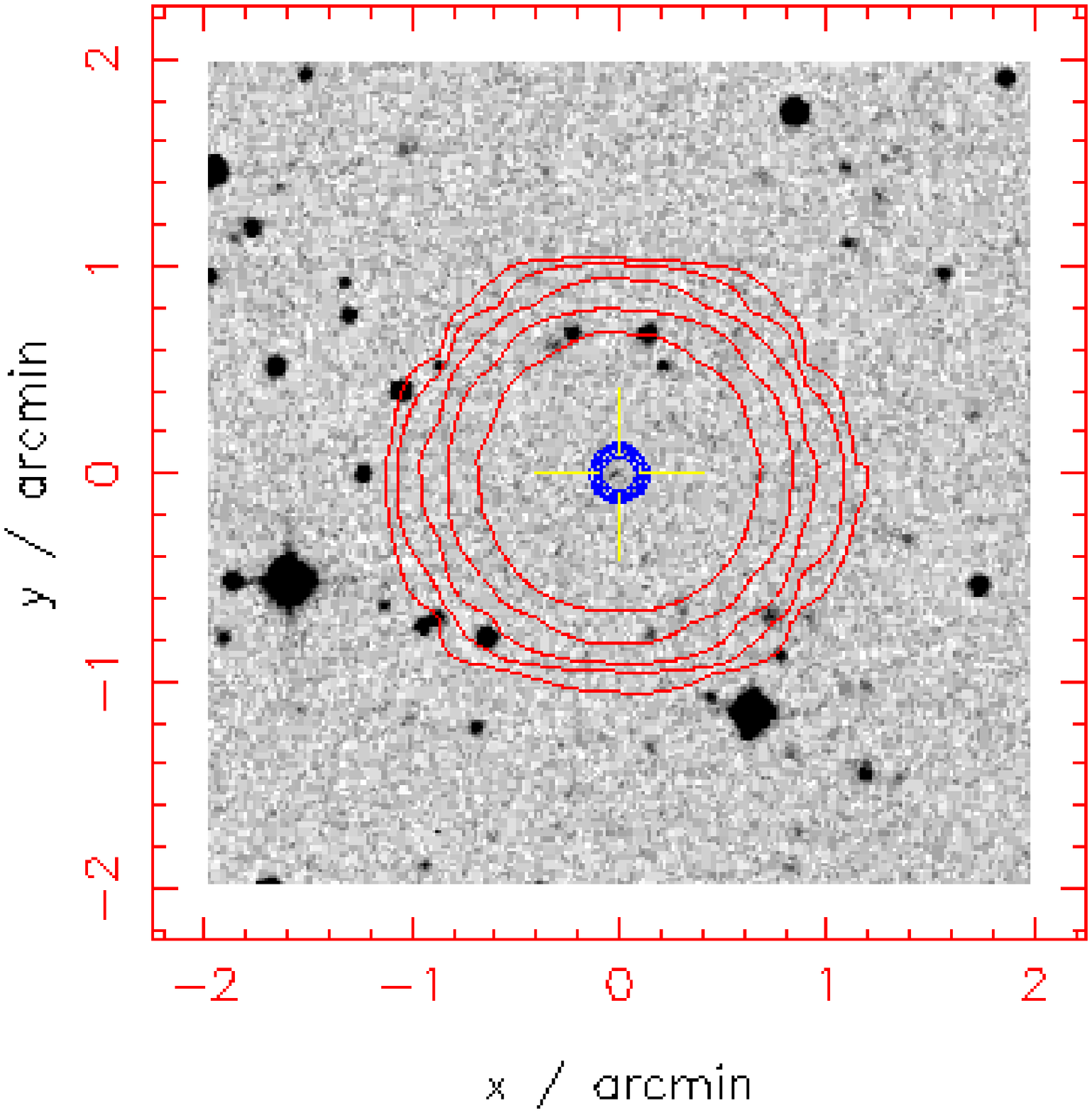}}
	\centerline{CoNFIG-001}
	\centerline{07 13 38.15 +43 49 17.20}
      \end{minipage}
      \hfill
      \begin{minipage}{5cm}
	\mbox{}
	\centerline{\includegraphics[scale=0.2,angle=270]{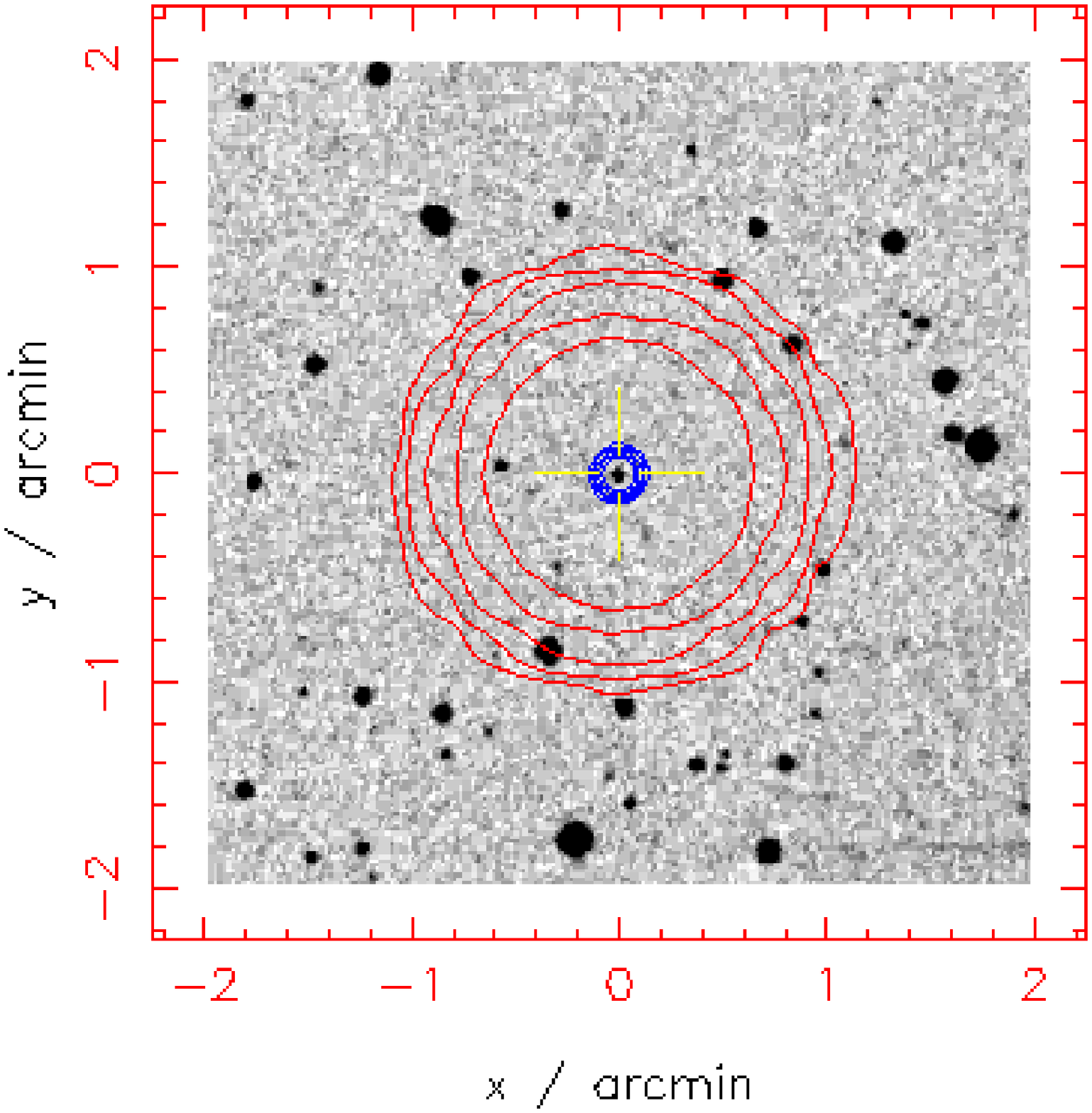}}
	\centerline{CoNFIG-002}
	\centerline{07 14 24.80 +35 34 39.90}
      \end{minipage}
      \hfill
      \begin{minipage}{5cm}
	\mbox{}
	\centerline{\includegraphics[scale=0.2,angle=270]{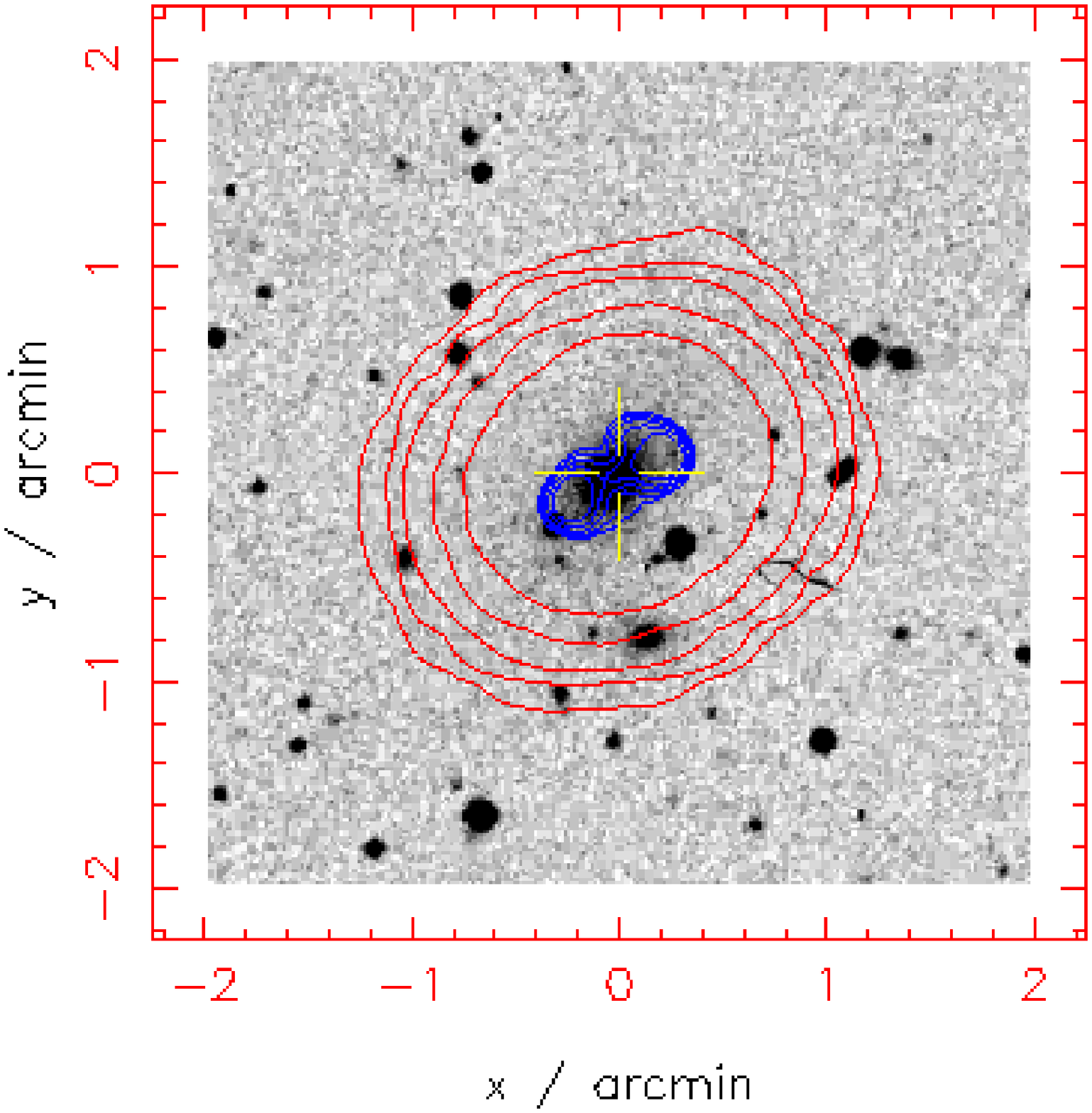}}
	\centerline{CoNFIG-003}
	\centerline{07 16 41.09 +53 23 10.30}
      \end{minipage}
  \end{minipage}
\end{figure*}

76 sources in the sample appear in the 3CRR catalogue \citep{Laing83}
in which the morphology type of each 3CRR 
source has been determined. These
sources are marked by a star next to their 3C name in
Appendix~\ref{details}.\\
\indent
For other sources, the morphology was determined by looking at the FIRST
and NVSS radio contour plots, obtained from the NASA virtual observatory
Skyview (\texttt{http://skyview.gsfc.nasa.gov}). The contour plots are
shown in Figure~\ref{contsamp} and Appendix~\ref{ContPlot}.\\
\indent
If the contour plot displays distinct hot spots at the edge of the lobes
(as in Figure~\ref{contour2}), and the lobes are aligned, the source was
classified as FRII. On the contrary, sources with collimated jets
showing hot spots near the core and jets were classified as FRI
(see Figure~\ref{contour1}). Most irregular-looking sources were also
classified as FRI \citep{Parma92}.\\
\indent
Contour plots were obtained for all but one (CoNFIG-225) of
the CoNFIG sources. 53 sources were identified as extended but the resolution
is too low to confirm their exact morphological
types. Additional VLA observations were obtained for 31 of these
sources (see \S\ref{proposal}) while for the other 22 sources, data
were found in the VLA archives.\\ 
\indent
As specified in \S\ref{samdef}, 7 sources were deleted from
the sample after examination of their contour plots. In one case, the
source is a globular cluster and in another case, the source has no
FIRST or NVSS contours. The other 5 NVSS `sources', when observed with
FIRST, actually consist of separate and independent sources, each with
\SoneG$<$1.3~Jy. The NVSS radio positions of these sources can be found
in Table~\ref{DissSour}. 

%------------------------------ 
%------------------------------
%------------------------------

%%%%%%%%%%%%%%%%%%%% VLA observations %%%%%%%%%%%%%%%%%%%%%%%%%%%%%%%%%
\subsection{VLA Observations}\label{proposal}

Radio observations of 31 extended sources with uncertain morphological
classification were taken at 8~GHz using the VLA
in A configuration on July 29, 2007. The observations used the
standard two IFs of 8435.1 and 8485.1~GHz, with a bandwidth
of 50~MHz. The A configuration is the configuration with the largest
spacing between antennas ($\sim$36 km), providing a synthesized beam of 0.24
arcsec FWHM at 8~GHz.\\
\indent
The 8~GHz flux density of each source was derived from the 1.4~GHz
NVSS flux density and spectral index (see details in \S\ref{lumdist}),
and the exposure time was computed for each source such 
as to provide a signal-to-noise ratio of at least 5. The sources were
placed accordingly into 4 different exposure time groups (5, 10, 20 and 30 min),
except for CoNFIG-227 (60 min) and CoNFIG-257 (40 min). For
observations with a 30 min exposure time, the exposure was split into
two separate 15 min scans to improve \textit{uv} coverage. The primary
calibrator 3C174 (0542+498) was observed at the beginning of the run
while 3C286 (1331+305) was observed twice during the run, once in the
middle and once at the end. These calibrate the flux
density scale assuming a flux density of 4.84~Jy at 8435.1~MHz and
4.81~Jy at 8485.1~MHz for 3C174 and 5.18~Jy at 8435.1~MHz and 4.99~Jy
at 8485.1~MHz for 3C286, based on the scale of \cite{Baars77}. Nearby
secondary calibrators were observed approximately every 30 min to
provide phase calibration. Details of the observations are shown in
Appendix~\ref{VLAobs}.

%------------------------------
\begin{table}
  \caption[Deleted sources]{\label{DissSour}Sources deleted from
  the original sample.}
  \medskip
  \medskip
  \centerline{
    \begin{tabular}{|l|l|}
      \hline
      Source & Reason deleted\\
      \hline
      08 14 43.59 +12 58 10.0 & 2 separate sources\\
      08 30 04.12 +07 45 45.0 & 2 separate sources\\
      10 34 05.09 +11 12 32.1 & 2 separate sources\\
      12 30 35.69 +12 19 18.2 & globular cluster\\
      13 23 02.33 +29 41 34.0 & 3 separate sources\\
      14 17 00.49 +07 10 50.2 & 2 separate sources\\
      15 40 30.42 $-$05 03 17.4 & no contours\\
      \hline
  \end{tabular}}
\end{table}
%------------------------------
\begin{figure*}
  \caption[]{Contour plots from FIRST 1.4GHz (outermost contours) and VLA 8 GHz observations
    (innermost contours). The complete set of images can be found in
    Appendix~\ref{VLAobs}.}
  \begin{minipage}{16cm}
    \begin{minipage}{5cm}
      \tiny
      \mbox{}
      \centerline{\includegraphics[scale=0.2,angle=270]{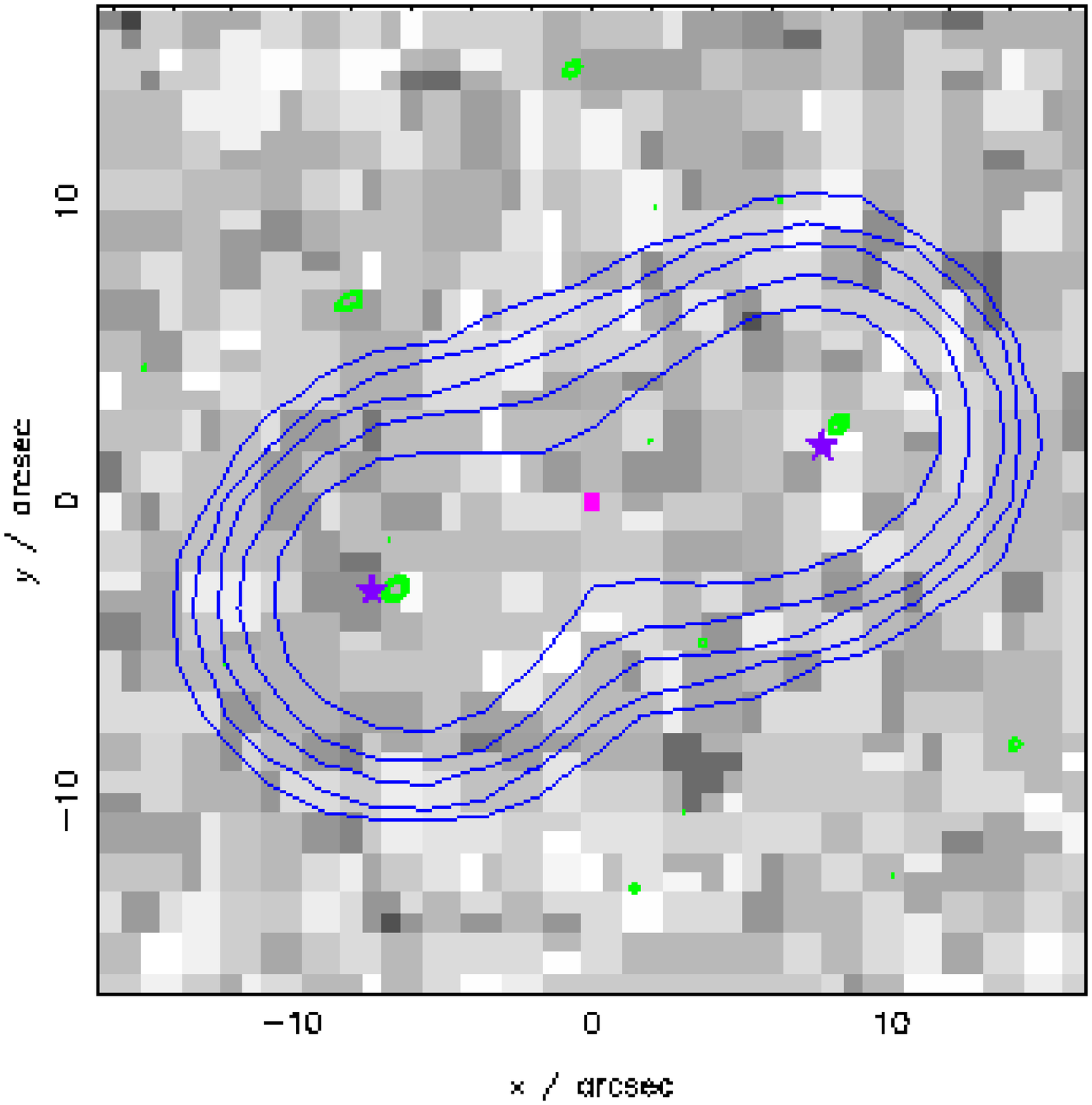}}
      \centerline{CoNFIG-009}
      \centerline{8 GHz, A configuration}
    \end{minipage}
    \hfill
    \begin{minipage}{5cm}
      \mbox{}
      \centerline{\includegraphics[scale=0.2,angle=270]{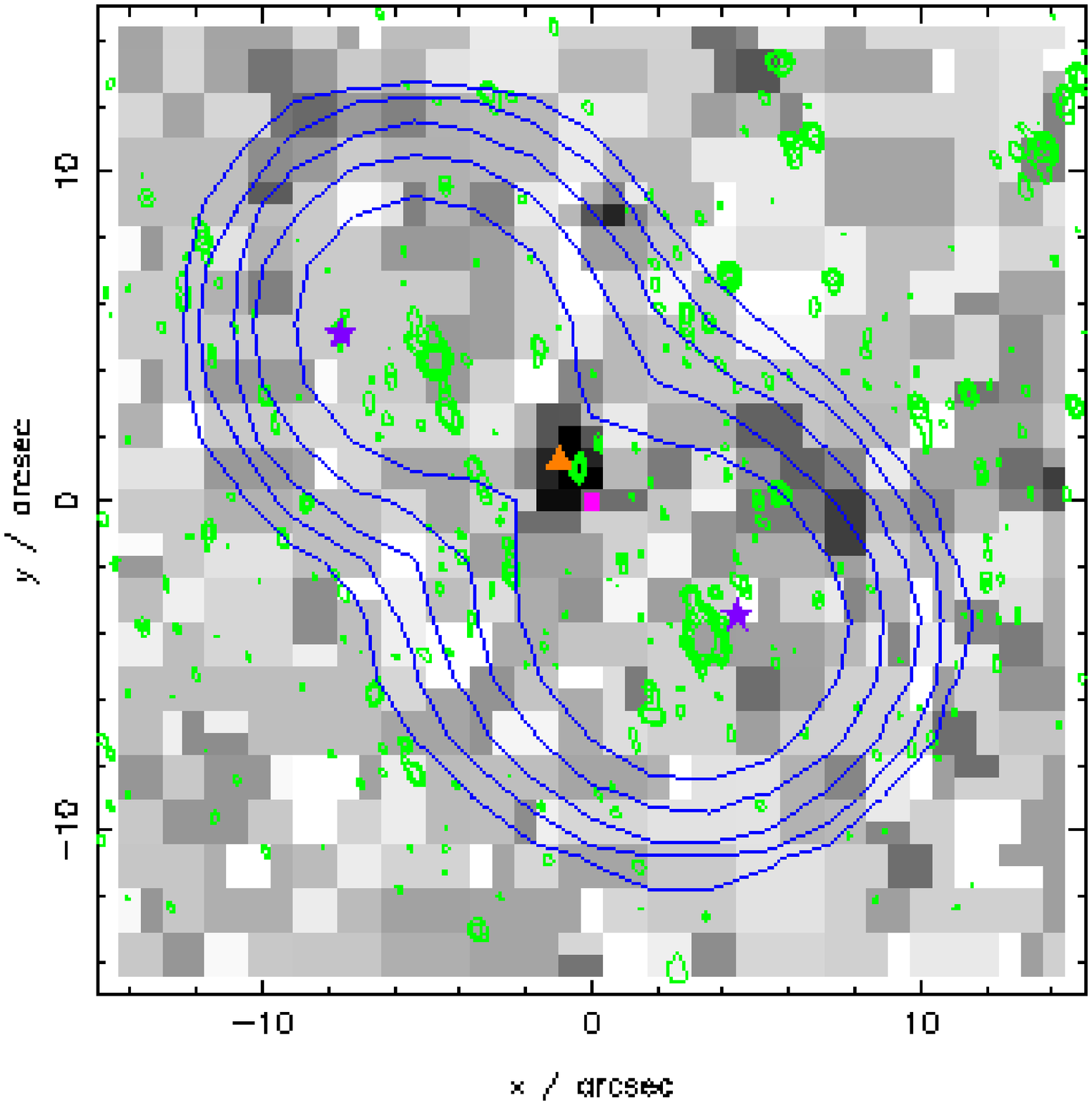}}
      \centerline{CoNFIG-010}
      \centerline{8 GHz, A configuration}
    \end{minipage}
    \begin{minipage}{5cm}
      \mbox{}
    \centerline{\includegraphics[scale=0.2,angle=270]{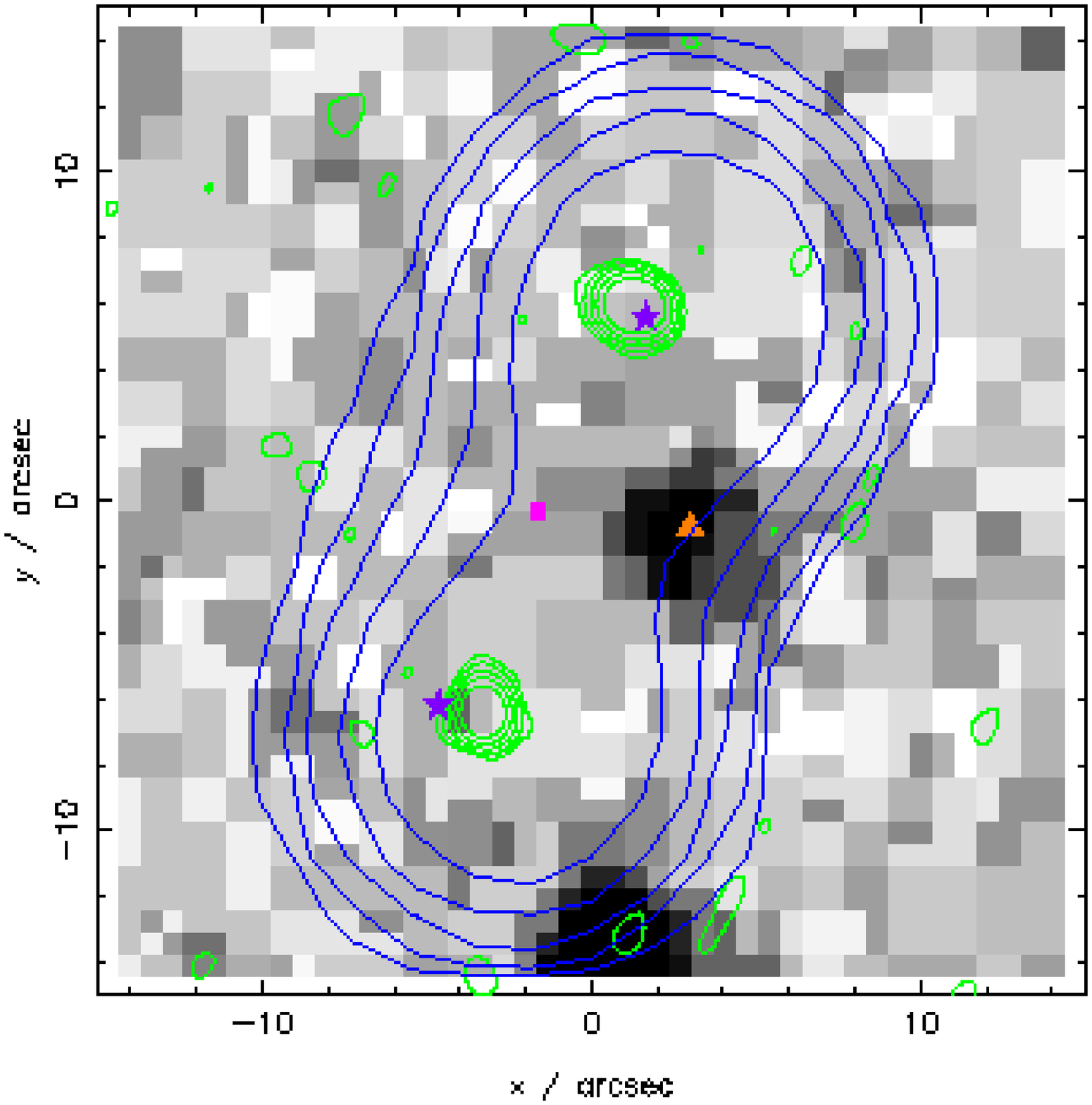}}
    \centerline{CoNFIG-012}
    \centerline{8 GHz, A configuration}
    \end{minipage}
  \end{minipage}
\end{figure*}
%------------------------------

\subsection{Final Classification}\label{final}

Data were extracted from the VLA archives at other frequencies and
configurations for some sources for which the 8~GHz contours were not
satisfactory. All data were reduced using standard procedures
incorporated within the AIPS software provided by NRAO, and the
resulting images are grouped in Appendix~\ref{VLAobs}.\\
\indent
Over 60\% of sources in the CoNFIG sample are classified either as FRI
or FRII. The rest of the sources are classified as compact or
deleted from the sample (see Table~\ref{DissSour}). Some
sources in this group will be unresolved high redshift FRIIs, and
some others are confirmed as truly compact from the VLBA calibrator list
\citep{Beasley02,Fomalont03,Petrov05,Petrov06,Kovalev07} or from the
Pearson-Readhead survey \citep{Pearson88}. Some of these confirmed
compact sources show a steep ($\alpha \le -0.6$) spectral index and
are possible Compact Steep Sources (CSS) \citep{Fanti94}.\\ 
\indent
After looking at optical counterparts from the SuperCOSMOS Sky Survey
(as discussed in \S\ref{IDandz}), two particular subclasses of
compact and FRII sources are identified. \\ 
(1) Some compact sources show no optical
counterpart. These sources can be classified either as CSS or as
unresolved FRII. When confirmed as a compact source from the VLBA
calibrator list
\citep[see][]{Beasley02,Fomalont03,Petrov05,Petrov06,Kovalev07} or the
Pearson-Readhead survey \citep{Pearson88}, the source is classified as
CSS and is included in the compact sources statistics. Otherwise, the
source is assumed to be an unresolved FRII. Their inclusion into the
FRII group has a low impact on the FRII statistics, as seen in
Figure~\ref{TypeLumDist}.\\ 
(2) Other sources are classified as FRII and present a stellar type
optical identification. These sources are - on the basis of the
unified model - steep-spectrum RQSOs, which occur when the line of
sight of the observer is oriented at less than 45$^{\circ}$ with
respect to the jets, enabling the observer to look inside the dusty
torus.\\ 
\indent
The final classification for each source is shown in Table~\ref{details} and
the distribution of morphological types is presented in
Table~\ref{Morphtab}. Errors in flux density measurements are well
defined for point sources in the catalogues, but corresponding error
estimates for the extended sources were too uncertain to be worth
including here.\\ 

%------------------------------
\begin{table}
  \caption[Morphology of CoNFIG sources]{\label{Morphtab}Morphology of
  the sources in the CoNFIG sample. The morphology
  of each source is determined by looking at FIRST and NVSS contour
  plots or from VLA observations as described in \S\ref{FirstID}
  and \S\ref{proposal} respectively.}
  \medskip
  \medskip
  \centerline{
    \begin{tabular}{|ccccc|}
      \hline
      FRI & FRII & Unres. & Comp. & CSS\\
      \hline
      36 & 135 & 22 & 75 & 6\\
      13.1\% & 49.3\% & 8.0\% & 27.4\% & 2.2\%\\
      \hline
  \end{tabular}}
\end{table}
%------------------------------

\subsection{CoNFIG-2, 3 and 4}\label{otherC}

In order to improve the FRI/FRII statistics, three more samples of
about 250 sources each, and complete to 0.8~Jy (CoNFIG-2), 0.2~Jy
(CoNFIG-3) and 50~mJy (CoNFIG-4), were constructed from the NVSS
catalogue over various areas, all included in the FIRST northern
region. For each sample, the morphologies of the sources were
determined as described in \S\ref{FirstID}. Details of the
samples are given in Table~\ref{SCsamples} while the data for the
samples are shown in Tables~\ref{CoNFIG2} and \ref{CoNFIG34}. Optical
identification and redshifts were retrieved for CoNFIG-2 only, as
described in the next section.\\ 
Because the morphology information is not complete for all sources in
the CoNFIG-2, 3 and 4 samples, these objects were only used to compute
the FRI/FRII source count (see \S\ref{compSC}).

%------------------------------
\begin{table}
  \caption[Source count samples]{\label{SCsamples}Details of the extra
  samples constructed to better define the FRI/FRII source counts. The
  total number of sources as well as the number FRI, FRII and unresolved
  extended, compact and Compact Steep Spectrum sources, are shown 
  here.}
  \medskip
  \medskip
  \centerline{
    \begin{tabular}{|lllllll|}
      \hline
       &$S_{lim}$ &  Area &\multicolumn{4}{|c|}{Number of sources}\\
       &\scriptsize{(mJy)} & \scriptsize{(\degsq)} & \scriptsize{FRI} & \scriptsize{FRII} & \scriptsize{C} & \scriptsize{Tot.} \\
      \hline
      CoNFIG-2&800 &2915.25 & 27 &149 &67 &243 \\
      \hline
      CoNFIG-3&200 &\phantom{0}370.00 & 41 & 205 & 45 & 291\\
      \hline
      CoNFIG-4&\phantom{0}50 &\phantom{00}64.00 & 42 &119 &73 &234\\
      \hline
  \end{tabular}}
\end{table}
%------------------------------

%%%%%%% Redshifts & Optical ID %%%%%%%%%%%%%%%%%%%%%%%%%%%%%%%%%%%%%%%%%%%%%%%%%%%%%%%%%%%%%%%%%%%
\section{Optical Identifications and Redshifts}\label{IDandz}

\begin{figure}
  \centerline{
    \includegraphics[scale=0.3,angle=270]{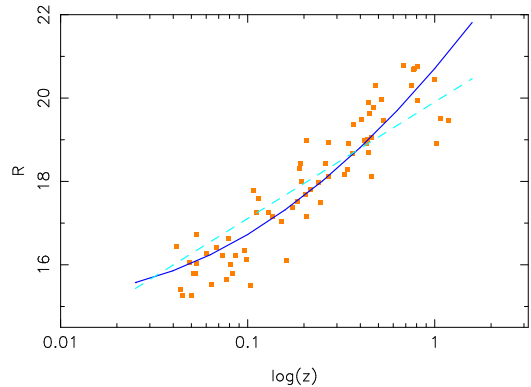}}
  \caption[R-z relation]{\label{Photoz}R-z relation from CoNFIG data
  for extended sources with $15\le R \le 21$. The squares represent
  objects classified as FRI or FRII (excluding unresolved sources)
  with optical ID and spectroscopic redshift information. The solid
  line shows the quadratic fit ($R = 20.70 + 5.24 \log z + 1.28 {(\log
  z)}^2$) from which photometric redshifts are estimated. For
  comparison, the linear fit is also shown (dashed line).}  
\end{figure}

To determine core coordinates in order to retrieve redshifts, optical
identifications (together with $B_j$, R1, R2 and I magnitudes) were
obtained for 79\% of the sources in the CoNFIG sample and 61\% of 
the sources in the CoNFIG-2 sample from the SuperCOSMOS Sky Survey
\citep[SSS,][]{Hambly01}.\\ 
\indent
Redshifts were retrieved for 230 CoNFIG sources and 161 CoNFIG-2 sources,
using the SIMBAD (\texttt{http://simbad.u-strasbg.fr/simbad}) and
NED (\texttt{http://nedwww.ipac.caltech.edu/}) databases.\\
Redshift and magnitude information are listed in Table~\ref{OptIDEx} and
Appendix~\ref{details} for CoNFIG, and in Table~\ref{CoNFIG2} for
CoNFIG~-2.\\ 
When spectroscopic redshifts were not available, we estimated
photometric redshifts via an empirical R - z relation derived from
extended sources in the CoNFIG sample with galaxy identifications and
$15\le R \le 21$, as shown in Figure~\ref{Photoz}. A
simple fit to these data is 
\begin{eqnarray}
R = 20.70 + 5.24 \log z + 1.28 {(\log z)}^2
\end{eqnarray}
Photometric redshift estimate is given to 14 CoNFIG objects falling
into this category, but with no spectroscopic redshift information,
bringing the redshift coverage to 244 sources (89\% of the sample).

%------------------------------
%------------------------------

\begin{figure}
  \centerline{
    \includegraphics[scale=0.4]{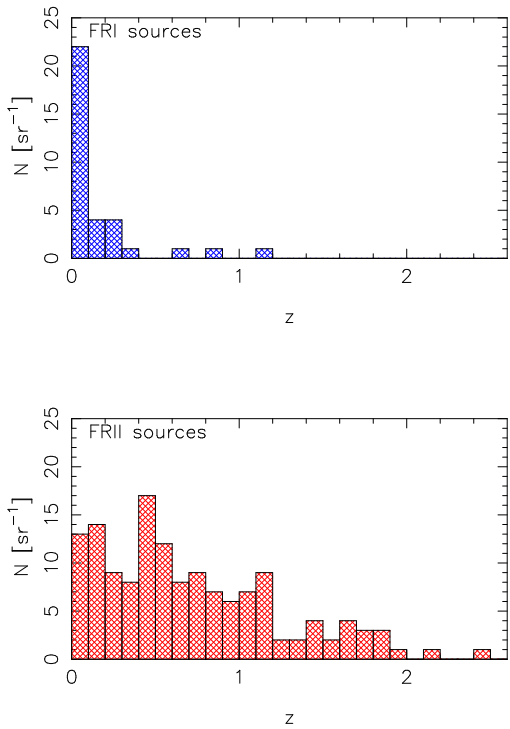}}
  \caption[Morphology dependent redshift
  distribution]{\label{TypeRedDist}Redshift distributions for the
  CoNFIG sample. Details are given in Table~\ref{redtab}.  The FRI
  distribution is concentrated at low redshifts ($z\le 0.5$) while the
  FRII distribution lies over a wider range, up to z$\sim$2.5.} 
\end{figure}

For the CoNFIG sample, redshifts range from z=0.0033 to z=3.522 with a
mean redshift of z=0.715 and a median redshift of z=0.580. The
morphology-dependent distribution is shown in Figure~\ref{TypeRedDist}
and details of the redshift distributions are given in
Tables~\ref{redtab} and \ref{redtab2}. The FRI distribution is
concentrated at low redshifts ($z\le 0.5$) while the FRII distribution
covers a wider range, up to z=2.5.\\ 
Note that for 38 CoNFIG and 34 CoNFIG-2 sources, redshift information
was retrieved from the literature using the radio position of the
object, but no counterpart was found in SSS.

%------------------------
\begin{table}
  \caption[Redshift completeness]{\label{redtab} Redshift
  completeness in the CoNFIG and CoNFIG-2 samples. Redshift information was
  retrieved from the SIMBAD database. CSS and unresolved sources from
  Table~\ref{Morphtab} are regrouped with Compact and FRII sources
  respectively.} 
  \medskip
  \medskip
  \centerline{
    \begin{tabular}{|c|ccc|}
      \hline
      \multicolumn{4}{|c|}{CoNFIG} \\
      & FRI & FRII & Compact\\
      \hline
      with z & 34 & 142 & 68\\
      \% of group with z & 94.4\% & 90.4\% &83.9\%\\
      \% of sample with z & 12.4\% & 51.8\% & 24.8\%\\
      \hline
      \hline
      \multicolumn{4}{|c|}{CoNFIG-2} \\
      & FRI & FRII & Compact\\
      \hline
      with z & 22 & 97 &52 \\
      \% of group with z & 81.5\% & 65.1\% &77.6\%\\
      \% of sample with z & 9.0\% & 39.9\% & 21.4\%\\
      \hline
  \end{tabular}}
\end{table}
%------------------------------
\begin{table}
  \caption[Redshift statistics]{\label{redtab2}Redshift
  distribution information for the CoNFIG sample.}
  \medskip
  \medskip
  \small
  \centerline{
    \begin{tabular}{|l|c|cc|cc|}
      \hline
      Type& \% of group & \multicolumn{4}{|c|}{redshift:} \\
       & with ID  & min & max & mean & med. \\
      \hline
      all & 79.6\% & 0.0034  & 3.522  &0.715  & 0.580 \\ 
      FRI & 97.2\%& 0.0034  & 1.191 & 0.157  & 0.053 \\ 
      FRII & 75.1\%& 0.0456  & 2.474  & 0.712  & 0.572 \\ 
      Comp. & 80.2\% & 0.0336 & 3.5220 & 1.001 & 0.909 \\
      \hline
      \normalsize
  \end{tabular}}
\end{table}
%------------------------------  EXAMPLE OF THE OPTICAL ID TABLE
 \begin{table*}
   \centering
   \small
   \caption[Optical identification table (Example)]{\label{OptIDEx}Optical
   identification table for CoNFIG (example). The complete table can be found in
   Table \ref{OptID}.}
   \medskip
  \begin{tabular}{|c|l|c|l|cccc|}
    \hline
    ID & \footnotesize{Rad.} & Optical Position & \footnotesize{Opt.}
    & $B_j$ & R1 & R2 & I \\
    & \footnotesize{Type}&\footnotesize{(J2000)} & \footnotesize{Type} & & & & \\
    \hline
  1 & C* & 07 13 38.15 +43 49 17.20 & g &  21.79 &  19.79 &  19.93 &  18.62\\
  2 & C* & 07 14 24.80 +35 34 39.90 & s &  18.74 &  17.59 &  18.42 &  17.50\\
  3 & II & 07 16 41.09 +53 23 10.30 & g &  15.43 &  15.54 &  14.22 &  13.30\\
  4 & C  & 07 35 55.54 +33 07 09.60 & s &  21.22 &        &  20.34 &  18.55\\
  5 & C* & 07 41 10.70 +31 12 00.40 & s &  16.52 &  16.28 &  16.32 &  15.83\\
  6 & II & 07 45 42.13 +31 42 52.60 & s &  15.76 &  15.55 &  15.56 &  15.11\\
  8 & I  & 07 58 28.60 +37 47 13.80 & g &  14.51 &  15.42 &  14.06 &  12.23\\
 10 & II & 08 01 35.32 +50 09 43.00 & s &  21.79 &        &  20.17 &       \\
 11 & II & 08 05 31.31 +24 10 21.30 & g &  17.28 &  16.28 &  15.60 &  15.16\\
 12 & II & 08 10 03.67 +42 28 04.00 & g &  21.08 &  19.47 &  18.91 &  17.38\\
 13 & II & 08 12 59.48 +32 43 05.60 & g &  21.19 &  19.76 &  19.19 &  18.64\\
 14 & II & 08 13 36.07 +48 13 01.90 & s &  18.45 &  17.79 &  17.82 &  17.54\\
 16 & II & 08 19 47.55 +52 32 29.50 & g &  20.37 &  18.31 &  18.26 &  17.80\\
 17 & II & 08 21 33.77 +47 02 35.70 & g &  18.55 &  17.26 &  17.07 &  16.26\\
 \hline
  \end{tabular}
 \end{table*}
%------------------------------
%------------------------  EXAMPLE OF THE DATA TABLE

\begin{sidewaystable*}[p]
Data table for CoNFIG (example). The complete table can be found in Appendix~\ref{details}.
  \centering
  \small
  \begin{tabular}{ll|ll|l|l|llllll|l|ll}
    CoNFIG & Name& RA & DEC & Type & Red. & \SoneM &\SthreeM & \SfourM &
    \SoneG & \StwoG & \SfiveG & $\alpha$ & \multicolumn{2}{|c|}{Ref. and}\\
    ID & & \multicolumn{2}{|c|}{(J2000)} & & & (Jy) & (mJy) & (Jy)& (mJy) & (Jy) & (Jy) & & \multicolumn{2}{|c|}{Comments}\\
    \hline
  1 & QSO B0710+439   & 07 13 38.15 & +43 49 17.20 & C*  &0.5180 &        &   \phantom{00}655.0 &   \phantom{00}0.75 &   \phantom{00}2011.4 &         &   \phantom{0}1.60 &  \phantom{$-$}0.82$^{\dag}$& 1 & \\    
  2 & QSO B0711+35    & 07 14 24.80 & +35 34 39.90 & C*  &1.6260 &        &   \phantom{00}840.0 &        &   \phantom{00}1467.1 &         &   \phantom{0}0.89 &  \phantom{$-$}0.41$^{\dag}$& 1 & \\    
  3 & 4C 53.16        & 07 16 41.09 & +53 23 10.30 & II  &0.0643 &    \phantom{000}5.4 &  \phantom{0}3913.0 &        &   \phantom{00}1501.4 &         &   \phantom{0}0.63 & $-$0.63& 2 & \\    
  4 & 4C 33.21        & 07 35 55.54 & +33 07 09.60 & C   &       &    \phantom{000}7.4 &  \phantom{0}6750.0 &        &   \phantom{00}2473.1 &         &   \phantom{0}0.92 & $-$0.56&   & \\    
  5 & QSO J0741+3111  & 07 41 10.70 & +31 12 00.40 & C*  &0.6300 &        &  \phantom{0}1365.0 &        &   \phantom{00}2284.3 &         &   \phantom{0}2.68 &  \phantom{$-$}0.38& 3 & \\    
  6 & 4C 31.30        & 07 45 42.13 & +31 42 52.60 & II  &0.4620 &    \phantom{000}4.4 &  \phantom{0}2448.0 &        &   \phantom{00}1357.8 &         &   \phantom{0}1.02 & $-$0.55& 3 & S\\   
  7 & 4C 56.16        & 07 49 48.10 & +55 54 21.00 & I   &0.0360 &    \phantom{000}4.2 &  \phantom{0}2636.0 &        &   \phantom{00}1660.4 &         &                   & $-$0.44& 4 & N\\   
  8 & NGC 2484        & 07 58 28.60 & +37 47 13.80 & I   &0.0438 &   \phantom{00}11.3 &         &   \phantom{00}5.39 &   \phantom{00}2717.9 &         &   \phantom{0}1.33 & $-$0.68& 5 & \\    
  9 & 4C 37.21        & 07 59 47.26 & +37 38 50.20 & II  &       &        &  \phantom{0}5374.0 &   \phantom{00}4.54 &   \phantom{00}1691.2 &         &   \phantom{0}0.46 & $-$0.84&  & NV\\   
 10 & TXS 0757+503    & 08 01 35.32 & +50 09 43.00 & II  &       &        &  \phantom{0}5810.0 &        &   \phantom{00}1471.7 &         &   \phantom{0}0.47 & $-$1.02&  & SV\\   
 11 & 3C 192*         & 08 05 31.31 & +24 10 21.30 & II  &0.0600 &   \phantom{00}21.0 &         &  \phantom{0}13.10 &   \phantom{00}5330.6 &    \phantom{00}3.30 &   \phantom{0}2.60 & $-$0.67& 3 & \\    
 12 & 3C 194          & 08 10 03.67 & +42 28 04.00 & II  &1.1840 &    \phantom{000}9.9 &  \phantom{0}6734.0 &   \phantom{00}5.68 &   \phantom{00}2056.6 &    \phantom{00}1.07 &   \phantom{0}0.61 & $-$0.78& 6 & A\\   
 13 & 4C 32.24        & 08 12 59.48 & +32 43 05.60 & II  &0.4700 &    \phantom{000}5.1 &  \phantom{0}4266.0 &        &   \phantom{00}1522.5 &         &   \phantom{0}0.47 & $-$0.68& 7 & \\    
 14 & 3C 196*         & 08 13 36.07 & +48 13 01.90 & II  &0.8710 &   \phantom{00}68.2 & 49023.0 &        &  \phantom{0}15010.0 &    \phantom{00}7.66 &   \phantom{0}4.36 & $-$0.75& 3 & S\\   
 16 & 4C 52.18        & 08 19 47.55 & +52 32 29.50 & II  &0.1890 &    \phantom{000}7.2 &  \phantom{0}6037.0 &        &   \phantom{00}2104.2 &         &   \phantom{0}0.80 & $-$0.62& 8 & V\footnotemark[1]\\
 17 & 3C 197.1        & 08 21 33.77 & +47 02 35.70 & II  &0.1300 &    \phantom{000}8.1 &  \phantom{0}5534.0 &   \phantom{00}4.68 &   \phantom{00}1787.1 &    \phantom{00}1.16 &   \phantom{0}0.86 & $-$0.75& 8 & A\\   
 18 & 4C 17.44        & 08 21 44.02 & +17 48 20.50 & C   &       &    \phantom{000}5.8 &  \phantom{0}4566.0 &   \phantom{00}4.56 &   \phantom{00}1875.1 &    \phantom{00}1.11 &   \phantom{0}0.68 & $-$0.57&  & \\	    
 19 & 4C 22.21        & 08 23 24.72 & +22 23 03.70 & C*  &0.9510 &    \phantom{000}4.5 &  \phantom{0}4086.0 &        &   \phantom{00}2272.4 &         &   \phantom{0}1.59 & $-$0.34& 9 & \\    
 20 & 4C 56.16A       & 08 24 47.27 & +55 52 42.60 & C*  &1.4170 &        &  \phantom{0}2028.0 &        &   \phantom{00}1449.4 &         &   \phantom{0}1.20 & $-$0.25& 10 & \\  
\hline
  \end{tabular}
\end{sidewaystable*}
%------------------------------ EXAMPLE OF THE CoNFIG-2 DATA
 \begin{table*}
   \centering
   \small
   \caption[]{\label{C234}Data for CoNFIG-2 sample (example). The complete table can be found in
   Table~\ref{CoNFIG2}. Data for CoNFIG-3 and CoNFIG-4 can be found in
   Table~\ref{CoNFIG34}}
   \medskip
  \begin{tabular}{|ll|l|l|ll|lr|}
    \hline
    RA & DEC & Radio & \SoneG & B-mag. & Opt. & z & Ref.\\
    \multicolumn{2}{|c|}{(J2000)} & Type &(mJy) &  &Type & & \\
    \hline
    09 20 11.16 & +17 53 25.00 & S  &     \phantom{00}1070.6 &       &   &       &   \\
    09 20 58.48 & +44 41 53.70 & C* &     \phantom{00}1017.2 & 17.20 & 2 & 2.180 &  2\\
    09 21 07.54 & +45 38 45.70 & II &     \phantom{00}8101.6 & 18.61 & 1 & 0.174 &  1\\
    09 21 47.05 & +37 54 16.10 & II &     \phantom{000}826.4 & 20.24 & 2 & 1.108 &  3\\
    09 22 49.93 & +53 02 21.20 & S  &     \phantom{00}1597.8 &       &   &       &   \\
    09 27 03.04 & +39 02 20.70 & C* &     \phantom{00}2884.6 & 17.06 & 2 & 0.698 &  1\\
    09 30 33.45 & +36 01 23.60 & II &     \phantom{00}1875.1 & 18.98 & 2 & 1.157 &  1\\
    09 30 54.27 & +58 55 16.60 & II &     \phantom{00}1082.9 &       &   &       &   \\
    09 34 15.80 & +49 08 21.00 & C* &     \phantom{000}800.5 &       &   & 2.582 &  2\\
    09 35 04.06 & +08 41 37.30 & S  &     \phantom{00}1037.6 &       &   &       &   \\
    09 35 06.62 & +39 42 07.60 & I  &     \phantom{00}1029.5 &       &   &       &   \\
    09 36 32.02 & +04 22 10.80 & S  &     \phantom{000}971.1 &       &   & 1.340 &  4\\
    09 39 50.20 & +35 55 53.10 & II &     \phantom{00}3719.0 & 18.59 & 1 & 0.137 &  1\\
    \hline
  \end{tabular}
 \end{table*}
%------------------------------
%------------------------------

%%%%%%%%%%%%%%%%%%%% Distributions %%%%%%%%%%%%%%%%%%%%%%%%%%%%%%%%%%%%%

\section{Luminosity Distributions}\label{lumdist}

In order to compute the radio luminosity, the spectral
index $\alpha$ (defined as $S_{\nu} \propto \nu^{\alpha}$) of
each source needs to be determined. To achieve this, flux densities at
different frequencies for each source were compiled and the spectral
index computed following the relation: 
\begin{eqnarray}
  \alpha=\frac{\Delta \log(S)}{\Delta \log(\nu)}
\end{eqnarray}
A summary of the different frequencies and corresponding surveys used
to retrieve the flux density information is given in
Table~\ref{SurveyTab}.\\ 
We made independent estimates for the low- and high-frequency spectral indices
(with $178~{\rm MHz} \le \nu \le 1.4~{\rm GHz}$ and $1.4~{\rm
  GHz} \le \nu \le 5~{\rm GHz}$ respectively). The low frequency
spectral indices are used to compute the luminosities; since
$\nu_{rest}=\nu_{obs}(z+1)$, the luminosity emitted at
$\nu_{rest}=1.4~{\rm GHz}$ will correspond to an observed flux at frequency
$\nu_{obs}< 1.4~{\rm GHz}$.\\

\begin{figure}
  \centerline{
    \includegraphics[angle=270,scale=0.35]{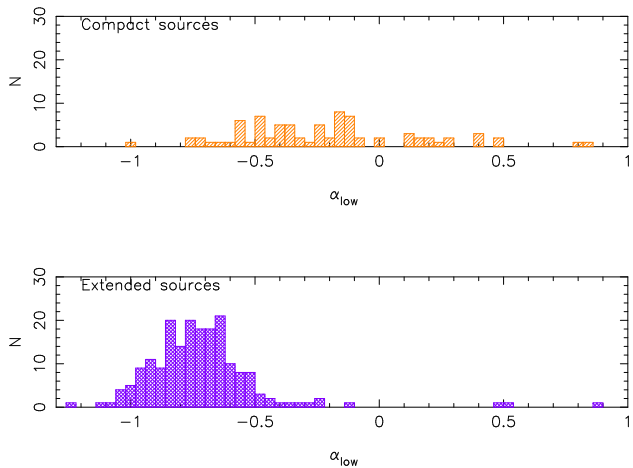}}
  \caption[Spectral index distribution]{\label{SpecDist}Spectral index
    distribution (using frequencies $178~{\rm MHz} \le \nu \le
    1.4~{\rm GHz}$). The distribution peaks around -0.7, which is the
    usual value for extended sources, but has a long tail to
    flat/inverted spectra.}
\end{figure}

\begin{figure}
  \centerline{
    \includegraphics[scale=0.4]{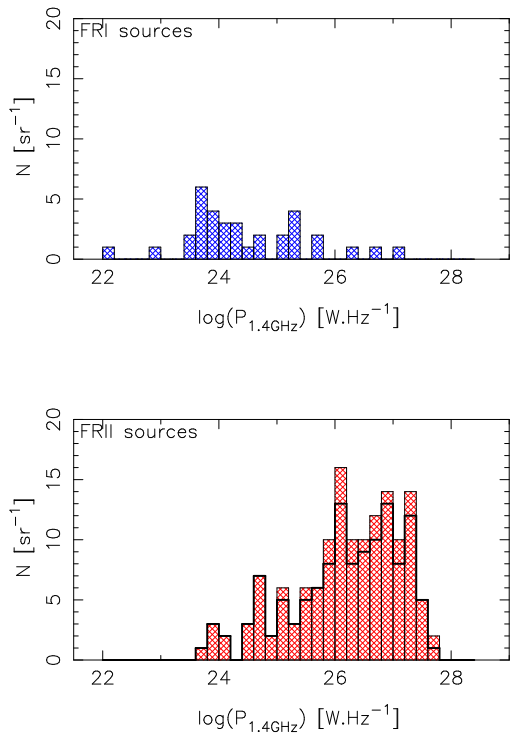}}
  \caption[Luminosity distribution by source
  type]{\label{TypeLumDist}Luminosity distributions for FRI and FRII
  sources. The solid-line histogram in the FRII luminosity
  distribution plot corresponds to the distribution we would get if the
  unresolved sources were excluded from the FRII grouping.}
\end{figure}

\indent
Values for spectral indices are listed in Table~\ref{details}. The
median value is $\alpha=-0.67$. However, the extended sources are
grouped around a median value of $\alpha=-0.73$ as seen in
Figure~\ref{SpecDist}, while the compact sources, as is well known,
show a broader distribution.\\
\indent
Figure~\ref{TypeLumDist} presents the luminosity distributions for FRI
and FRII objects. The distribution of FRII sources only (excluding
unresolved sources) is shown as a solid-line distribution on the
plot. As stated in \S\ref{final}, the inclusion of the unresolved
sources with the FRII sources has very little effect on the structure
of the distribution. 

%------------------------------
%------------------------------
%------------------------------
%------------------------------
%------------------------------
\begin{table}
  \caption[Surveys used to retrieve flux density
  information]{\label{SurveyTab}Surveys used to retrieve
  flux density information.}
  \medskip
  \medskip
  \centerline{
    \begin{tabular}{|lll|}
      \hline
      Frequency & Survey & Reference \\
      \hline
      178~MHz & 3C & \cite{KPW69}\\
      & 4C &\cite{Pil65}\\
      &&\\
      365~MHz &Texas & \cite{Douglas96}\\
      &&\\
      408~MHz &Parkes & \cite{Wright90}\\
      & B3 &\cite{Ficarra85}\\
      &&\\
      2.7~GHz & 3C & \cite{KPW69}\\
      & Parkes &\cite{Wright90}\\
      &&\\
      5.0~GHz & 3C &\cite{KPW69}\\
      & Parkes &\cite{Wright90}\\
      & MIT-Greenbank &\cite{Bennett86}\\
      \hline
  \end{tabular}}
\end{table}
%------------------------------

%%%%%%%%%%%%%%%%% Source Count %%%%%%%%%%%%%%%%%%%%%%%%%%%%%%%%%%%%%%%%%%%%%%%%
\section{FRI/FRII Source Counts}\label{extSC}  %%%%%%%% FRI/FRII SC %%%%%%%%%%%%%%%%%%

One of the challenges in modelling the space densities of FRI and FRII
sources is to compute an accurate source count for each morphological
type. This implies determining the morphological type of
all sources used.\\ 
\indent
To increase the number of sources with morphology information, five
samples at different flux density limits were combined with the CoNFIG
sample. These samples include the CENSORS \citep{Best02}and BDFL
\citep{BDFL} samples, as well as the 3 extra CoNFIG samples (see
details in \S\ref{otherC} and Table~\ref{SCsamples}). The
morphologies of the sources in each sample were determined by looking
at the FIRST and NVSS contour plots as described in \S\ref{morpho},
(with the exception of the CENSORS sources and some of the BDFL
sources, as described in the next section).  

\subsection{The CENSORS and BDFL samples}\label{othersamp}

\subsubsection{The CENSORS sample}\label{CENSORS}  %%%%%%%% CENSORS %%%%%%%%%%%%%%

The CENSORS (Combined EIS-NVSS Survey Of Radio Sources) sample
\citep{Best02,Brookes05,PaperIII} contains 150 sources complete to
\SoneG=7.2~mJy, selected from NVSS over the ESO Imaging Survey (EIS)
Patch D. The sample has a median flux density of \SoneG$\sim$15~mJy and
optical identifications for 68\% of the sources.\\
\indent
The wide field EIS comprises a relatively wide-angle survey of four
distinct patches of sky up to 6 \degsq each \citep{Nonino99}. Patch D
is the most northerly, with a limiting magnitude of I$\sim$23
\citep{Ben99} and an area of 2$\times$3 \degsq centred on $09^h 51^m
36.0^s $-$21^{\circ} 00' 00''$ (J2000).\\
\indent
The goals of the CENSORS sample \citep{Best02} are to constrain evolution of the top
end of the black hole mass function, study the environments around
radio sources at different luminosities, test the K-z relation for
radio sources and test the dual-population models.\\

\begin{figure}
  \centerline{
    \includegraphics[scale=0.35,angle=270]{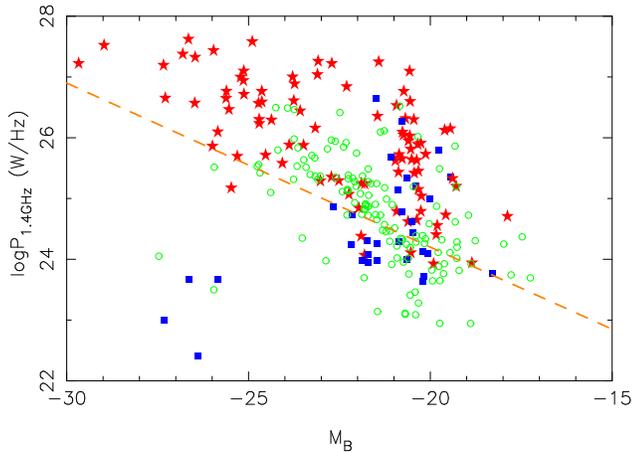}}
  \caption[FR sep]{\label{LedOw} Radio power-optical B-magnitude relation for
  the CoNFIG and CoNFIG-2 sources. FRIs (filled squares) seem to take
  over FRIIs (stars) below the dash line
  $logP_{1.4~GHz}=-0.27M_B+18.8$. This relation is used to determine
  the FRI/FRII classification for non-QSO sources from CENSORS (open
  circles).} 
\end{figure}

Little classification of the CENSORS sources has been done, although
\cite{PaperIII} presented a list of the possible RQSOs. To attempt a
preliminary morphological classification, we used the NVSS images to
determine the  morphology of the  sources; in most cases though, the
resolution is far too low to determine the morphology of the  extended
sources. In the case of the CoNFIG sample, this was solved by using
FIRST images together with supplementary VLA observations. However,
the CENSORS sample does not overlap with the FIRST survey, or any
other higher resolution samples.   \cite{Ledlow96} showed that FRI and
FRII sources are separated on a radio power-optical magnitude
diagram. Following this idea, we used radio flux densities and
B-magnitude data from CoNFIG and CoNFIG-2 to determine a line in the
radio power-optical B-magnitude diagram below which FRI morphology
dominates. The line is defined as (as shown in Figure~\ref{LedOw}):
\begin{eqnarray}
logP_{1.4~GHz}=-0.27M_B+18.8
\end{eqnarray}
This relation was then applied to the 136 sources from CENSORS
using the flux density and magnitude information from \cite{Best02},
using a K-correction value of $\rm K_{corr}=1.122\ z$. As a result, 49
sources were classified as FRI and 87 as FRII. Note that, for 63
CENSORS sources, B-magnitude information were not available and a
B-z relation similar to the R-z relation defined in \S\ref{IDandz} was
then used to determine $M_B$. 

%------------------------------
%------------------------------

\subsubsection{The BDFL sample}\label{BDFLs}

The BDFL sample \citep{BDFL} contains 424 sources and is complete to
\SoneG$\ge$1.7~Jy in the area of sky $-5^{\circ}<\delta <+70^{\circ}$,
$|b|>5^{\circ}$.\\
\indent
To improve definition of the high flux-density ends of the
morphological source counts, sources from this sample with
\SoneG$\ge$4.0~Jy were selected, yielding a total number of 90
sources. The morphology of each source was determined from either
\cite{Laing83} or \cite{Kharb04}, or by looking at the NVSS
contours. As a result, of the 90 sources, 21 sources are classified as
FRI, 38 as FRII and 31 as compact.

%%%%%%%%%%%%%%%%%%%%%%%%%%%%%%%%%%%%%%%%%%%%%%%%%%%%
\subsection{Compiling the Source Count}\label{compSC}  

\begin{figure}
  \centerline{
    \includegraphics[scale=0.35,angle=270]{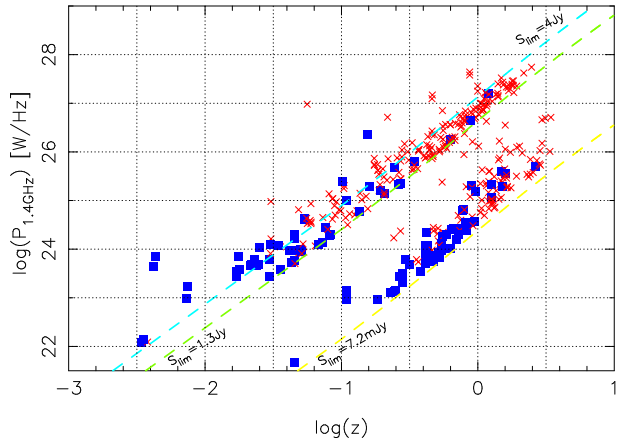}}
  \caption[P-z plane for CoNFIG sources]{\label{PzPlane}P-z plane for
  the CoNFIG sources. FRIs and FRIIS are represented by squares
  and crosses respectively. The flux-density limit lines for the BDFL
  (down to 4.0~Jy), CENSORS (at 7.2~mJy) and CoNFIG (at 1.3~Jy) samples
  are represented as dashed lines respectively. The plane is divided
  into sections of $\Delta log(P)=1$ and $\Delta log(z)=0.5$ to
  illustrate the poor coverage of some regions.} 
\end{figure}

In total, there are $\sim$500 sources from the combined BDFL (down to
4.0~Jy), CENSORS (at 7.2~mJy) and CoNFIG (at 1.3~Jy) samples. Still, as
illustrated in Figure~\ref{PzPlane}, large areas of the P-z plane are
not covered.

%------------------------------
%------------------------------

\begin{figure}
  \centerline{
    \includegraphics[scale=0.35,angle=270]{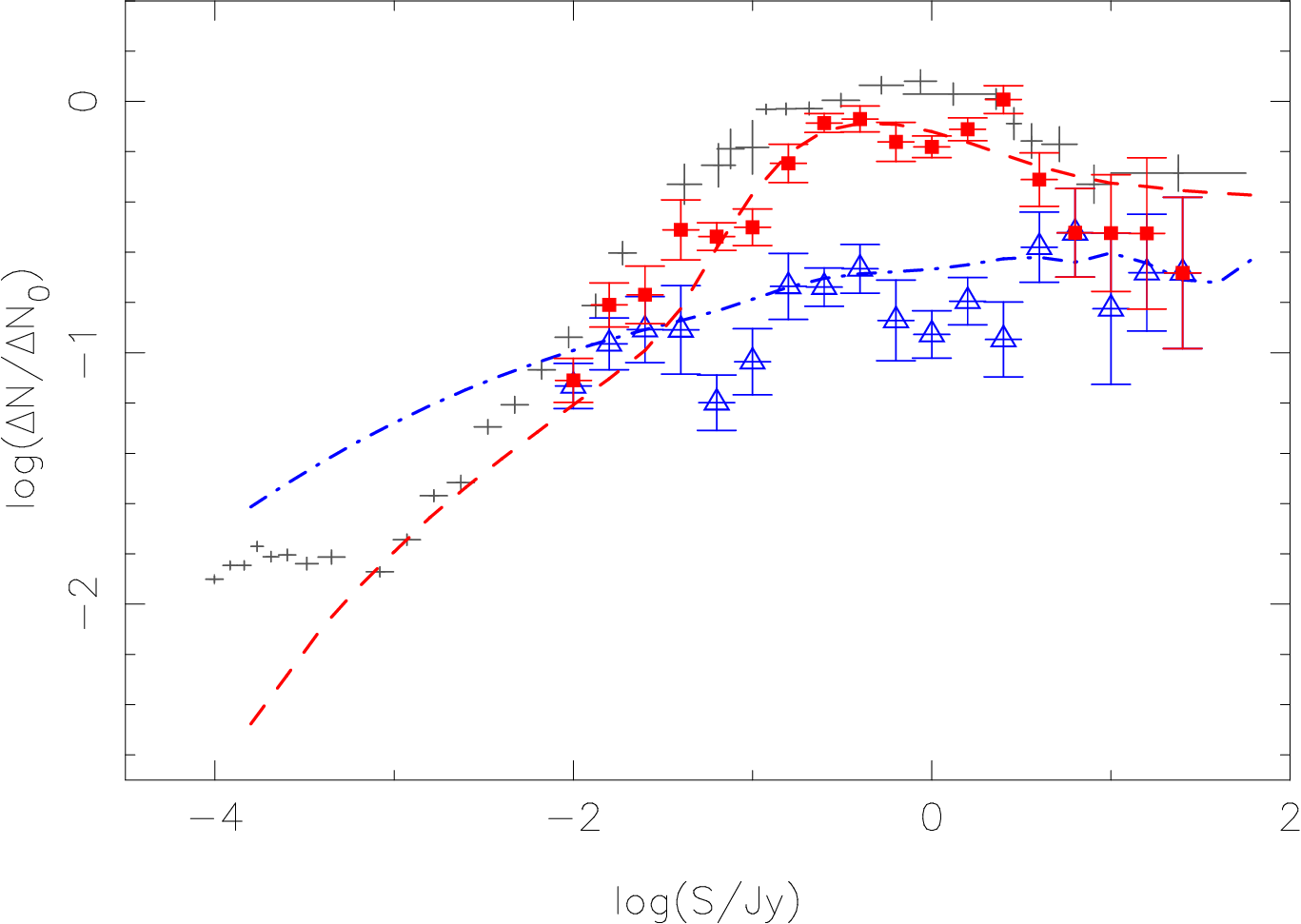}}
  \caption[Relative differential source count for FRI and FRII
  sources]{\label{RelSC}Relative differential source count 
  for FRI (triangles) and FRII (squares) sources. A 1.4~GHz source
  count, compiled from the data of \cite{BDFL}, \cite{Machalski78},
  \cite{Hopkins03} and \cite{Prandoni01}, is represented by the grey
  crosses for comparison. Here, $\Delta N_0 = 200 \Delta (S^{-1.5})$ 
  and the error bars correspond to $\sqrt{N}$ where N is the number of
  objects in each bin. The counts are fitted by an illustrative model
  (FRI: dot-dashed line, FRII: dash line), the details of which are described in
  \S\ref{compSC}. The FRII sources mostly dominate the count, except
  at low flux densities, where FRI sources take over. Our results show
  that FRI objects must undergo some mild evolution and that many of
  the mJy sources are radio galaxies, mostly FRIs.} 
\end{figure}

To improve coverage, the three other CoNFIG
samples described in \ref{otherC} were used. In all three of these
samples, sources with \SoneG$\ge$1.3~Jy were removed as they are
already present in the CoNFIG sample.\\
The relative differential source counts $\Delta N/\Delta N_0$ for FRI
and FRII sources were then computed from the combined sample of 244
FRI and 736 FRII sources. Data for the source count presented in
Figure~\ref{RelSC} can be found in Table~\ref{SCdata}.\\ 

\indent
It is seen that FRII sources dominate the total count, except at low flux densities
(log\SoneG$\lesssim-1.6$), where the FRI sources suddenly take over. Since most of
the FRI count at low flux densities is composed of low-luminosity
sources at low redshift, our results show that FRI objects must
undergo some mild evolution. This is consistent with
the results of \cite{Sadler07}, who studied low power sources in the
2SLAQ survey \citep{Richards05} and found evidence that FRIs undergo
significant evolution over $0<z<0.7$. Our results also show that
FRIs undergo less evolution than FRIIs, and they do not participate
much in the source-count ``evolution bump'' around
\SoneG$\sim$1Jy. This is in agreement with previous investigations
stretching back to \cite{Long66}.\\
\indent
An illustrative exponential evolution model of space density, as
described by \cite{WPL80}, was used to fit the data, as shown in
Figure~\ref{RelSC}. In the case of both FRI and FRII sources, the
luminosity distribution obtained for the CoNFIG sample was used as
starting points. This approach, as detailed in \cite{WPL80}, assumes
that the luminosity function $\rho(P,z)$ can be factorized as
$\rho(P,z)=F(P,z)\rho_0(P)$, where $\rho_0(P)$ is the local luminosity
function and $F(P,z)$ an evolution function transforming $\rho_0(P)$
into the redshift and luminosity-dependent $\rho(P,z)$. The details of
the evolution function $F(P,z)$ successfully fitting the two
morphology counts (with the same set of parameters) are as follows: 
\begin{eqnarray}
F(P,z)= \left\{ 
\begin{array}{ll}  
e^{M(P)\tau} & z \le z_c\\
0 & z > z_c\\
\end{array} 
\right.
\end{eqnarray}
where $z_c$ is the redshift cutoff (maximum redshift at which a
population exists) and $\tau$ is the look-back time 
\begin{eqnarray}
\tau=\frac{1}{H_0}\int_0^z{\frac{dz'}{(1+z)\sqrt{\Omega_M (1+z)^3
      +\Omega_K (1+z)^2 + \Omega_{\Lambda}}}} 
\end{eqnarray}
with M(P) defined as
\begin{eqnarray}
M= \left\{ 
\begin{array}{ll}  
M_1 &P < P_1\\
\frac{(M_2-M_1)(logP_1-logP)}{logP_1-logP_2}+M_1 &P_1 \le P \le P_2\\
M_2 &P > P_2\\
\end{array} 
\right.
\end{eqnarray}
The best fit values for this model are $z_c=4.37$, $M_1=2.38$,
$M_2=11.63$, $logP_1=26.0$ and $logP_2=26.9$.\\
This is only an illustrative model. Accurate modelling of the FRI/FRII
source counts and luminosity functions will be the subject of a future
study.\\

\indent
The morphological counts and the model fit raise two points:
\begin{enumerate}
\item[(1)] If a single model, albeit one with differential evolution, can describe both
populations, are the populations separate, or necessarily one
\citep{Snellen01,Rigby07}? At present the question is a semantic one:
we do not have an immediate  hypothesis to test requiring FRIs and
FRIIs to be one population or two.  What is of interest in this is
whether FRIs and FRIIs at the same radio luminosity show exactly the
same evolution, and our single model successfully fitting the data
suggests that they do. This question will be examined in more detail
in subsequent analyses. 
\item[(2)] Our model is a hands-off best-fit to the composite and morphologically-divided
source counts at 1.4~GHz -- down to 10~mJy. It is not a valid model, because it
predicts far too many sources in total at lower flux densities. A valid model needs
a rapid downturn to be achieved at about 0.01~Jy, and it must be that further
modelling details need introducing to achieve this. What
seems clear however is that a large proportion of sources at a level of 1~mJy (i.e.
at "mJy corner") must be FRIs. In fact starburst galaxies are usually thought to
account for the sub-mJy part of the source count. However, analysis of several radio
sources from the VLA-CDFS survey by \cite{Padovani07} shows that over two
thirds of the sub-mJy sources are faint radio galaxies, mostly FRIs. Our results
support this finding. The next paper of this short series will examine the combined
data and space density models in more detail.
\end{enumerate}

%------------------------------
%------------------------------

%%%%%%%%%%%%%%%%% Conclusion %%%%%%%%%%%%%%%%%%%%%%%%%%%%%%%%%%%%%%%%%%%%%%%%
\section{Summary}

The CoNFIG sample is constructed as a sample of 274 radio sources
from NVSS with $S\ge 1.3$~Jy. Redshift information is available for
$\sim$80\% of the sample, and morphological classifications
were obtained for all of the sources, either from NVSS and FIRST contour
plots, or, for 46 sources, from 8~GHz VLA observations. These data
allow us to compute morphology-dependent luminosity
distributions and source counts.\\ 
\indent
To increase the number of sources with morphology information, three more
samples were constructed in sub-areas of the main region with flux
density limits of 0.8~Jy, 0.2~Jy and 50~mJy. Morphological
identifications were obtained only from NVSS and FIRST contour plots
for those sources. Morphological information for the CENSORS and BDFL
sample were obtained from the Ledlow-Owen relation between radio power
and optical magnitude and from published classification
respectively. Combining these six samples allowed us to compile source
counts for FRI and FRII sources separately. A simple, single evolution
model for space density was then fitted to these data.\\  
\indent
Our data show mild evolution of the FRI sources at low redshift;
however, they do not participate in the ``evolution bump'' around
\SoneG$\sim$1~Jy. The results also support the observation that a
large number of mJy sources are FRIs galaxies and not starburst
galaxies as previously assumed.\\

%%%%%%%%%%%%%%%%%%%%%%%%%%%%%%%%%%%%%%%%%%%%%%%%%%%%%%%%%%%%%%%%%%%%%%%%%%%

%Acknowledgments
\section*{Acknowledgements}
We are very grateful to Jim Dunlop for many helpful suggestions, and
to Rick Perley and Eric Greison for their precious help with the VLA
observation and data reduction. We also thank the referee for very
helpful comments.\\
This work was supported by the National Sciences and Engineering
Research Council of Canada.\\
The National Radio Astronomy Observatory
is a facility of the National Science Foundation operated under
cooperative agreement by Associated Universities, Inc. This research
has made use of the SIMBAD database, operated at CDS, Strasbourg,
France.\\

%%%%%%%%%%%%%%%%% Bibliography %%%%%%%%%%%%%%%%%%%%%%%%%%%%%%%%%%%%%%%%%%%%

%%%%%%%%%%%%%%%%%%%%%%%%%%%%%%%%%%%%%%%%%%%%%%%%%%%%%%%%%%%%%%%%%%%%%%%%%%%%%
\onecolumn
\appendix
%%%% TABLES OF SAMPLE DETAILS WITH CoNFIG NAMES
\section{CoNFIG Sample}\label{details}

Data for CoNFIG sample. The 3CRR sources from \cite{Laing83} are
  indicated by stars. The RA and DEC gives the NVSS position of the 
  source. Note that 3 sources (CoNFIG-015, 076 and 225) were
  deleted later in the sample construction process and therefore do
  not appear here.\\ 
The sources of C* type are confirmed compact sources from the VLBA
  calibrator list \citep[see ][]{Beasley02,Fomalont03,Petrov05,Petrov06,Kovalev07} or the
Pearson-Readhead survey \citep{Pearson88}. Sources of S* type 
are confirmed compact sources which show a steep ($\alpha
\le -0.6$) spectral index. These are probably CSS sources. Finally,
sources of S type look compact with a steep spectral index, but
are not confirmed compact. They are probably unresolved FRII
sources. Designations I and II are \cite{FR74} types.\\
The values for \SoneM, \SthreeM, \SfourM, \StwoG and \SfiveG  were
retrieved from the samples listed in Table~\ref{SurveyTab}. References
for the redshift information and comments are given in the last two
column.\\
The spectral index $\alpha$ (where $S^{\alpha}_{\nu} \propto
\nu^{\alpha}$) corresponds to $\alpha_{low}$ as defined in
\S\ref{lumdist}. The $\alpha$ values with $^{\dag}$ sign correspond
to sources for which the value is derived from a non-satisfactory
single-power-low fit of the $logS - log\nu$ relation.\\  
Comments are as follow:\\
N - No optical counterpart identified above the SSS limit.\\
V - One of the 52 sources with uncertain morphology from the 1.4 GHz FIRST contours. Contours were obtained from our VLA A$-$configuration observations.\\
A - One of the 52 sources with uncertain morphology from the 1.4 GHz FIRST contours. Contours were obtained from NRAO archives.\\
S - Steep-spectrum QSO, since the source has a stellar-type optical counterpart and is classified as FRII.\\
{\footnotesize $^1$} Because of the very low quality of the data for
  this source, its morphology remains uncertain.\\
{\footnotesize $^2$} At first, this source looked like a steep-spectrum source with an
  FRI morphology. However, it is actually a well known radio loud QSO
  used as a VLBA calibrator.\\ 
{\footnotesize $^3$} This source's contours probably show a QSO with a suggestion of a
  jet.\\
{\footnotesize $^p$} photometric redshifts derived from the R-z relation; see \S\ref{IDandz}

\begin{sidewaystable}
  \centering
  \small
\begin{tabular}{ll|ll|l|l|llllll|l|ll}
    CoNFIG & Name& RA & DEC & Type & z & \SoneM &\SthreeM & \SfourM &
    \SoneG & \StwoG & \SfiveG & $\alpha_{low}$ & \multicolumn{2}{|c|}{Ref. and}\\
    ID & & \multicolumn{2}{|c|}{(J2000)} & & & (Jy) & (mJy) & (Jy)& (mJy) & (Jy) & (Jy) & & \multicolumn{2}{|c|}{Comments}\\
    \hline
  1 & QSO B0710+439   & 07 13 38.15 & +43 49 17.20 & C*  &0.5180 &        &   \phantom{00}655.0 &   \phantom{00}0.75 &   \phantom{00}2011.4 &         &   \phantom{0}1.60 &  \phantom{$-$}0.82$^{\dag}$& 1 & \\    
  2 & QSO B0711+35    & 07 14 24.80 & +35 34 39.90 & C*  &1.6260 &        &   \phantom{00}840.0 &        &   \phantom{00}1467.1 &         &   \phantom{0}0.89 &  \phantom{$-$}0.41$^{\dag}$& 1 & \\    
  3 & 4C 53.16        & 07 16 41.09 & +53 23 10.30 & II  &0.0643 &    \phantom{000}5.4 &  \phantom{0}3913.0 &        &   \phantom{00}1501.4 &         &   \phantom{0}0.63 & $-$0.63& 2 & \\    
  4 & 4C 33.21        & 07 35 55.54 & +33 07 09.60 & C   &       &    \phantom{000}7.4 &  \phantom{0}6750.0 &        &   \phantom{00}2473.1 &         &   \phantom{0}0.92 & $-$0.56&   & \\    
  5 & QSO J0741+3111  & 07 41 10.70 & +31 12 00.40 & C*  &0.6300 &        &  \phantom{0}1365.0 &        &   \phantom{00}2284.3 &         &   \phantom{0}2.68 &  \phantom{$-$}0.38& 3 & \\    
  6 & 4C 31.30        & 07 45 42.13 & +31 42 52.60 & II  &0.4620 &    \phantom{000}4.4 &  \phantom{0}2448.0 &        &   \phantom{00}1357.8 &         &   \phantom{0}1.02 & $-$0.55& 3 & S\\
\hline
  \end{tabular}
\end{sidewaystable}
\begin{sidewaystable}
  \centering
  \small
  \begin{tabular}{ll|ll|l|l|llllll|l|ll}
    CoNFIG & Name& RA & DEC & Type & z & \SoneM &\SthreeM & \SfourM &
    \SoneG & \StwoG & \SfiveG & $\alpha_{low}$ & \multicolumn{2}{|c|}{Ref. and}\\
    ID & & \multicolumn{2}{|c|}{(J2000)} & & & (Jy) & (mJy) & (Jy)& (mJy) & (Jy) & (Jy) & & \multicolumn{2}{|c|}{Comments}\\
    \hline   
 41 & 3C 215*         & 09 06 31.88 & +16 46 13.00 & II  &0.4115 &   \phantom{00}11.4 &  \phantom{0}5064.0 &   \phantom{00}6.30 &   \phantom{00}1586.2 &    \phantom{00}0.70 &   \phantom{0}0.42 & $-$0.95& 17 & S\\  
 42 & 4C 41.19        & 09 07 34.92 & +41 34 53.80 & II  &0.5272$^p$ &    \phantom{000}5.4 &  \phantom{0}5070.0 &   \phantom{00}4.25 &   \phantom{00}1394.5 &         &   \phantom{0}0.46 & $-$0.70&  & \\	    
 43 & 3C 217*         & 09 08 50.56 & +37 48 20.20 & II  &0.8980 &   \phantom{00}14.5 &  \phantom{0}8372.0 &   \phantom{00}7.09 &   \phantom{00}2086.4 &    \phantom{00}1.00 &   \phantom{0}0.56 & $-$0.95& 8 & N\\   
 44 & 3C 216*         & 09 09 33.53 & +42 53 47.40 & S*  &0.6700 &   \phantom{00}20.1 & 13932.0 &  \phantom{0}11.90 &   \phantom{00}4233.8 &    \phantom{00}3.39 &   \phantom{0}1.81 & $-$0.77& 11 & \\   
 45 & 4C 16.27        & 09 12 04.00 & +16 18 29.70 & II  &       &    \phantom{000}6.4 &  \phantom{0}4429.0 &   \phantom{00}4.48 &   \phantom{00}1374.6 &    \phantom{00}0.60 &   \phantom{0}0.35 & $-$0.77&  & NV\\   
 46 & 4C 17.48        & 09 14 04.83 & +17 15 52.40 & II  &       &    \phantom{000}6.7 &  \phantom{0}4612.0 &   \phantom{00}4.44 &   \phantom{00}1527.3 &    \phantom{00}0.84 &   \phantom{0}0.45 & $-$0.74&  & \\
\hline
  \end{tabular}
\end{sidewaystable}
\begin{sidewaystable}
  \centering
  \small
  \begin{tabular}{ll|ll|l|l|llllll|l|ll}
    CoNFIG & Name& RA & DEC & Type & z & \SoneM &\SthreeM & \SfourM &
    \SoneG & \StwoG & \SfiveG & $\alpha_{low}$ & \multicolumn{2}{|c|}{Ref. and}\\
    ID & & \multicolumn{2}{|c|}{(J2000)} & & & (Jy) & (mJy) & (Jy)& (mJy) & (Jy) & (Jy) & & \multicolumn{2}{|c|}{Comments}\\
    \hline   
 81 & 3C 245*         & 10 42 44.54 & +12 03 31.80 & S*  &1.0285 &   \phantom{00}14.4 &  \phantom{0}9449.0 &   \phantom{00}8.90 &   \phantom{00}3305.7 &    \phantom{00}2.00 &   \phantom{0}1.46 & $-$0.73& 11 & \\   
 82 & 4C 20.23        & 10 52 26.06 & +20 29 48.00 & C   &       &    \phantom{000}2.3 &  \phantom{0}2771.0 &        &   \phantom{00}1727.5 &         &   \phantom{0}0.69 & $-$0.17$^{\dag}$&  & \\	    
 83 & 4C 20.24        & 10 58 17.46 & +19 52 09.50 & II  &1.1100 &    \phantom{000}9.5 &  \phantom{0}5535.0 &        &   \phantom{00}2143.0 &    \phantom{00}1.65 &   \phantom{0}1.13 & $-$0.72& 11 & NA\\ 
 84 & 4C 01.28        & 10 58 29.62 & +01 33 58.20 & C*  &0.8880 &    \phantom{000}4.2 &  \phantom{0}4275.0 &   \phantom{00}4.47 &   \phantom{00}3220.2 &    \phantom{00}3.15 &   \phantom{0}3.16 & $-$0.14$^{\dag}$& 23 & A\\  
 85 & 3C 247*         & 10 58 58.69 & +43 01 23.70 & II  &0.7490 &   \phantom{00}16.8 &  \phantom{0}7612.0 &   \phantom{00}7.01 &   \phantom{00}2875.1 &    \phantom{00}1.63 &   \phantom{0}0.95 & $-$0.84& 19 & N\\  
 86 & 3C 249          & 11 02 03.91 & $-$01 16 18.30 & II  &0.3110 &   \phantom{00}16.9 &  \phantom{0}9990.0 &   \phantom{00}8.04 &   \phantom{00}2799.6 &    \phantom{00}1.35 &   \phantom{0}0.68 & $-$0.88& 24 & NV\\
\hline
  \end{tabular}
\end{sidewaystable}
\begin{sidewaystable}
  \centering
  \small
  \begin{tabular}{ll|ll|l|l|llllll|l|ll}
    CoNFIG & Name& RA & DEC & Type & z & \SoneM &\SthreeM & \SfourM &
    \SoneG & \StwoG & \SfiveG & $\alpha_{low}$ & \multicolumn{2}{|c|}{Ref. and}\\
    ID & & \multicolumn{2}{|c|}{(J2000)} & & & (Jy) & (mJy) & (Jy)& (mJy) & (Jy) & (Jy) & & \multicolumn{2}{|c|}{Comments}\\
    \hline   
121 & 3C 268.2        & 12 00 59.77 & +31 33 57.90 & II  &0.3620 &    \phantom{000}9.7 &  \phantom{0}2886.0 &        &   \phantom{00}1301.6 &    \phantom{00}0.73 &   \phantom{0}0.38 & $-$0.93& 8 & \\    
122 & 4C $-$04.40       & 12 04 02.13 & $-$04 22 43.90 & II  &       &    \phantom{000}8.3 &  \phantom{0}4264.0 &   \phantom{00}4.57 &   \phantom{00}2141.3 &    \phantom{00}1.32 &   \phantom{0}0.97 & $-$0.64&  & SV\\   
123 & 4C 04.40        & 12 06 19.93 & +04 06 12.20 & I   &0.6285$^p$ &    \phantom{000}8.3 &  \phantom{0}4362.0 &   \phantom{00}4.67 &   \phantom{00}1501.2 &    \phantom{00}0.87 &   \phantom{0}0.43 & $-$0.83&  & \\	    
124 & 3C 268.4*       & 12 09 13.52 & +43 39 18.70 & II  &1.4000 &   \phantom{00}10.3 &  \phantom{0}6585.0 &   \phantom{00}5.69 &   \phantom{00}1979.9 &    \phantom{00}1.06 &   \phantom{0}0.60 & $-$0.81& 18 & S\\  
125 & 4C 20.27        & 12 12 56.06 & +20 32 37.90 & S   &0.4649$^p$ &    \phantom{000}6.5 &  \phantom{0}4680.0 &        &   \phantom{00}1417.9 &    \phantom{00}0.80 &   \phantom{0}0.36 & $-$0.76&  & \\	    
126 & 4C 13.46        & 12 13 32.13 & +13 07 20.40 & C*  &1.1410 &    \phantom{000}3.1 &  \phantom{0}2515.0 &   \phantom{00}2.38 &   \phantom{00}1344.2 &    \phantom{00}0.95 &   \phantom{0}0.88 & $-$0.41& 10 & \\
\hline
  \end{tabular}
\end{sidewaystable}
\begin{sidewaystable}
  \centering
  \small
  \begin{tabular}{ll|ll|l|l|llllll|l|ll}
    CoNFIG & Name& RA & DEC & Type & z & \SoneM &\SthreeM & \SfourM &
    \SoneG & \StwoG & \SfiveG & $\alpha_{low}$ & \multicolumn{2}{|c|}{Ref. and}\\
    ID & & \multicolumn{2}{|c|}{(J2000)} & & & (Jy) & (mJy) & (Jy)& (mJy) & (Jy) & (Jy) & & \multicolumn{2}{|c|}{Comments}\\
    \hline   
161 & 4C $-$00.50       & 13 19 38.73 & $-$00 49 40.90 & C*  &0.8920 &    \phantom{000}4.4 &  \phantom{0}3749.0 &   \phantom{00}3.37 &   \phantom{00}1468.9 &    \phantom{00}1.01 &   \phantom{0}0.63 & $-$0.56& 11 & \\   
162 & 4C 17.56        & 13 20 21.45 & +17 43 12.40 & II  &       &    \phantom{000}5.4 &  \phantom{0}4084.0 &   \phantom{00}3.83 &   \phantom{00}1573.2 &    \phantom{00}1.10 &   \phantom{0}0.62 & $-$0.62&  & NV\\   
163 & 4C 11.45        & 13 21 18.84 & +11 06 49.40 & II  &2.1710 &    \phantom{000}8.1 &  \phantom{0}6214.0 &   \phantom{00}5.90 &   \phantom{00}2238.0 &    \phantom{00}1.38 &   \phantom{0}0.75 & $-$0.65& 1 & SA\\  
164 & 3C 285*         & 13 21 21.28 & +42 35 15.20 & II  &0.0794 &   \phantom{00}11.4 &         &   \phantom{00}5.03 &   \phantom{00}2085.0 &    \phantom{00}1.23 &   \phantom{0}0.76 & $-$0.81& 44 & \\   
165 & 4C 03.27        & 13 23 21.04 & +03 08 02.80 & I   &0.2690&    \phantom{000}5.9 &  \phantom{0}2659.0 &   \phantom{00}3.68 &   \phantom{00}1385.2 &    \phantom{00}0.79 &   \phantom{0}0.41 & $-$0.68& 11 & \\	    
166 & 4C 32.44        & 13 26 16.51 & +31 54 09.70 & C*  &0.3700 &    \phantom{000}5.8 &  \phantom{0}7486.0 &        &   \phantom{00}4861.9 &         &   \phantom{0}2.15 & $-$0.12$^{\dag}$& 45 & \\
\hline
  \end{tabular}
\end{sidewaystable}
\begin{sidewaystable}
  \centering
  \small
  \begin{tabular}{ll|ll|l|l|llllll|l|ll}
    CoNFIG & Name& RA & DEC & Type & z & \SoneM &\SthreeM & \SfourM &
    \SoneG & \StwoG & \SfiveG & $\alpha_{low}$ & \multicolumn{2}{|c|}{Ref. and}\\
    ID & & \multicolumn{2}{|c|}{(J2000)} & & & (Jy) & (mJy) & (Jy)& (mJy) & (Jy) & (Jy) & & \multicolumn{2}{|c|}{Comments}\\
    \hline   
201 & 4C 00.52        & 14 48 39.98 & +00 18 17.90 & II  &0.4381 &    \phantom{000}6.2 &  \phantom{0}4154.0 &   \phantom{00}4.12 &   \phantom{00}1651.5 &    \phantom{00}1.00 &   \phantom{0}0.55 & $-$0.65& 66 & V\\    
202 & 3C 305*         & 14 49 21.74 & +63 16 13.90 & I   &0.0416 &   \phantom{00}15.7 &  \phantom{0}9820.0 &        &   \phantom{00}3006.0 &    \phantom{00}1.64 &   \phantom{0}0.92 & $-$0.81& 50 & \\   
203 & 4C $-$04.53       & 14 55 01.43 & $-$04 20 22.50 & II  &0.4403 &   \phantom{00}11.4 &  \phantom{0}5414.0 &   \phantom{00}6.72 &   \phantom{00}2104.3 &    \phantom{00}0.90 &   \phantom{0}0.53 & $-$0.81& 51 & \\   
204 & 3C 311          & 15 04 09.27 & +60 00 55.50 & S   &1.0220 &    \phantom{000}7.4 &  \phantom{0}5224.0 &        &   \phantom{00}1553.1 &         &   \phantom{0}0.44 & $-$0.78& 13 & \\   
205 & B2 1502+28      & 15 04 19.50 & +28 35 34.30 & I   &0.0432$^p$ &        &   \phantom{00}499.0 &        &   \phantom{00}1626.5 &         &   \phantom{0}0.37 &  \phantom{$-$}0.88$^{\dag}$&  & \\	    
206 & PKS 1502+106    & 15 04 25.03 & +10 29 38.50 & C*  &1.8390 &        &  \phantom{0}1388.0 &   \phantom{00}1.55 &   \phantom{00}1774.2 &    \phantom{00}1.74 &   \phantom{0}2.04 & \phantom{$-$}0.15& 10 & \\
\hline
  \end{tabular}
\end{sidewaystable}
\begin{sidewaystable}
  \centering
  \small
  \begin{tabular}{ll|ll|l|l|llllll|l|ll}
    CoNFIG & Name& RA & DEC & Type & z & \SoneM &\SthreeM & \SfourM &
    \SoneG & \StwoG & \SfiveG & $\alpha_{low}$ & \multicolumn{2}{|c|}{Ref. and}\\
    ID & & \multicolumn{2}{|c|}{(J2000)} & & & (Jy) & (mJy) & (Jy)& (mJy) & (Jy) & (Jy) & & \multicolumn{2}{|c|}{Comments}\\
    \hline   
241 & 3C 331          & 16 12 19.02 & +22 22 15.60 & S   &0.6362$^p$ &    \phantom{000}7.1 &  \phantom{0}5360.0 &        &   \phantom{00}1401.9 &    \phantom{00}0.70 &   \phantom{0}0.37 & $-$0.81&  & \\	    
242 & QSO B1611+3420  & 16 13 41.08 & +34 12 47.70 & C*  &1.4010 &        &  \phantom{0}2928.0 &        &   \phantom{00}4024.1 &         &   \phantom{0}2.54 &  \phantom{$-$}0.24$^{\dag}$& 10 & \\   
243 & PKS 1614+26     & 16 16 38.29 & +26 47 01.60 & C   &       &        &  \phantom{0}1711.0 &        &   \phantom{00}1484.4 &    \phantom{00}1.20 &   \phantom{0}0.96 & $-$0.11$^{\dag}$&  & N\\    
244 & 3C 333          & 16 17 15.75 & +21 07 29.40 & II  &       &    \phantom{000}6.9 &  \phantom{0}5309.0 &        &   \phantom{00}1748.5 &    \phantom{00}1.00 &   \phantom{0}0.55 & $-$0.69&  & NV\\   
245 & NGC 6109*       & 16 17 38.89 & +35 00 48.00 & I   &0.0296 &    \phantom{000}4.3 &  \phantom{0}1513.0 &        &   \phantom{00}1706.2 &         &   \phantom{0}0.66 & $-$0.38$^{\dag}$& 11 & \\   
246 & 3C 332          & 16 17 43.28 & +32 23 02.40 & II  &0.1517 &    \phantom{000}9.5 &         &        &   \phantom{00}2598.5 &    \phantom{00}1.46 &   \phantom{0}0.83 & $-$0.63& 37 & \\
\hline
  \end{tabular}
\end{sidewaystable}

\medskip
 References: (1) \cite{Liu2002}; (2) \cite{Cappi2003}; (3) \cite{Enya02}; (4)
\cite{Sari86}; (5) \cite{Marzke96}; (6) \cite{Strom90}; (7)
\cite{Machalski98}; (8) \cite{Nilsson98}; (9) \cite{Stickel93}; (10)
\cite{Snel02}; (11) \cite{Aba04}; (12) \cite{Falco99}; (13)
\cite{Hewitt89}; (14) \cite{Spinrad85}; (15) \cite{Bark01}; (16)
\cite{Hewitt91}; (17) \cite{Ellingson94}; (18) \cite{Aba05}; (19)
\cite{Daly04}; (20) \cite{Gan06}; (21) \cite{Pol03}; (22)
\cite{Best99}; (23) \cite{Croom04}; (24) \cite{Mathez69}; (25)
\cite{Tinti06}; (26) \cite{Madore92}; (27) \cite{Wegner01}; (28)
\cite{Singal93}; (29) \cite{Roche98}; (30) \cite{Max95}; (31)
\cite{Lara01}; (32) \cite{Hew01}; (33) \cite{Sti94}; (34)
\cite{Ryab03}; (35) \cite{Vermu95}; (36) \cite{Trag00}; (37)
\cite{Adel06}; (39) \cite{Beck06}; (40) \cite{Smith00}; (41)
\cite{Owen95}; (42) \cite{Rines01}; (43) \cite{Pin00}; (44)
\cite{van93}; (45) \cite{Herbig92}; (46) \cite{Coll01}; (47)
\cite{Thom90}; (48) \cite{Holt03}; (49) \cite{Thimm94}; (50)
\cite{Miller02}; (51) \cite{Woo05}; (52) \cite{Greg82}; (53)
\cite{Schn01}; (54) \cite{Willott00}; (55) \cite{Heck94}; (56)
\cite{Gri92}; (57) \cite{Grimes05}; (58) \cite{Hill96}; (59)
\cite{Sti942}; (60) \cite{Pol95}; (61) \cite{Jones01}; (62)
\cite{SSS76}; (63) \cite{WW76}; (64) \cite{Aba03}; (65) \cite{Bon05};
(66) \cite{Can06}

\begin{table}
  \small
  \centering
  \small
  \caption[Optical Identifications]{\label{OptID}Optical
  identifications for the CoNFIG sources. The optical type (g
  for galaxy and s for stellar) as well as the $B_j$, R1, R2
  and I magnitudes were retrieved from the SuperCosmos Sky Survey
  \citep{Hambly01}. The radio type is as described in the
  previous table. Sources marked by a $^z$ have an optical
  identification but no spectroscopic redshift.}
  \medskip
  \begin{tabular}{|c|l|c|l|cccc|}
    \hline
    ID & \footnotesize{Rad.} & Optical Position & \footnotesize{Opt.} & $B_j$ & R1 & R2 & I \\
    & \footnotesize{Type}&\footnotesize{(J2000)} & \footnotesize{Type} & & & & \\
    \hline
  1 & C* & 07 13 38.15 +43 49 17.20 & g &  21.79 &  19.79 &  19.93 &  18.62\\
  2 & C* & 07 14 24.80 +35 34 39.90 & s &  18.74 &  17.59 &  18.42 &  17.50\\
  3 & II & 07 16 41.09 +53 23 10.30 & g &  15.43 &  15.54 &  14.22 &  13.30\\
  4$^z$ & C  & 07 35 55.54 +33 07 09.60 & s &  21.22 &        &  20.34 &  18.55\\
  5 & C* & 07 41 10.70 +31 12 00.40 & s &  16.52 &  16.28 &  16.32 &  15.83\\
  6 & II & 07 45 42.13 +31 42 52.60 & s &  15.76 &  15.55 &  15.56 &  15.11\\
  8 & I  & 07 58 28.60 +37 47 13.80 & g &  14.51 &  15.42 &  14.06 &  12.23\\
 10$^z$ & II & 08 01 35.32 +50 09 43.00 & s &  21.79 &        &  20.17 &       \\
 11 & II & 08 05 31.31 +24 10 21.30 & g &  17.28 &  16.28 &  15.60 &  15.16\\
 12 & II & 08 10 03.67 +42 28 04.00 & g &  21.08 &  19.47 &  18.91 &  17.38\\
 13 & II & 08 12 59.48 +32 43 05.60 & g &  21.19 &  19.76 &  19.19 &  18.64\\
 14 & II & 08 13 36.07 +48 13 01.90 & s &  18.45 &  17.79 &  17.82 &  17.54\\
 16 & II & 08 19 47.55 +52 32 29.50 & g &  20.37 &  18.31 &  18.26 &  17.80\\
 17 & II & 08 21 33.77 +47 02 35.70 & g &  18.55 &  17.26 &  17.07 &  16.26\\
 18$^z$ & C  & 08 21 44.02 +17 48 20.50 & g &  20.40 &  17.92 &  18.01 &  17.58\\
 19 & C* & 08 23 24.72 +22 23 03.70 & s &  20.77 &  18.31 &  20.01 &  19.17\\
 20 & C* & 08 24 47.27 +55 52 42.60 & s &  18.22 &  17.88 &  17.93 &  17.92\\
 21 & C* & 08 24 55.43 +39 16 41.80 & s &  18.37 &  18.29 &  17.55 &  17.55\\
 22 & II & 08 27 25.40 +29 18 44.80 & g &  20.96 &  19.04 &  18.89 &  18.82\\
 23 & S  & 08 31 10.00 +37 42 09.90 & s &  18.10 &  18.52 &  17.78 &  18.63\\
 25$^z$ & II & 08 34 48.37 +17 00 46.10 & g &  22.06 &  20.09 &  20.08 &       \\
 26 & C  & 08 34 54.91 +55 34 21.00 & g &  18.92 &  17.36 &  16.86 &  16.21\\
 27 & I  & 08 37 53.51 +44 50 54.60 & g &  18.99 &  17.15 &  17.30 &  16.45\\
 28 & II & 08 39 06.50 +57 54 13.40 & s &  18.15 &  17.62 &  17.23 &  16.90\\
 29 & II & 08 40 47.70 +13 12 23.90 & s &  18.09 &  17.49 &  17.62 &  16.67\\
 31 & I  & 08 47 53.83 +53 52 36.80 & g &  18.18 &  10.37 &  16.58 &  15.16\\
 32 & II & 08 47 57.00 +31 48 40.50 & g &  15.85 &  10.28 &  14.50 &  13.28\\
 33 & II & 08 53 08.83 +13 52 55.30 & s &  18.88 &  16.87 &  18.05 &  17.37\\
 34 & II & 08 54 39.35 +14 05 52.10 & s &  20.04 &  20.30 &  19.22 &  18.46\\
 35 & C* & 08 54 48.87 +20 06 30.70 & s &  15.54 &  13.98 &  15.19 &  13.59\\
 36 & II & 08 57 40.64 +34 04 06.40 & g &        &        &  20.31 &  19.10\\
 38 & II & 08 58 41.51 +14 09 43.80 & s &  20.04 &  19.22 &  18.79 &  18.04\\
 39 & I  & 09 01 05.40 +29 01 45.70 & g &  19.31 &  18.01 &  17.77 &  17.07\\
 40 & C* & 09 03 04.04 +46 51 04.70 & s &  19.13 &  18.54 &  18.66 &  18.48\\
 41 & II & 09 06 31.88 +16 46 13.00 & s &  17.99 &  18.60 &  18.24 &  17.63\\
 42$^z$ & II & 09 07 34.92 +41 34 53.80 & g &  21.51 &  19.34 &  19.06 &  18.72\\
 44 & S* & 09 09 33.53 +42 53 47.40 & g &  19.52 &  18.54 &  17.43 &  17.55\\
 46$^z$ & II & 09 14 04.83 +17 15 52.40 & g &        &  19.28 &  19.39 &  18.78\\
 47 & II & 09 21 07.54 +45 38 45.70 & g &  18.61 &  17.38 &  16.86 &  16.16\\
 49 & C* & 09 27 03.04 +39 02 20.70 & s &  17.06 &  17.53 &  16.49 &  16.35\\
 50 & II & 09 30 33.45 +36 01 23.60 & s &  18.98 &  18.67 &  18.23 &  17.74\\
 53 & C  & 09 42 08.40 +13 51 52.20 & g &  20.08 &  19.49 &  18.82 &       \\
 54 & S  & 09 42 15.35 +13 45 49.60 & s &        &        &  19.95 &       \\
 56 & II & 09 44 16.40 +09 46 19.20 & s &  22.45 &  20.21 &        &       \\
 58 & C* & 09 48 55.36 +40 39 44.80 & s &  19.00 &  17.80 &  18.13 &  17.66\\
 60 & II & 09 51 58.83 $-$00 01 26.80 & s &  15.88 &        &  14.88 &  14.41\\
 6$^z$1 & II & 09 52 00.52 +24 22 29.70 & g &        &  18.67 &  17.78 &  17.25\\
 62$^z$ & C  & 09 52 06.14 +28 28 33.20 & s &  21.59 &  19.62 &  19.78 &       \\
 63 & C* & 09 57 38.18 +55 22 57.40 & s &  18.28 &  17.31 &  17.36 &  16.98\\
 64 & II & 10 01 46.73 +28 46 56.50 & g &  18.64 &  17.52 &  17.32 &  16.15\\
 65 & II & 10 06 01.74 +34 54 10.40 & g &  17.27 &  16.12 &  15.64 &  15.04\\
 67 & C  & 10 08 00.04 +07 30 16.20 & g &  21.53 &        &        &       \\
 70 & S  & 10 17 14.15 +39 01 24.00 & g &  20.83 &  18.97 &  18.72 &       \\
 71 & II & 10 20 49.61 +48 32 04.20 & g &  17.95 &  16.04 &  16.02 &  15.64\\
 73$^z$ & S  & 10 23 38.71 +59 04 49.50 & s &  20.95 &  19.39 &  20.02 &  18.93\\
 74$^z$ & I  & 10 27 14.97 +46 03 21.90 & g &        &  19.58 &  19.71 &       \\
 75 & II & 10 33 33.87 +58 14 37.90 & g &  20.89 &  18.92 &  19.21 &  17.93\\
    \hline
  \end{tabular}
\end{table}

\begin{table}
  \small
  \centering  
  \begin{tabular}{|c|l|c|l|cccc|}  
    \hline
    ID & \footnotesize{Rad.} & Optical Position & \footnotesize{Opt.} & $B_j$ & R1 & R2 & I  \\
    & \footnotesize{Type}&\footnotesize{(J2000)} & \footnotesize{Type} & & & & \\
    \hline
 77$^z$ & S  & 10 34 17.86 +50 13 30.20 & s &        &  20.78 &  20.28 &       \\
 79 & C* & 10 41 17.16 +06 10 16.50 & s &  17.40 &  15.06 &  16.49 &  16.69\\
 80 & II & 10 41 39.01 +02 42 33.00 & g &  21.65 &        &  19.47 &  19.04\\
 81 & S* & 10 42 44.54 +12 03 31.80 & s &  18.03 &  17.10 &  17.01 &  17.00\\
 82$^z$ & C  & 10 52 26.06 +20 29 48.00 & g &  21.62 &  18.96 &  18.90 &  18.15\\
 84 & C* & 10 58 29.62 +01 33 58.20 & s &  18.75 &  17.60 &  17.63 &  16.71\\
 89 & I  & 11 09 52.06 +37 38 43.90 & g &  21.28 &  18.90 &  18.65 &  18.18\\
 91 & II & 11 12 38.36 +43 26 27.10 & s &  22.15 &  19.93 &  20.40 &  18.36\\
 92$^z$ & II & 11 13 32.13 $-$02 12 55.20 & g &  20.84 &  19.15 &  19.51 &  18.05\\
 93 & II & 11 14 38.43 +40 37 20.80 & s &  18.27 &  16.97 &  17.33 &  17.24\\
 94 & II & 11 16 34.70 +29 15 20.50 & g &  16.72 &  16.06 &  15.43 &  13.90\\
 96 & C  & 11 20 27.81 +14 20 54.40 & s &        &  19.45 &  19.49 &       \\
 97 & S  & 11 20 43.07 +23 27 55.30 & s &  22.63 &        &        &       \\
100 & S  & 11 31 38.90 +45 14 51.50 & g &  21.58 &  19.48 &  19.34 &  18.97\\
101 & II & 11 34 38.46 +43 28 00.50 & s &  22.52 &  20.35 &  20.62 &       \\
103 & II & 11 37 16.95 +61 20 38.40 & g &  18.53 &  17.25 &  16.78 &  16.17\\
104$^z$ & I  & 11 40 27.69 +12 03 07.60 & g &  16.88 &  16.00 &  14.98 &  14.65\\
106 & II & 11 41 08.23 +01 14 17.70 & g &  22.04 &  19.90 &  19.34 &  18.81\\
107 & II & 11 43 25.04 +22 06 56.00 & s &        &  19.43 &  20.60 &       \\
108$^z$ & II & 11 44 34.45 +37 10 16.90 & g &  18.95 &        &  17.59 &  16.67\\
109 & I  & 11 45 05.23 +19 36 37.80 & g &   8.98 &  14.10 &  13.04 &   6.55\\
110 & II & 11 45 31.03 +31 33 37.00 & g &  20.86 &  19.93 &  19.28 &       \\
113 & C* & 11 50 43.88 $-$00 23 54.30 & s &  17.24 &        &  16.94 &  16.34\\
114 & S* & 11 53 24.51 +49 31 09.50 & s &  17.02 &  16.49 &  16.23 &  15.79\\
115 & II & 11 54 13.01 +29 16 08.50 & g &  20.56 &  18.16 &  18.54 &  18.36\\
116 & I  & 11 55 26.63 +54 54 13.60 & g &  16.03 &  15.25 &  14.09 &  13.67\\
11$^z$7 & S  & 11 56 03.67 +58 47 05.40 & g &  20.39 &  18.15 &  18.35 &  18.08\\
118 & C  & 11 56 18.74 +31 28 05.00 & g &  19.18 &  18.55 &  18.45 &  17.63\\
120 & C* & 11 59 31.80 +29 14 44.30 & s &  17.49 &  16.56 &  17.65 &  13.10\\
121 & II & 12 00 59.77 +31 33 57.90 & g &  20.39 &  18.68 &  18.61 &  17.81\\
122$^z$ & II & 12 04 02.13 $-$04 22 43.90 & s &  17.75 &  16.17 &  16.69 &  15.72\\
123$^z$ & I  & 12 06 19.93 +04 06 12.20 & g &  21.72 &  19.70 &  19.09 &       \\
124 & II & 12 09 13.52 +43 39 18.70 & s &  18.49 &  17.54 &  17.22 &  17.10\\
125$^z$ & S  & 12 12 56.06 +20 32 37.90 & g &  21.69 &        &  19.10 &  18.75\\
126 & C* & 12 13 32.13 +13 07 20.40 & s &  17.87 &  17.58 &  17.10 &  17.08\\
127 & C* & 12 14 04.08 +33 09 45.50 & s &  17.91 &  17.10 &  17.64 &  17.05\\
128 & II & 12 15 29.80 +53 35 54.10 & s &  18.67 &  17.77 &  18.19 &  17.86\\
129 & C* & 12 15 55.60 +34 48 15.10 & s &  20.08 &  19.58 &  19.17 &  19.35\\
130 & I  & 12 17 29.83 +03 36 44.00 & g &  17.16 &  15.65 &  15.57 &  14.43\\
131 & I  & 12 19 15.33 +05 49 40.40 & g &   5.16 &  14.72 &   4.51 &  10.37\\
132 & II & 12 20 33.88 +33 43 10.90 & s &  18.96 &  18.13 &  18.12 &  17.49\\
134 & C* & 12 24 54.62 +21 22 47.20 & s &  16.01 &  18.03 &  15.94 &  15.62\\
135 & I  & 12 25 03.78 +12 52 35.20 & g &        &        &  16.31 &  14.23\\
136 & C* & 12 27 58.78 +36 35 11.60 & s &  22.11 &  19.86 &        &       \\
137 & C* & 12 29 06.41 +02 03 05.10 & s &  11.74 &   9.82 &  11.12 &  10.58\\
138 & I  & 12 29 51.84 +11 40 24.20 & g &  16.06 &  15.78 &  14.87 &  14.08\\
139 & I  & 12 30 49.46 +12 23 21.60 & g &  11.08 &        &  15.49 &  14.76\\
140 & C* & 12 32 00.13 $-$02 24 04.10 & s &  17.35 &        &  16.32 &  16.17\\
141 & II & 12 35 22.97 +21 20 18.30 & s &  22.08 &  18.89 &  19.08 &  19.02\\
142 & I  & 12 36 29.13 +16 32 32.10 & g &  17.31 &  16.42 &  15.30 &  14.41\\
143 & S  & 12 42 19.68 $-$04 46 19.70 & g &  21.28 &        &  20.31 &       \\
144 & II & 12 43 57.63 +16 22 52.70 & s &  18.63 &  18.77 &  18.50 &  17.44\\
145 & C* & 12 44 49.18 +40 48 06.50 & s &  20.37 &  19.49 &  19.80 &       \\
146$^z$ & II & 12 51 44.47 +08 56 27.80 & g &  20.17 &  18.15 &  18.01 &  17.84\\
147 & S* & 12 52 26.33 +56 34 19.70 & s &  17.93 &  16.57 &  17.09 &  17.09\\
148$^z$ & II & 12 53 03.55 +02 38 22.30 & g &  20.50 &  18.53 &  18.33 &  17.18\\
149 & II & 12 53 32.70 +15 42 27.30 & g &  22.46 &        &  20.69 &       \\
150 & II & 12 54 11.68 +27 37 32.70 & g &  17.27 &  16.22 &  15.93 &  15.14\\
151 & C* & 12 56 11.15 $-$05 47 20.10 & s &  17.57 &  15.36 &  15.62 &  13.67\\
152 & II & 12 56 57.38 +47 20 19.80 & g &  22.29 &        &  20.44 &  19.57\\
153 & II & 13 00 32.87 +40 09 09.20 & s &  18.74 &  19.34 &  18.58 &  17.49\\
155 & II & 13 09 49.66 $-$00 12 36.60 & g &  21.17 &        &  18.97 &  18.18\\
156 & C* & 13 10 28.70 +32 20 44.30 & s &  20.08 &  19.20 &  17.68 &  17.79\\
157 & II & 13 11 08.56 +27 27 56.50 & g &  19.32 &  17.97 &  17.51 &  17.07\\
159 & I  & 13 16 20.51 +07 02 54.30 & g &  14.99 &  14.70 &  14.52 &  13.08\\
\hline
\end{tabular}
\end{table}

\begin{table}
  \small
  \centering  
  \begin{tabular}{|c|l|c|l|cccc|} 
    \hline
    ID & \footnotesize{Rad.} & Optical Position & \footnotesize{Opt.} & $B_j$ & R1 & R2 & I  \\
    & \footnotesize{Type}&\footnotesize{(J2000)} & \footnotesize{Type} & & & & \\
    \hline
160 & I  & 13 19 06.83 +29 38 33.80 & g &  17.30 &  16.22 &  15.52 &  14.86\\
161 & C* & 13 19 38.73 $-$00 49 40.90 & s &  18.20 &  17.40 &  16.96 &  17.23\\
163 & II & 13 21 18.84 +11 06 49.40 & s &  19.45 &  18.78 &  18.65 &  18.11\\
164 & II & 13 21 21.28 +42 35 15.20 & g &  17.87 &  16.62 &  16.20 &  15.65\\
165$^z$ & I  & 13 23 21.04 +03 08 02.80 & g &  21.03 &        &  18.94 &  18.00\\
166 & C* & 13 26 16.51 +31 54 09.70 & g &  21.37 &  19.42 &  18.88 &  18.33\\
167 & I  & 13 27 31.71 +31 51 27.30 & g &  19.74 &  18.43 &  17.81 &  17.18\\
168 & C* & 13 30 37.69 +25 09 11.00 & g &  19.12 &  17.86 &  17.45 &  17.24\\
169 & C* & 13 31 08.31 +30 30 32.40 & s &  17.55 &  17.20 &  17.11 &  16.66\\
170 & II & 13 32 56.37 +02 00 46.50 & g &  19.51 &  17.80 &  17.40 &  16.79\\
171 & II & 13 38 08.07 $-$06 27 11.20 & s &  17.77 &  16.88 &  18.03 &  17.57\\
172 & I  & 13 38 49.67 +38 51 11.10 & g &  19.18 &  17.49 &  17.03 &  16.57\\
173 & II & 13 42 13.13 +60 21 42.30 & s &  18.48 &  17.83 &  17.45 &  17.52\\
174 & I  & 13 42 43.57 +05 04 31.50 & g &  18.13 &  17.15 &  16.56 &  15.65\\
177 & C* & 13 47 33.42 +12 17 24.10 & g &  16.61 &  16.50 &  15.72 &  14.78\\
178 & I  & 13 52 17.81 +31 26 46.70 & g &  11.06 &  15.27 &  14.28 &  13.52\\
181 & II & 13 57 04.37 +19 19 08.10 & s &  16.37 &  15.44 &  16.18 &  15.79\\
183 & C* & 14 00 28.65 +62 10 38.60 & g &  22.14 &  19.83 &  19.53 &  18.48\\
185 & II & 14 11 20.63 +52 12 09.00 & g &  21.17 &  18.13 &  18.47 &  17.78\\
188 & I  & 14 16 53.50 +10 48 40.20 & g &   8.43 &  14.20 &  16.17 &   6.39\\
189 & S  & 14 17 23.95 $-$04 00 46.60 & g &  21.88 &        &        &       \\
190 & S* & 14 19 08.18 +06 28 36.30 & s &  16.97 &  16.43 &  16.36 &  15.60\\
191 & S  & 14 21 05.73 +41 44 49.70 & g &  20.50 &  19.36 &  18.43 &  18.42\\
192 & II & 14 23 00.81 +19 35 22.80 & g &  19.95 &  18.13 &  18.21 &  17.92\\
193 & II & 14 24 56.93 +20 00 22.70 & s &  17.73 &  17.07 &  17.02 &  16.71\\
194 & II & 14 25 50.67 +24 04 06.70 & s &  17.88 &  17.22 &  17.15 &  16.64\\
196$^z$ & I  & 14 30 03.34 +07 15 01.30 & g &   5.61 &  15.12 &  14.30 &  13.49\\
197 & C  & 14 36 57.07 +03 24 12.30 & g &  21.98 &        &        &  19.31\\
198 & C* & 14 38 44.71 +62 11 54.50 & s &  19.45 &  18.80 &  18.65 &  18.43\\
199 & II & 14 43 01.45 +52 01 38.20 & g &        &  20.97 &  20.60 &       \\
200 & C  & 14 45 16.48 +09 58 36.00 & s &  18.93 &  17.76 &  17.72 &  17.27\\
201$^z$ & II & 14 48 39.98 +00 18 17.90 & g &  21.36 &  19.01 &  18.52 &  17.86\\
202 & I  & 14 49 21.74 +63 16 13.90 & g &  14.82 &  16.43 &  14.17 &  13.28\\
203 & II & 14 55 01.43 $-$04 20 22.50 & g &  20.89 &  18.70 &  18.96 &  18.16\\
204 & S  & 15 04 09.27 +60 00 55.50 & g &  19.95 &  18.90 &  19.24 &  19.60\\
205$^z$ & I  & 15 04 19.50 +28 35 34.30 & g &  16.19 &  15.93 &  15.06 &  15.10\\
206 & C* & 15 04 25.03 +10 29 38.50 & s &  19.56 &  19.46 &  18.76 &  18.46\\
207 & I  & 15 04 58.98 +25 59 49.00 & g &  16.33 &  16.73 &  15.15 &  14.72\\
208 & I  & 15 10 53.55 $-$05 43 07.10 & s &  17.28 &        &  16.77 &  16.52\\
209 & II & 15 10 57.03 +07 51 24.80 & s &  20.29 &  18.05 &  17.91 &  16.74\\
211 & I  & 15 13 39.90 +26 07 33.70 & g &  18.38 &  17.78 &  16.91 &  16.31\\
212$^z$ & C* & 15 13 40.20 +23 38 35.30 & s &        &        &  20.84 &       \\
213 & II & 15 16 40.21 +00 15 02.40 & g &  16.25 &  15.78 &  15.15 &  14.18\\
214 & I  & 15 16 44.58 +07 01 18.10 & g &  14.16 &  13.96 &  16.14 &  12.15\\
215$^z$ & S  & 15 16 56.61 +18 30 21.60 & g &  21.22 &  19.53 &  19.93 &  18.49\\
216 & S* & 15 20 05.50 +20 16 05.70 & g &  20.57 &  18.73 &  19.40 &  18.53\\
218 & II & 15 24 05.64 +54 28 18.40 & g &  20.30 &  18.42 &  18.44 &  17.68\\
220 & II & 15 31 25.36 +35 33 40.60 & g &  20.90 &  18.28 &  18.14 &  17.33\\
221 & II & 15 31 50.71 +24 02 43.30 & g &        &  16.35 &  15.99 &  15.35\\
222 & C* & 15 34 52.45 +01 31 03.30 & s &  19.44 &  18.70 &  18.15 &  16.27\\
224$^z$ & I  & 15 37 32.39 +13 44 47.70 & g &  21.91 &        &  20.43 &  19.17\\
226 & C* & 15 40 49.51 +14 47 46.70 & s &  17.66 &  15.55 &  17.43 &  16.81\\
227 & II & 15 41 45.64 +60 15 36.20 & g &        &        &  20.79 &       \\
228 & C* & 15 46 09.50 +00 26 24.60 & g &        &        &  19.92 &  19.04\\
229 & II & 15 47 44.23 +20 52 41.00 & s &  16.35 &  16.14 &  15.86 &  15.11\\
231 & II & 15 49 58.54 +62 41 20.90 & s &  21.36 &  19.71 &  19.59 &  19.05\\
232 & C* & 15 50 35.26 +05 27 10.60 & s &  20.07 &  17.59 &  18.56 &  18.63\\
235$^z$ & II & 15 56 36.35 +42 57 09.60 & g &        &        &  20.68 &       \\
237 & II & 16 02 17.21 +01 58 19.40 & g &  16.42 &        &  15.52 &  15.13\\
238 & C* & 16 08 46.13 +10 29 08.20 & s &  18.95 &  18.05 &  18.16 &  17.36\\
239 & C  & 16 09 13.31 +26 41 29.20 & g &        &        &  20.38 &  19.17\\
241$^z$ & S  & 16 12 19.02 +22 22 15.60 & g &  22.04 &  19.72 &  20.18 &       \\
242 & C* & 16 13 41.08 +34 12 47.70 & s &  18.06 &  17.60 &  17.02 &  16.63\\
245 & I  & 16 17 38.89 +35 00 48.00 & g &  10.43 &  14.25 &  13.71 &  12.77\\
246 & II & 16 17 43.28 +32 23 02.40 & g &  18.58 &  17.04 &  16.81 &  16.06\\
\hline
\end{tabular}
\end{table}

\begin{table}
  \small
  \centering  
  \begin{tabular}{|c|l|c|l|cccc|}  
    \hline
    ID & \footnotesize{Rad.} & Optical Position & \footnotesize{Opt.} & $B_j$ & R1 & R2 & I  \\
    & \footnotesize{Type}&\footnotesize{(J2000)} & \footnotesize{Type} & & & & \\
    \hline
247 & II & 16 20 21.40 +17 36 29.30 & s &  16.72 &  16.17 &  16.70 &  16.16\\
248 & II & 16 24 39.42 +23 45 17.50 & s &  18.57 &  18.39 &  18.06 &  17.88\\
250 & II & 16 28 03.57 +27 41 36.10 & g &  22.17 &  19.63 &  19.76 &       \\
251 & I  & 16 28 38.34 +39 33 04.70 & g &  13.92 &  17.08 &  16.30 &   6.17\\
253 & II & 16 29 37.52 +23 20 13.40 & g &  22.62 &        &  20.70 &       \\
254 & C  & 16 31 45.29 +11 56 03.30 & s &  18.73 &  18.29 &  18.20 &  17.52\\
255 & C* & 16 34 33.86 +62 45 35.70 & g &  21.42 &  20.00 &  20.25 &  19.36\\
256 & C* & 16 35 15.51 +38 08 04.80 & s &  18.37 &  17.80 &  17.43 &  16.49\\
258 & C* & 16 38 28.22 +62 34 43.90 & s &        &  20.62 &  20.55 &       \\
259 & C* & 16 42 58.77 +39 48 37.00 & s &  17.72 &  15.71 &  16.89 &  16.59\\
260 & II & 16 43 05.93 +37 29 34.40 & g &  21.33 &  19.98 &  20.17 &       \\
261 & I  & 16 43 48.69 +17 15 48.80 & g &  18.26 &  16.11 &  16.36 &  15.93\\
262 & C* & 16 47 41.83 +17 20 11.50 & g &  20.41 &  18.39 &  18.32 &  17.79\\
263 & C  & 16 53 52.24 +39 45 36.60 & g &  15.03 &  14.03 &  13.29 &  12.97\\
264 & II & 16 59 27.57 +47 03 13.10 & g &  19.93 &  17.68 &  17.69 &  16.74\\
265 & S  & 17 04 07.21 +29 46 59.50 & s &  19.42 &  19.11 &  18.96 &  18.73\\
266 & II & 17 04 43.03 +60 44 49.60 & s &  15.10 &  14.74 &  14.53 &  14.34\\
267$^z$ & II & 17 05 06.57 +38 40 37.60 & g &  21.58 &  19.67 &  19.71 &       \\
268 & II & 17 10 44.11 +46 01 30.30 & g &        &        &  20.75 &       \\
269 & II & 17 23 20.85 +34 17 57.30 & s &  15.92 &  15.25 &  15.83 &  14.95\\
270 & II & 17 24 18.40 +50 57 54.00 & g &        &  19.52 &  20.71 &       \\
271 & II & 17 42 51.84 +61 45 51.00 & s &  18.20 &  18.19 &  18.11 &  17.85\\
272$^z$ & II & 07 14 35.25 +45 40  0.10 & g &  18.43 &  17.30 &  16.79 &  16.20\\
273 & II & 14 54 20.30 +16 20 55.80 & g &  14.91 &  14.71 &  13.59 &  12.80\\
274$^z$ & II & 08 31 20.33 +32 18 37.00 & g &  16.22 &  15.80 &  15.11 &  14.15\\
275$^z$ & II & 08 48 41.94 +05 55 35.00 & g &  22.03 &  20.20 &        &       \\
276 & II & 12 28 11.77 +20 23 19.10 & s &  18.40 &  18.00 &  17.77 &  17.56\\
277 & S  & 07 44 17.50 +37 53 16.90 & s &  18.07 &  17.59 &  17.64 &  17.76\\
\hline
  \end{tabular}
\end{table}

\begin{table}
  \section{CoNFIG-2, 3 and 4 Samples}
  \centering
  \small
  \caption[]{\label{CoNFIG2}Data for CoNFIG-2 sample. The RA and DEC gives the NVSS position of the source. The sources of
  C* type are confirmed compact sources from the VLBA calibrator list
  \citep[see ][]{Beasley02,Fomalont03,Petrov05,Petrov06,Kovalev07} or
  the Pearson-Readhead survey \citep{Pearson88}. Unresolved sources
  (type S) are probably unresolved FRII sources. Designations I and II
  are \cite{FR74} types.\\ 
  References:\\
  (1) See Table~\ref{details}; (2) \cite{Snel02}; (3) \cite{Lahulla91};
  (4) \cite{Heck94}; (5) \cite{Hewitt91}; (6) \cite{Bauer00};
  (7) \cite{Enya02}; (8) \cite{Aba03}; (9) \cite{Brink95}; (10)
  \cite{Madore92}; (11) \cite{Machalski98}; (12) \cite{Willis76}; (13)
  \cite{Xu94}; (14) \cite{Kovalev99}; (15) \cite{Liu2002}; (16)
  \cite{Willott02}; (17) \cite{Tinti06}; (18) \cite{Bark01}; (19)
  \cite{Pih03}; (20) \cite{Hewitt89}; (21) \cite{Aba05}; (22)
  \cite{Strom90}; (26) \cite{Lu93}; (27) \cite{Schn07}; (28) \cite{Zen02};
  (29) \cite{Max95}; (30) \cite{Pin00}; (31) \cite{Schn05}; (32)
  \cite{Hewitt89}; (33) \cite{Nilsson98}; (35) \cite{Bon05}; (38)
  \cite{Gopal05}\\}
  \medskip
  \begin{tabular}{|ll|l|l|ll|lr|}
    \hline
    RA & DEC & Radio & \SoneG & B-mag. & Opt. & z & Ref.\\
    \multicolumn{2}{|c|}{(J2000)} & Type &(mJy) &  &Type & & \\
    \hline
09 20 11.16 & +17 53 25.00 & S  &     \phantom{00}1070.6 &       &   &       &   \\
09 20 58.48 & +44 41 53.70 & C* &     \phantom{00}1017.2 & 17.20 & 2 & 2.180 &  2\\
09 21 07.54 & +45 38 45.70 & II &     \phantom{00}8101.6 & 18.61 & 1 & 0.174 &  1\\
09 21 47.05 & +37 54 16.10 & II &     \phantom{000}826.4 & 20.24 & 2 & 1.108 &  3\\
09 22 49.93 & +53 02 21.20 & S  &     \phantom{00}1597.8 &       &   &       &   \\
09 27 03.04 & +39 02 20.70 & C* &     \phantom{00}2884.6 & 17.06 & 2 & 0.698 &  1\\
09 30 33.45 & +36 01 23.60 & II &     \phantom{00}1875.1 & 18.98 & 2 & 1.157 &  1\\
09 30 54.27 & +58 55 16.60 & II &     \phantom{00}1082.9 &       &   &       &   \\
09 34 15.80 & +49 08 21.00 & C* &     \phantom{000}800.5 &       &   & 2.582 &  2\\
09 35 04.06 & +08 41 37.30 & S  &     \phantom{00}1037.6 &       &   &       &   \\
09 35 06.62 & +39 42 07.60 & I  &     \phantom{00}1029.5 &       &   &       &   \\
09 36 32.02 & +04 22 10.80 & S  &     \phantom{000}971.1 &       &   & 1.340 &  4\\
09 39 50.20 & +35 55 53.10 & II &     \phantom{00}3719.0 & 18.59 & 1 & 0.137 &  1\\
09 41 22.70 &$-$01 43 01.00& II &     \phantom{000}830.0 &       &   & 0.382 &  5\\
09 41 23.62 & +39 44 14.10 & II &     \phantom{00}2064.7 & 18.00 & 1 & 0.107 &  1\\
09 42 08.40 & +13 51 52.20 & C  &     \phantom{00}1338.5 & 20.08 & 1 & 1.565 &  1\\
09 42 15.35 & +13 45 49.60 & S  &     \phantom{00}3336.4 &       &   & 0.580 &  1\\
09 43 12.74 & +02 43 27.50 & II &     \phantom{00}1331.5 &       &   & 0.592 &  8\\
09 43 19.16 &$-$00 04 22.30& S  &     \phantom{00}1188.9 & 21.27 & 1 &       &   \\
09 44 16.40 & +09 46 19.20 & II &     \phantom{00}2393.7 & 22.45 & 2 & 0.818 &  1\\
09 45 13.81 & +16 55 21.70 & II &     \phantom{00}1062.1 &       &   &       &   \\
09 47 47.27 & +07 25 13.80 & II &     \phantom{00}7617.0 & 17.65 & 1 & 0.086 &  1\\
09 47 44.60 & +00 04 37.20 & II &     \phantom{000}935.8 & 20.62 & 2 &       &   \\
09 48 55.36 & +40 39 44.80 & C* &     \phantom{00}1599.5 & 19.00 & 2 & 1.252 &  1\\
09 50 10.77 & +14 19 57.30 & II &     \phantom{00}3711.6 &       &   & 0.552 &  1\\
09 51 58.83 &$-$00 01 26.80& II &     \phantom{00}3152.1 & 15.88 & 2 & 1.487 &  1\\
09 52 00.52 & +24 22 29.70 & II &     \phantom{00}1788.6 &       &   &       &   \\
09 52 06.14 & +28 28 33.20 & C  &     \phantom{00}1362.7 &       &   &       &   \\
09 53 38.99 & +25 16 24.60 & S  &     \phantom{000}993.1 &       &   & 0.146 &  2\\
09 54 56.81 & +17 43 31.50 & C* &     \phantom{00}1158.5 & 16.94 & 2 & 1.472 &  2\\
09 54 04.03 & +21 22 28.10 & II &     \phantom{000}948.8 & 17.66 & 2 & 0.295 &  6\\
09 56 49.88 & +25 15 15.90 & C* &     \phantom{00}1080.1 & 18.13 & 2 & 0.712 &  2\\
09 57 38.18 & +55 22 57.40 & C* &     \phantom{00}3079.2 & 18.28 & 2 & 0.909 &  1\\
09 58 20.92 & +32 24 01.60 & S* &     \phantom{00}1247.1 & 15.57 & 2 & 0.530 &  7\\
09 58 28.78 &$-$01 39 59.30& S  &     \phantom{00}1213.7 &       &   &       &   \\
10 00 17.51 & +00 05 23.00 & II &     \phantom{000}923.7 & 17.92 & 2 & 0.907 &  8\\
10 00 21.95 & +22 33 18.20 & II &     \phantom{00}1116.3 & 16.98 & 2 & 0.419 &  9\\
10 00 28.11 & +14 01 34.10 & II &     \phantom{00}1166.4 &       &   &       &   \\
10 01 24.45 &$-$00 26 02.80& II &     \phantom{00}1221.0 &       &   &       &   \\
10 01 46.73 & +28 46 56.50 & II &     \phantom{00}5597.0 & 18.64 & 1 & 0.185 &  1\\
10 02 57.12 & +19 51 53.50 & I  &     \phantom{00}1226.5 & 16.73 & 1 & 0.168 & 1 \\
10 04 32.94 & +31 51 51.50 & II &     \phantom{00}1263.8 &       &   & 0.900 & 11\\
10 06 01.74 & +34 54 10.40 & II &     \phantom{00}3236.6 & 17.27 & 1 & 0.099 &  1\\
10 07 18.92 & +44 25 01.40 & II &     \phantom{00}1413.7 &       &   &       &   \\
10 07 27.54 & +12 48 40.90 & II &     \phantom{00}1216.1 & 14.73 & 2 & 0.240 &  7\\
10 07 41.51 & +13 56 29.30 & C* &     \phantom{000}936.3 & 18.27 & 2 & 2.707 & 12\\
10 07 42.54 & +59 08 09.90 & II &     \phantom{00}1082.8 & 19.46 & 2 &       &   \\
10 08 00.04 & +07 30 16.20 & C  &     \phantom{00}6522.1 & 21.53 & 1 & 0.880 &  1\\
    \hline
  \end{tabular}
\end{table}
\begin{table}
  \centering
  \begin{tabular}{|ll|l|l|ll|lr|}
    \hline
    RA & DEC & Radio & \SoneG & B-mag. & Opt. & z & Ref.\\
    \multicolumn{2}{|c|}{(J2000)} & Type &(mJy) &  &Type & & \\
    \hline
10 09 55.50 & +14 01 54.10 & S  &     \phantom{000}994.6 & 17.50 & 1 & 0.215 & 12\\
10 11 00.36 & +06 24 40.20 & II &     \phantom{00}2964.2 &       &   & 1.405 &  1\\
10 11 45.46 & +46 28 20.10 & II &     \phantom{00}1557.2 &       &   & 1.790 &  1\\
10 14 16.03 & +10 51 06.30 & II &     \phantom{000}902.2 &       &   & 0.388 & 21\\
10 14 47.05 & +23 01 12.70 & I  &     \phantom{00}1095.5 & 13.00 & 2 & 0.565 &  7\\
10 14 48.92 & +08 52 58.80 & C  &     \phantom{000}877.4 &       &   &       &   \\
10 16 01.90 & +40 46 57.40 & II &     \phantom{00}1078.5 & 16.69 & 1 & 0.128 & 21\\
10 17 14.15 & +39 01 24.00 & S  &     \phantom{00}1392.2 & 20.83 & 1 & 0.206 &  1\\
10 17 49.77 & +27 32 07.70 & II &     \phantom{00}1274.7 & 17.90 & 2 & 0.469 & 12\\
10 21 54.58 & +21 59 30.90 & II &     \phantom{00}1686.2 &       &   & 1.617 &  1\\
10 22 30.31 & +30 41 05.80 & C* &     \phantom{000}967.8 & 17.61 & 2 & 1.316 &  2\\
10 23 11.60 & +39 48 17.20 & C* &     \phantom{00}1122.6 & 17.02 & 2 & 1.254 & 13\\
10 23 38.71 & +59 04 49.50 & S  &     \phantom{00}1609.3 &       &   &       &   \\
10 24 29.63 &$-$00 52 55.20& C* &     \phantom{000}986.2 & 17.95 & 2 & 2.552 & 14\\
10 25 20.72 & +20 10 21.30 & I  &     \phantom{00}1250.3 &       &   &       &   \\
10 25 29.87 & +42 57 43.10 & II &     \phantom{000}852.4 &       &   &       &   \\
10 26 31.96 & +06 27 32.70 & II &     \phantom{000}851.5 & 17.93 & 2 & 1.699 & 15\\
10 27 14.64 & +46  2 20.10 & I  &     \phantom{00}1437.4 &       &   &       &   \\
10 27 32.89 & +48 17 51.40 & II &     \phantom{000}985.3 & 18.76 & 1 & 0.280 & 11\\
10 28 20.08 & +15 11 29.50 & S  &     \phantom{000}824.5 &       &   &       &   \\
10 30 09.91 & +00 37 40.20 & S  &     \phantom{00}1077.1 &       &   &       &   \\
10 31 43.55 & +52 25 37.90 & I  &     \phantom{000}921.6 & 17.21 & 1 & 0.166 &  8\\
10 33 28.31 & +17 42 44.70 & II &     \phantom{000}929.0 &       &   &       &   \\
10 33 33.87 & +58 14 37.90 & II &     \phantom{00}4187.9 & 20.89 & 1 & 0.430 &  1\\
10 34 17.86 & +50 13 30.20 & S  &     \phantom{00}1545.2 &       &   &       &   \\
10 35 07.04 & +56 28 47.30 & C* &     \phantom{00}1801.9 &       &   & 0.460 &  1\\
10 41 17.16 & +06 10 16.50 & C* &     \phantom{00}1405.2 & 17.40 & 2 & 1.270 &  1\\
10 41 39.01 & +02 42 33.00 & II &     \phantom{00}2710.1 & 21.65 & 1 & 0.535 &  1\\
10 42 36.53 & +29 49 45.60 & C  &     \phantom{000}841.2 &       &   &       &   \\
10 42 44.54 & +12 03 31.80 & S* &     \phantom{00}3305.7 & 18.03 & 2 & 1.028 &  1\\
10 46 18.04 & +54 59 37.50 & II &     \phantom{00}1032.7 &       &   &       &   \\
10 46 34.99 & +15 43 47.20 & S  &     \phantom{00}1071.9 &       &   &       &   \\
10 48 34.23 & +34 57 25.50 & S  &     \phantom{00}1034.4 & 20.70 & 2 & 1.594 & 16\\
10 49 26.18 &$-$02 54 52.70& I  &     \phantom{000}956.5 &       &   &       &   \\
10 51 48.80 & +21 19 52.80 & C* &     \phantom{00}1253.1 & 17.40 & 2 & 1.300 & 12\\
10 52 26.06 & +20 29 48.00 & C  &     \phantom{00}1727.5 & 21.62 & 1 &       &  1\\
10 57 15.77 & +00 12 03.80 & C* &     \phantom{000}898.1 &       &   & 0.650 & 17\\
10 58 17.46 & +19 52 09.50 & II &     \phantom{00}2143.0 &       &   & 1.110 &  1\\
10 58 29.62 & +01 33 58.20 & C* &     \phantom{00}3220.2 & 18.75 & 2 & 0.888 &  1\\
10 58 58.69 & +43 01 23.70 & II &     \phantom{00}2875.1 &       &   & 0.749 &  1\\
11 00 02.02 & +30 27 42.00 & II &     \phantom{000}991.4 &       &   &       &   \\
11 02 03.91 &$-$01 16 18.30& II &     \phantom{00}2799.6 &       &   & 0.311 &  1\\
11 02 24.97 &$-$02 35 34.10& S  &     \phantom{000}838.4 &       &   &       &   \\
11 02 26.19 & +55 50 03.30 & I  &     \phantom{00}1206.4 & 18.47 & 1 &       &   \\
11 05 26.17 & +20 52 17.40 & S  &     \phantom{000}985.6 &       &   &       &   \\
11 06 31.77 &$-$00 52 51.50& II &     \phantom{00}1065.6 & 16.25 & 2 & 0.426 &  7\\
11 07 15.02 & +16 28 01.50 & I  &     \phantom{000}868.6 & 16.53 & 2 & 0.632 &  7\\
11 08 08.31 & +14 35 35.80 & C  &     \phantom{00}1348.7 &       &   &       &   \\
11 08 36.10 & +38 58 58.10 & II &     \phantom{000}919.7 & 18.69 & 2 & 0.781 & 18\\
11 08 51.79 & +25 00 52.10 & II &     \phantom{00}1090.0 &       &   &       &   \\
11 09 28.86 & +37 44 31.40 & C* &     \phantom{00}1221.6 &       &   & 2.290 & 14\\
11 09 46.04 & +10 43 43.40 & C  &     \phantom{00}1481.3 &       &   & 0.550 &  1\\
11 09 47.74 & +37 38 16.50 & I  &     \phantom{00}2214.1 & 21.28 & 1 & 0.346 &  1\\
11 11 20.09 & +19 55 36.10 & C* &     \phantom{00}1194.8 & 18.56 & 2 & 0.299 & 19\\
11 11 22.71 & +03 09 10.40 & II &     \phantom{00}1017.7 &       &   &       &   \\
11 11 31.56 & +35 40 45.50 & II &     \phantom{00}1336.3 &       &   & 1.105 &  1\\
11 11 38.98 & +40 50 15.30 & II &     \phantom{000}819.2 & 15.69 & 1 & 0.074 & 15\\
11 12 38.36 & +43 26 27.10 & II &     \phantom{00}1433.0 & 22.15 & 2 & 1.680 &  1\\
11 13 32.13 &$-$02 12 55.20& II &     \phantom{00}1595.6 & 20.84 & 1 & 0.125 & 35\\
11 14 38.43 & +40 37 20.80 & II &     \phantom{00}3127.9 & 18.27 & 2 & 0.734 &  1\\
11 16 34.70 & +29 15 20.50 & II &     \phantom{00}1927.9 & 16.72 & 1 & 0.049 &  1\\
11 17 33.85 &$-$02 36 00.60& S  &     \phantom{000}996.1 &       &   &       &   \\
11 18 11.85 & +53 19 44.70 & II &     \phantom{000}919.2 & 17.92 & 2 & 0.235 & 2 \\
11 18 57.28 & +12 34 42.20 & S* &     \phantom{00}1112.2 & 18.12 & 2 & 2.118 & 21\\
    \hline
  \end{tabular}
\end{table}
\begin{table}
  \centering
  \begin{tabular}{|ll|l|l|ll|lr|}
    \hline
    RA & DEC & Radio & \SoneG & B-mag. & Opt. & z & Ref.\\
    \multicolumn{2}{|c|}{(J2000)} & Type &(mJy) &  &Type & & \\
    \hline
11 19 25.22 &$-$03 02 51.60& S  &     \phantom{00}1730.4 &       &   & 1.355 &  1\\
11 20 27.81 & +14 20 54.40 & C  &     \phantom{00}2446.9 &       &   & 0.362 &  1\\
11 20 43.07 & +23 27 55.30 & S  &     \phantom{00}1362.0 & 22.63 & 2 & 1.819 &  1\\
11 23 09.10 & +05 30 20.30 & S  &     \phantom{00}1721.1 &       &   & 2.474 &  1\\
11 24 37.45 & +04 56 18.80 & II &     \phantom{00}1145.4 & 17.86 & 1 & 0.283 &  8\\
11 24 43.90 & +19 19 29.70 & S  &     \phantom{000}874.7 & 18.82 & 1 & 0.165 & 22\\
11 25 53.70 & +26 10 20.10 & C* &     \phantom{000}921.2 & 18.79 & 2 & 2.341 &  2\\
11 26 08.53 & +30 03 36.50 & II &     \phantom{000}933.6 & 18.19 & 2 &       &   \\
11 26 23.65 & +33 45 27.10 & C  &     \phantom{00}1376.8 &       &   & 1.230 &  1\\
11 26 26.88 & +12 20 37.40 & II &     \phantom{00}1110.5 & 19.35 & 2 &       &   \\
11 29 35.97 & +00 15 17.50 & II &     \phantom{000}979.5 & 18.75 & 1 & 0.211 &   \\
11 29 47.93 & +50 25 51.90 & S  &     \phantom{000}926.9 &       &   &       &   \\
11 31 38.90 & +45 14 51.50 & S  &     \phantom{00}2048.8 & 21.58 & 1 & 0.404 &  1\\
11 32 59.49 & +10 23 42.70 & S  &     \phantom{000}879.1 & 16.60 & 2 & 0.540 &  8\\
11 33 09.56 & +33 43 12.60 & II &     \phantom{000}886.6 & 17.70 & 1 & 0.190 & 11\\
11 33 13.18 & +50 08 39.90 & II &     \phantom{000}844.9 & 18.28 & 1 & 0.310 & 11\\
11 34 38.46 & +43 28 00.50 & II &     \phantom{00}1567.1 & 22.52 & 2 & 0.572 &  1\\
11 34 54.61 & +30 05 25.20 & II &     \phantom{00}1145.5 & 19.28 & 1 & 0.614 &  7\\
11 35 13.04 &$-$00 21 19.40& S  &     \phantom{00}1267.8 & 16.85 & 2 & 0.160 & 17\\
11 35 55.93 & +42 58 44.80 & C  &     \phantom{00}1448.8 &       &   &       &   \\
11 37 29.68 & +01 16 13.30 & S  &     \phantom{00}1059.7 & 19.32 & 1 & 0.430 & 1 \\
11 40 17.03 & +17 43 39.00 & I  &     \phantom{00}1143.3 &  7.05 & 1 & 0.011 & 26\\
11 40 27.69 & +12 03 07.60 & I  &     \phantom{00}1527.0 & 16.88 & 1 & 0.081 & 21\\
11 40 49.54 & +59 12 26.00 & C  &     \phantom{00}2179.4 &       &   &       &   \\
11 41 08.23 & +01 14 17.70 & II &     \phantom{00}2690.8 & 22.04 & 1 & 0.443 &  1\\
11 42 57.23 & +21 29 12.50 & II &     \phantom{000}921.0 & 17.91 & 1 &       &   \\
11 42 58.78 & +48 51 19.70 & II &     \phantom{000}958.6 & 19.91 & 2 &       &   \\
11 43 25.04 & +22 06 56.00 & II &     \phantom{00}3128.7 &       &   & 0.366 &  1\\
11 43 39.63 & +46 21 20.70 & II &     \phantom{000}863.2 & 16.66 & 1 & 0.117 & 21\\
11 44 21.23 & +29 58 25.30 & II &     \phantom{000}837.3 &       &   &       &   \\
11 44 34.45 & +37 10 16.90 & II &     \phantom{00}1998.9 & 18.95 & 1 & 0.115 & 21\\
11 44 54.01 &$-$00 31 36.50& II &     \phantom{000}947.6 &       &   &       &   \\
11 45 05.23 & +19 36 37.80 & I  &     \phantom{00}5689.0 &  8.98 & 1 & 0.021 &  1\\
11 45 31.03 & +31 33 37.00 & II &     \phantom{00}2890.9 & 20.86 & 1 & 0.811 &  1\\
11 45 43.41 & +49 46 08.40 & II &     \phantom{00}1424.5 &       &   & 1.275 &  1\\
11 47 14.71 & +25 23 20.20 & II &     \phantom{000}879.0 &       &   &       &   \\
11 48 47.51 & +04 55 27.70 & II &     \phantom{000}827.9 &       &   & 0.420 & 38\\
11 49 55.54 & +12 47 15.90 & II &     \phantom{00}2519.9 &       &   & 1.144 &  1\\
11 50 43.88 &$-$00 23 54.30& C* &     \phantom{00}2773.9 & 17.24 & 2 & 1.976 &  1\\
11 53 03.11 & +11 07 20.40 & II &     \phantom{000}853.5 &       &   &       &   \\
11 53 12.54 & +09 14 02.50 & C* &     \phantom{000}809.2 & 17.36 & 2 & 0.698 &  8\\
11 53 24.51 & +49 31 09.50 & S* &     \phantom{00}1572.2 & 17.02 & 2 & 0.334 &  1\\
11 53 54.65 & +40 36 52.90 & C* &     \phantom{00}1135.6 & 20.16 & 1 & 0.916 & 27\\
11 54 13.01 & +29 16 08.50 & II &     \phantom{00}1620.3 & 20.56 & 1 & 0.329 &  1\\
11 55 13.61 & +54 52 50.40 & II &     \phantom{00}1233.3 &       &   &       &   \\
11 55 34.69 & +54 53 40.40 & I  &     \phantom{00}2226.9 & 16.03 & 1 & 0.050 &  1\\
11 56 03.67 & +58 47 05.40 & S  &     \phantom{00}1591.7 & 20.39 & 1 &       &   \\
11 56 18.74 & +31 28 05.00 & C  &     \phantom{00}2978.3 & 19.18 & 1 & 0.418 &  1\\
11 57 34.91 & +16 38 59.30 & C* &     \phantom{000}813.2 & 17.26 & 2 & 1.050 &  2\\
11 58 25.80 & +24 50 17.70 & C* &     \phantom{00}1021.2 & 18.19 & 1 & 0.202 & 28\\
11 59 13.79 & +53 53 07.40 & C  &     \phantom{00}1740.6 &       &   &       &   \\
11 59 31.80 & +29 14 44.30 & C* &     \phantom{00}2030.8 & 17.49 & 2 & 0.729 &  1\\
11 59 48.83 & +58 20 20.80 & C  &     \phantom{00}1191.3 & 19.56 & 1 &       &   \\
12 00 31.19 & +45 48 43.20 & S  &     \phantom{00}1156.5 &       &   & 0.743 & 29\\
12 00 57.78 & +31 32 52.90 & II &     \phantom{00}1301.6 & 20.39 & 1 & 0.362 &  1\\
12 01 25.01 & +25 20 24.00 & II &     \phantom{000}862.9 &       &   &       &   \\
12 02 04.19 & +58 02 01.90 & I  &     \phantom{000}843.2 & 16.10 & 1 & 0.102 & 3 \\
12 02 32.27 &$-$02 40 03.20& S  &     \phantom{000}906.1 &       &   &       &   \\
12 03 21.95 & +04 14 17.70 & C* &     \phantom{00}1146.1 & 18.39 & 2 & 1.212 & 31\\
12 04 02.13 &$-$04 22 43.90& II &     \phantom{00}2141.3 & 17.75 & 2 &       &   \\
12 04 18.46 & +52 02 18.80 & C  &     \phantom{000}962.3 &       &   &       &   \\
12 04 52.31 & +29 29 12.40 & II &     \phantom{00}1230.6 & 18.57 & 1 &       &   \\
12 06 19.93 & +04 06 12.20 & I  &     \phantom{00}1501.2 & 21.72 & 1 &       &   \\
12 09 13.52 & +43 39 18.70 & II &     \phantom{00}1979.9 & 18.49 & 2 & 1.400 &  1\\
    \hline
  \end{tabular}
\end{table}
\begin{table}
  \centering
  \begin{tabular}{|ll|l|l|ll|lr|}
    \hline
    RA & DEC & Radio & \SoneG & B-mag. & Opt. & z & Ref.\\
    \multicolumn{2}{|c|}{(J2000)} & Type &(mJy) &  &Type & & \\
    \hline
12 12 56.06 & +20 32 37.90 & S  &     \phantom{00}1417.9 & 21.69 & 1 &       &   \\
12 13 32.13 & +13 07 20.40 & C* &     \phantom{00}1344.2 & 17.87 & 2 & 1.141 &  1\\
12 14 04.08 & +33 09 45.50 & C* &     \phantom{00}1403.6 & 17.91 & 2 & 1.598 &  1\\
12 14 47.73 &$-$01 00 12.10& II &     \phantom{00}1092.5 &       &   &       &   \\
12 15 14.69 & +17 30 02.20 & S* &     \phantom{00}1010.2 &       &   &       &   \\
12 15 29.80 & +53 35 54.10 & II &     \phantom{00}2755.0 & 18.67 & 2 & 1.065 &  1\\
12 15 55.60 & +34 48 15.10 & C* &     \phantom{00}1506.8 & 20.08 & 2 & 0.857 &  1\\
12 17 29.83 & +03 36 44.00 & I  &     \phantom{00}2411.5 & 17.16 & 1 & 0.077 &  1\\
12 17 56.90 & +25 29 27.20 & S  &     \phantom{000}866.8 & 17.81 & 2 &       &   \\
12 18 59.22 & +19 55 28.90 & II &     \phantom{00}1047.0 &       &   &       &   \\
12 19 15.33 & +05 49 40.40 & I  &     \phantom{0}10445.0 &  5.16 & 1 & 0.007 &  1\\
12 20 28.08 & +09 28 26.90 & II &     \phantom{00}1064.4 & 18.33 & 2 & 1.082 &  8\\
12 20 33.88 & +33 43 10.90 & II &     \phantom{00}2845.9 & 18.96 & 2 & 1.519 &  1\\
12 21 52.92 & +31 30 56.70 & II &     \phantom{000}805.8 &       &   &       &   \\
12 22 22.59 & +04 13 17.30 & C* &     \phantom{000}800.3 & 16.91 & 2 & 0.965 & 32\\
12 24 27.29 & +42 06 10.30 & II &     \phantom{00}1352.3 &       &   & 0.944 &  1\\
12 24 52.44 & +03 30 50.10 & C* &     \phantom{00}1280.3 & 18.05 & 2 & 0.957 &  2\\
12 24 54.62 & +21 22 47.20 & C* &     \phantom{00}2094.4 & 16.01 & 2 & 0.435 &  1\\
12 25 03.78 & +12 52 35.20 & I  &     \phantom{00}6012.8 &       &   & 0.003 &  1\\
12 26 25.60 & +09 40 05.70 & S  &     \phantom{00}1008.3 &       &   &       &   \\
12 27 58.78 & +36 35 11.60 & C* &     \phantom{00}2098.4 & 22.11 & 2 & 1.973 &  1\\
12 28 11.77 & +20 23 19.10 & II &     \phantom{00}1269.5 & 18.00 & 2 & 0.680 & 1 \\
12 29 06.41 & +02 03 05.10 & C* &     \phantom{0}54991.2 & 11.74 & 2 & 0.158 &  1\\
12 29 32.62 & +17 50 20.90 & C  &     \phantom{000}931.0 & 17.87 & 1 &       &   \\
12 29 51.84 & +11 40 24.20 & I  &     \phantom{00}1519.0 & 16.06 & 1 & 0.083 &  1\\
12 30 49.46 & +12 23 21.60 & I  &               141949.3 & 11.08 & 1 & 0.004 &  1\\
12 32 00.13 &$-$02 24 04.10& C* &     \phantom{00}1646.7 & 17.35 & 2 & 1.045 &  1\\
12 32 05.31 &$-$01 34 55.10& I  &     \phantom{000}807.4 &       &   &       &   \\
12 35 22.97 & +21 20 18.30 & II &     \phantom{00}2918.5 & 22.08 & 2 & 0.422 &  1\\
12 36 28.18 & +16 31 35.40 & I  &     \phantom{00}1383.9 & 17.31 & 1 & 0.068 &  1\\
12 36 49.96 & +36 55 18.00 & II &     \phantom{000}920.2 &       &   &       &   \\
12 39 07.16 & + 5 19 24.50 & II &     \phantom{000}938.2 &       &   &       &   \\
12 39 09.05 & +32 30 27.30 & S  &     \phantom{000}827.8 &       &   &       &   \\
12 41 49.84 & +57 30 36.90 & II &     \phantom{00}1299.2 &       &   &       &   \\
12 42 19.68 &$-$04 46 19.70& S  &     \phantom{00}3672.1 & 21.28 & 1 & 0.480 &  1\\
12 43 57.63 & +16 22 52.70 & II &     \phantom{00}2895.8 & 18.63 & 2 & 0.557 &  1\\
12 44 49.18 & +40 48 06.50 & C* &     \phantom{00}1341.8 & 20.37 & 2 & 0.813 &  1\\
12 47 07.40 & +49 00 18.20 & C  &     \phantom{00}1204.6 & 17.43 & 1 & 0.206 & 15\\
12 49 48.76 & +33 23 15.80 & II &     \phantom{00}1289.2 &       &   & 0.350 & 11\\
12 51 35.43 & +50 34 01.40 & II &     \phantom{00}1155.8 &       &   & 0.414 & 33\\
12 51 40.63 & +08 55 58.60 & II &     \phantom{00}1684.4 & 20.17 & 1 &       &   \\
12 52 08.60 & +52 45 30.80 & S  &     \phantom{000}958.5 &       &   &       &   \\
12 52 16.81 & +47 15 39.00 & II &     \phantom{000}982.9 &       &   &       &   \\
12 52 22.78 & +03 15 50.40 & I  &     \phantom{000}935.3 & 16.74 & 1 & 0.099 &  8\\
12 52 26.33 & +56 34 19.70 & S* &     \phantom{00}2288.3 & 17.93 & 2 & 0.320 &  1\\
12 53 03.55 & +02 38 22.30 & II &     \phantom{00}1604.9 & 20.50 & 1 &       &   \\
12 53 32.70 & +15 42 27.30 & II &     \phantom{00}1952.2 & 22.46 & 1 & 0.766 &  1\\
12 54 11.68 & +27 37 32.70 & II &     \phantom{00}2923.9 & 17.27 & 1 & 0.086 &  1\\
12 56 11.15 &$-$05 47 20.10& C* &     \phantom{00}9711.2 & 17.57 & 2 & 0.536 &  1\\
12 58 01.96 & +44 35 20.60 & I  &     \phantom{000}929.8 & 18.11 & 1 &       &   \\
13 00 32.87 & +40 09 09.20 & II &     \phantom{00}1368.9 & 18.74 & 2 & 1.667 &  1\\
13 02 28.56 & +58 18 46.90 & C  &     \phantom{000}898.0 &       &   &       &   \\
13 04 28.87 & +53 50 02.50 & II &     \phantom{000}908.9 &       &   &       &   \\
13 04 43.69 &$-$03 46 02.30& S* &     \phantom{000}826.0 & 18.61 & 2 & 1.250 & 18\\
13 05 36.05 & +08 55 15.90 & II &     \phantom{00}1461.8 &       &   & 1.409 &  1\\
13 07 54.01 & +06 42 15.90 & II &     \phantom{00}1122.0 & 16.54 & 2 & 0.599 &  7\\
13 09 33.94 & +11 54 24.20 & C* &     \phantom{000}855.0 & 18.52 & 2 &       &   \\
13 09 49.66 &$-$00 12 36.60& II &     \phantom{00}1636.7 & 21.17 & 1 & 0.419 &  1\\
13 10 28.70 & +32 20 44.30 & C* &     \phantom{00}1686.6 & 20.08 & 2 & 0.997 &  1\\
13 11 08.56 & +27 27 56.50 & II &     \phantom{00}2044.6 & 19.32 & 1 & 0.239 &  1\\
13 13 37.88 & +54 58 24.30 & C* &     \phantom{00}1304.6 &       &   &       &   \\
13 15 01.28 & +20 44 30.10 & S  &     \phantom{00}1098.7 &       &   &       &   \\
13 15 09.94 & + 8 41 44.60 & II &     \phantom{000}932.8 &       &   &       &   \\
13 16 14.55 & +07 02 19.90 & I  &     \phantom{00}2006.4 & 14.99 & 1 & 0.051 &  1\\
    \hline
  \end{tabular}
\end{table}
\begin{table}
  \centering
  \begin{tabular}{|ll|l|l|ll|lr|}
    \hline
    RA & DEC & Radio & \SoneG & B-mag. & Opt. & z & Ref.\\
    \multicolumn{2}{|c|}{(J2000)} & Type &(mJy) &  &Type & & \\
    \hline
13 19 06.83 & +29 38 33.80 & I  &     \phantom{00}1372.5 & 17.30 & 1 & 0.073 &  1\\
13 19 38.73 &$-$00 49 40.90& C* &     \phantom{00}1468.9 & 18.20 & 2 & 0.892 &  1\\
13 19 46.40 & +51 48 06.70 & II &     \phantom{00}1092.6 & 16.59 & 2 & 1.060 & 15\\
    \hline
  \end{tabular}
\end{table}

%%%%%%%%%%%%%%%%%%%%%%%%%%%%%%%%%%%%%%%%%%%%%%%%%%%%%%%%%%%%%%%%%%%%%%%%%%%%%%%%%%%%%%%%%%%

\begin{table}
  \centering
  \small
  \caption[]{\label{CoNFIG34}Data for CoNFIG-3 and CoNFIG-4 sample. The RA and DEC gives the NVSS position of the source. The sources of
  C* type are confirmed compact sources from the VLBA calibrator list
  \citep[see ][]{Beasley02,Fomalont03,Petrov05,Petrov06,Kovalev07} or
  the Pearson-Readhead survey \citep{Pearson88}. Unresolved sources
  (type S) are probably unresolved FRII sources. Designations I and II
  are \cite{FR74} types.\\ 
}
  \medskip
\begin{tabular}{llll|c|llll}
    \hline
    \multicolumn{9}{|c|}{CoNFIG-3} \\
    RA & DEC & Radio & \SoneG & &RA & DEC & Radio & \SoneG \\
    \multicolumn{2}{|c|}{(J2000)} & Type &(mJy) & &\multicolumn{2}{|c|}{(J2000)} & Type &(mJy) \\
    \hline    
14 41 28.26 & +10 55 24.70 & S  &      422.7 & & 14 55  7.32 & +14 12 22.20 & II &      246.9 \\
14 42  3.43 & +25  5  4.30 & II &      204.7 & & 14 55 55.36 & +11 51 44.70 & I  &      385.3 \\
14 42  3.93 & +13 29 17.60 & C  &      706.0 & & 14 56  5.65 & +16 26 52.80 & I  &     1290.3 \\
14 42 22.99 & +14 54 58.20 & S  &      202.6 & & 14 56 21.94 & +26 35 56.40 & II &      345.4 \\
14 42 27.11 & +14 31 39.90 & II &      224.1 & & 14 56 28.69 & +13 02 41.10 & S  &      260.7 \\
\hline
  \end{tabular}
\end{table}
\begin{table}
  \centering
  \begin{tabular}{llll|c|llll}
    \hline
    \multicolumn{9}{|c|}{CoNFIG-3} \\
    RA & DEC & Radio & \SoneG & &RA & DEC & Radio & \SoneG \\
    \multicolumn{2}{|c|}{(J2000)} & Type &(mJy) & &\multicolumn{2}{|c|}{(J2000)} & Type &(mJy) \\
    \hline        
15 09 15.60 & +14 06 14.80 & S  &      402.7 & & 15 23 58.49 & +16 14 47.60 & C  &      234.8 \\
15 09 16.35 & +16 58 57.20 & II &      233.5 & & 15 24  5.26 & +29 00 23.60 & C  &      279.1 \\
15 09 42.80 & +29 39  2.80 & II &      205.6 & & 15 24 12.71 & +19 23 59.70 & S  &      263.9 \\
15 09 42.80 & +15 56 59.10 & C  &      284.9 & & 15 24 41.60 & +15 21 21.70 & S  &      445.0 \\
15 09 50.53 & +15 57 25.70 & S  &      386.0 & & 15 24 54.42 & +27 57  7.70 & I  &      209.6 \\
\hline
  \end{tabular}
\end{table}
\begin{table}
  \centering
  \begin{tabular}{llll|c|llll}
    \hline
    \multicolumn{9}{|c|}{CoNFIG-3} \\
    RA & DEC & Radio & \SoneG & &RA & DEC & Radio & \SoneG \\
    \multicolumn{2}{|c|}{(J2000)} & Type &(mJy) & &\multicolumn{2}{|c|}{(J2000)} & Type &(mJy) \\
    \hline
15 46  2.56 & +17 54 38.40 & II &      229.9 & & 15 53 19.50 & +17 49 43.10 & II &      383.7 \\
15 46  5.99 & +27 49 17.30 & S  &      217.1 & & 15 53 27.68 & +27 30 55.80 & S  &      228.5 \\
15 46 31.47 & +21 57 41.30 & II &      427.2 & & 15 53 32.77 & +12 56 50.80 & C* &      947.7 \\
15 47  7.52 & +11 42 49.80 & II &      367.1 & & 15 53 43.61 & +23 48  4.70 & I  &      732.1 \\
15 47 12.96 & +18 04 10.80 & I  &      419.9 & & 15 53 50.45 & +25 01 25.30 & S  &      367.8 \\
\hline
  \end{tabular}
\end{table}

\begin{table}
  \centering
  \small
  \medskip
  \begin{tabular}{llll|c|llll}
    \hline
    \multicolumn{9}{|c|}{CoNFIG-4} \\
    RA & DEC & Radio & \SoneG & &RA & DEC & Radio & \SoneG \\
    \multicolumn{2}{|c|}{(J2000)} & Type &(mJy) & &\multicolumn{2}{|c|}{(J2000)} & Type &(mJy) \\
    \hline
14 08  7.86 & -04 51 28.40 & II &      190.1 & & 14 11  7.85 & +00 36  7.80 & C* &      242.0 \\
14 08 18.51 & +00 13  9.50 & C  &      113.8 & & 14 11  8.29 & +01 24 41.10 & II &      187.5 \\
14 08 28.14 & +02 25 48.70 & I  &      243.7 & & 14 11 10.28 & -00 35 59.10 & II &      103.3 \\
14 08 32.29 & +00 31 33.90 & II &       61.2 & & 14 11 14.61 & +02 17 22.50 & I  &       86.5 \\
14 08 32.70 & -01 31 19.50 & C  &       73.3 & & 14 11 20.30 & -04 32 36.70 & II &      188.3 \\
\hline
  \end{tabular}
\end{table}
\begin{table}
  \centering
  \small
  \medskip
  \begin{tabular}{llll|c|llll}
    \hline
    \multicolumn{9}{|c|}{CoNFIG-4} \\
    RA & DEC & Radio & \SoneG & &RA & DEC & Radio & \SoneG \\
    \multicolumn{2}{|c|}{(J2000)} & Type &(mJy) & &\multicolumn{2}{|c|}{(J2000)} & Type &(mJy) \\
    \hline
14 15 28.72 & +01 05 54.20 & II &      107.2 & & 14 26 30.42 & +01 42 36.10 & II &       90.8 \\
14 15 30.47 & +02 23  1.30 & II &      153.9 & & 14 26 49.60 & -00 47 18.30 & C  &       87.7 \\
14 15 53.96 & +02 10 31.90 & C  &       86.5 & & 14 26 49.84 & +00 55 59.90 & II &       57.9 \\
14 16  4.11 & +00 29 15.40 & I  &      107.2 & & 14 26 55.09 & -02 49 20.90 & C  &       60.3 \\
14 16  9.42 & -01 25 13.60 & I  &      114.2 & & 14 26 55.47 & -02 15 45.10 & C  &      167.1 \\
\hline
  \end{tabular}
\end{table}
\begin{table}
  \centering
  \small
  \medskip
  \begin{tabular}{llll|c|llll}
    \hline
    \multicolumn{9}{|c|}{CoNFIG-4} \\
    RA & DEC & Radio & \SoneG & &RA & DEC & Radio & \SoneG \\
    \multicolumn{2}{|c|}{(J2000)} & Type &(mJy) & &\multicolumn{2}{|c|}{(J2000)} & Type &(mJy) \\
    \hline
14 34 10.14 & -01 23 24.80 & II &      122.2 & & 14 37 41.15 & +02 17 26.70 & I  &       67.7 \\
14 34 32.06 & -03 54 40.20 & C  &      123.8 & & 14 37 42.80 & -00 15  4.20 & I  &       67.0 \\
14 34 47.71 & -04 27 44.90 & II &      105.1 & & 14 37 48.30 & -01 47  9.40 & C  &       52.1 \\
14 34 49.27 & -02 15  9.20 & II &      112.5 & & 14 37 56.68 & -00 41 16.10 & II &       61.6 \\
14 34 52.84 & +02 36  3.00 & II &      322.1 & & 14 37 56.76 & +01 56 38.30 & II &       86.7 \\
\hline
  \end{tabular}
\end{table}

\begin{table}
  \caption[Source count details for all the CoNFIG
  sources]{\label{SCdata} Source counts. $S_{min}$ and $S_{max}$
  define the flux density bins and $S_{av}$ corresponds to the center
  of the bin.}
  \medskip
  \medskip
  \centerline{
    \begin{tabular}{|ccccc|}
      \hline
      $logS_{min}$ & $logS_{max}$ & $logS_{av}$ & N & Area (sr)\\
      \hline
      \multicolumn{5}{|c|}{All Sources} \\
      \hline
      $-$2.10 & $-$1.90 & $-$2.00 &  \phantom{0}39 &  0.0018 \\
      $-$1.90 & $-$1.70 & $-$1.80 &  \phantom{0}34 &  0.0018 \\
      $-$1.70 & $-$1.50 & $-$1.60 &  \phantom{0}19 &  0.0018 \\
      $-$1.50 & $-$1.30 & $-$1.40 &  \phantom{0}15 &  0.0213 \\
      $-$1.30 & $-$1.10 & $-$1.20 & 107 &  0.0213 \\
      $-$1.10 & $-$0.90 & $-$1.00 &  \phantom{0}52 &  0.0213 \\
\hline
  \end{tabular}}
\end{table}

%%% CONTOUR PLOTS PANEL
\begin{figure}
\section{Contour Plots}\label{ContPlot}
\subsection{Contours}
Contour plots from NVSS (outermost contours) and FIRST (innermost contours) of
all sources in the CoNFIG sample. The contours are
displayed in logarithmic intervals from 5 mJy/beam (unless specified
otherwise) to the peak flux per beam of the source. The grey-scale
background corresponds to the SSS optical R-band image. Sources marked
with a $^n$ or a $^v$ correspond to objects for which higher
resolution images were used to determine the morphology, either from
the NRAO archives ($^n$) or our VLA 8 GHz observations ($^v$). The 3
sources marked by a $^R$ were removed from the sample after a closer
inspection of their contour plots. They are part of the 7 deleted
sources listed in Table~\ref{DissSour}. The full version of these figures can
be found online on MNRAS.\\  

\centerline{
  \hspace{5mm}
  \begin{minipage}{4cm}
    \tiny
    \mbox{}
    \centerline{\includegraphics[scale=0.2,angle=270]{source001.ps}}
    \centerline{CoNFIG-001}
    \centerline{07 13 38.15 +43 49 17.20}
    \vfill
    \mbox{}
    \centerline{\includegraphics[scale=0.2,angle=270]{source002.ps}}
    \centerline{CoNFIG-002}
    \centerline{07 14 24.80 +35 34 39.90}
    \vfill
    \mbox{}
    \centerline{\includegraphics[scale=0.2,angle=270]{source003.ps}}
    \centerline{CoNFIG-003}
    \centerline{07 16 41.09 +53 23 10.30}
    \vfill
    \mbox{}
    \centerline{\includegraphics[scale=0.2,angle=270]{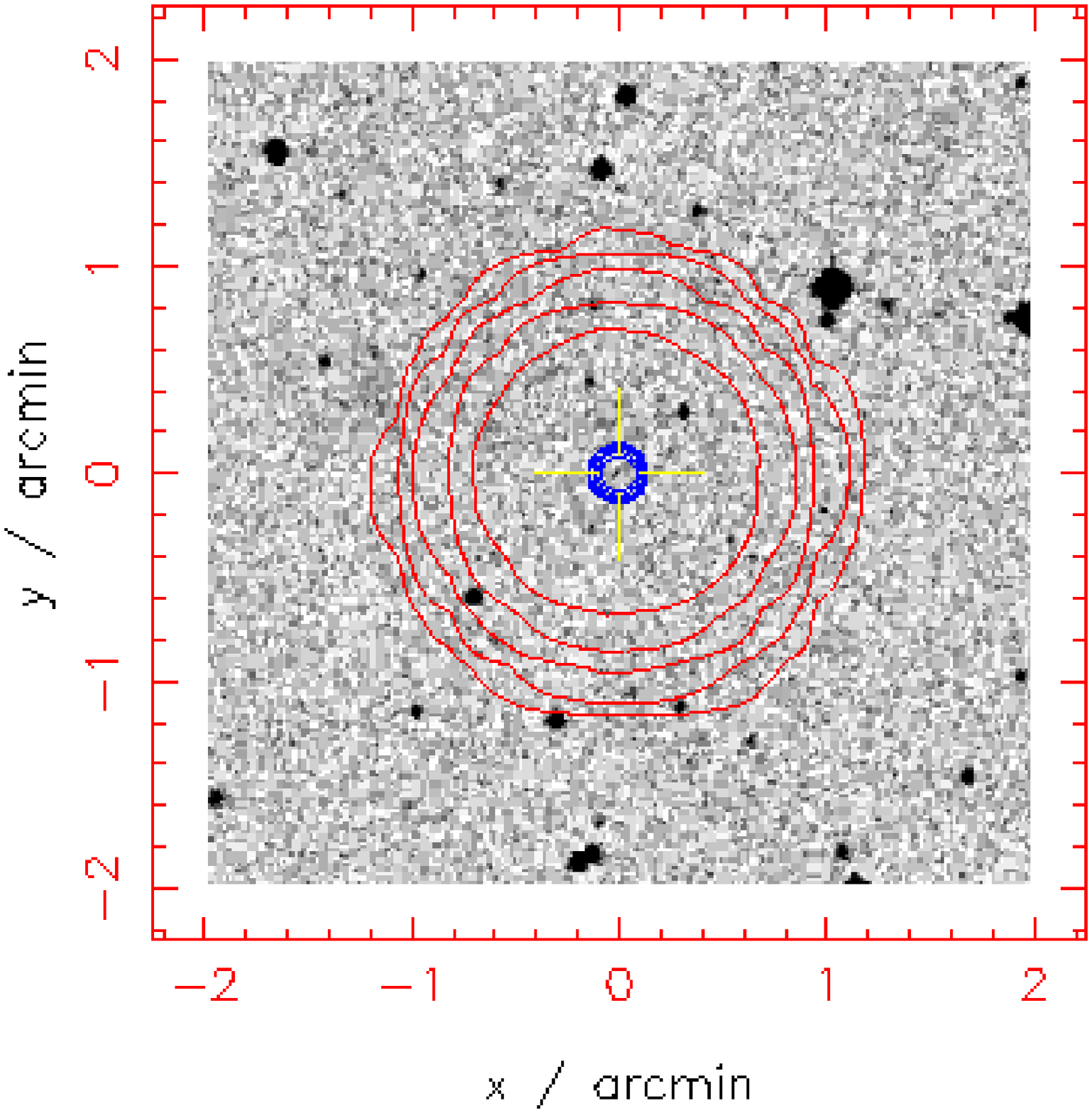}}
    \centerline{CoNFIG-004}
    \centerline{07 35 55.54 +33 07  9.60}
  \end{minipage}
  \hspace{1.5cm}
    \begin{minipage}{4cm}
    \tiny
    \mbox{}
    \centerline{\includegraphics[scale=0.2,angle=270]{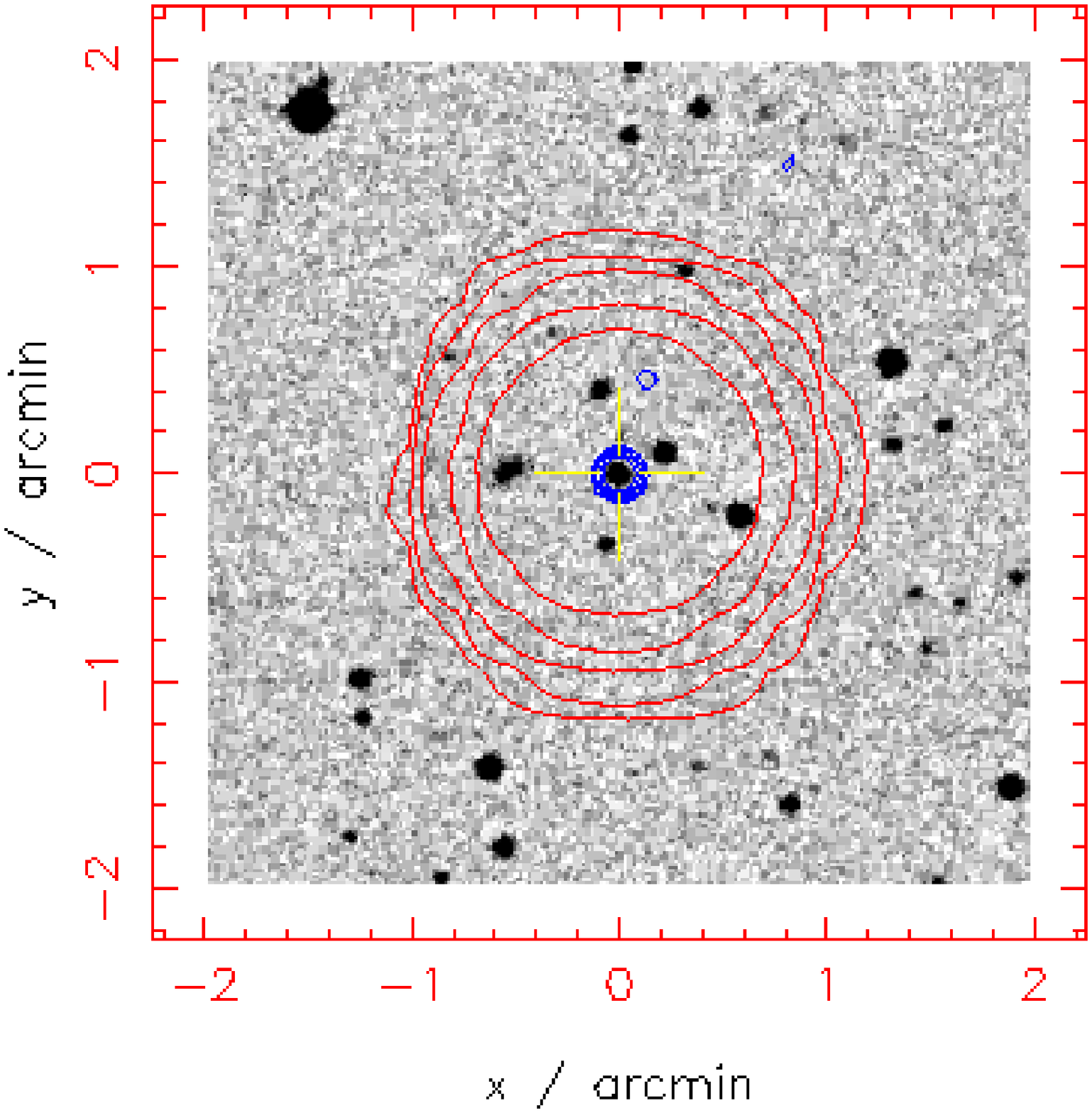}}
    \centerline{CoNFIG-005}
    \centerline{07 41 10.70 +31 12  0.40}
    \vfill
    \mbox{}
    \centerline{\includegraphics[scale=0.2,angle=270]{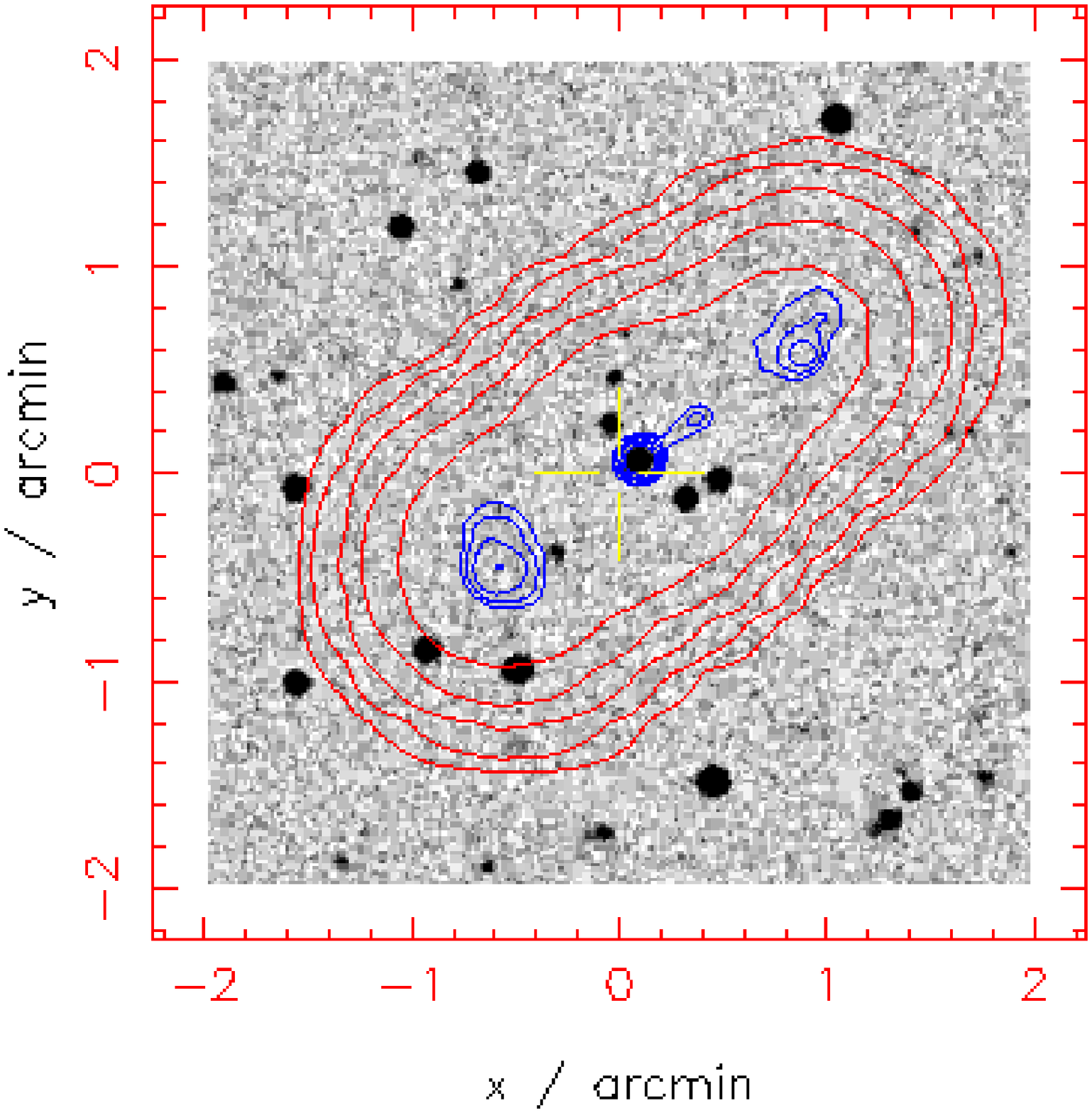}}
    \centerline{CoNFIG-006}
    \centerline{07 45 42.13 +31 42 52.60}
    \vfill
    \mbox{}
    \centerline{\includegraphics[scale=0.2,angle=270]{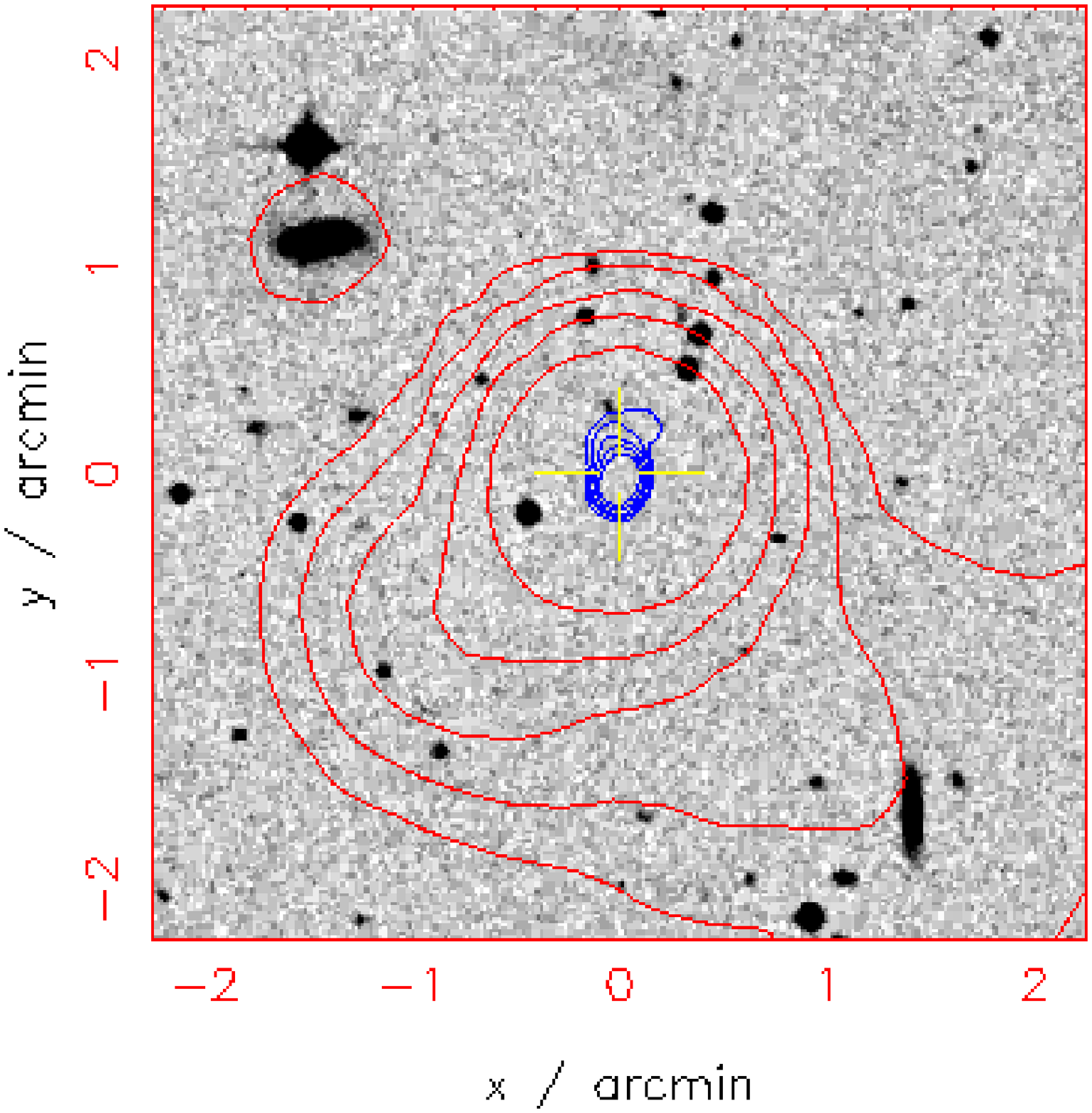}}
    \centerline{CoNFIG-007 (from 10 mJy/beam)}
    \centerline{07 49 48.10 +55 54 21.00}
    \vfill
    \mbox{}
    \centerline{\includegraphics[scale=0.2,angle=270]{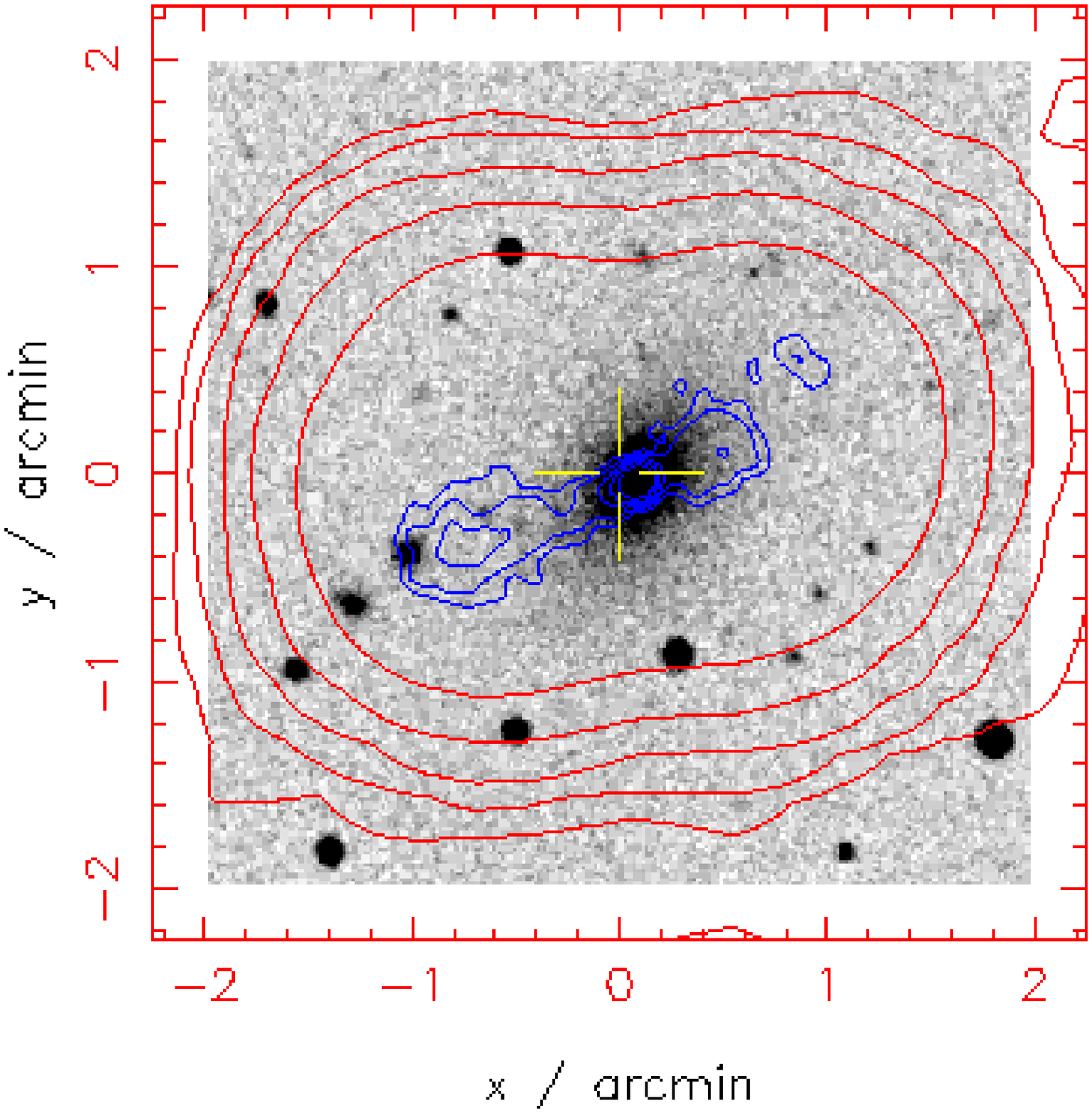}}
    \centerline{CoNFIG-008 (from 1 mJy/beam)}
    \centerline{07 58 28.60 +37 47 13.80}
  \end{minipage}
  \hspace{1.5cm}
  \begin{minipage}{4cm}
    \tiny
    \mbox{}
    \centerline{\includegraphics[scale=0.2,angle=270]{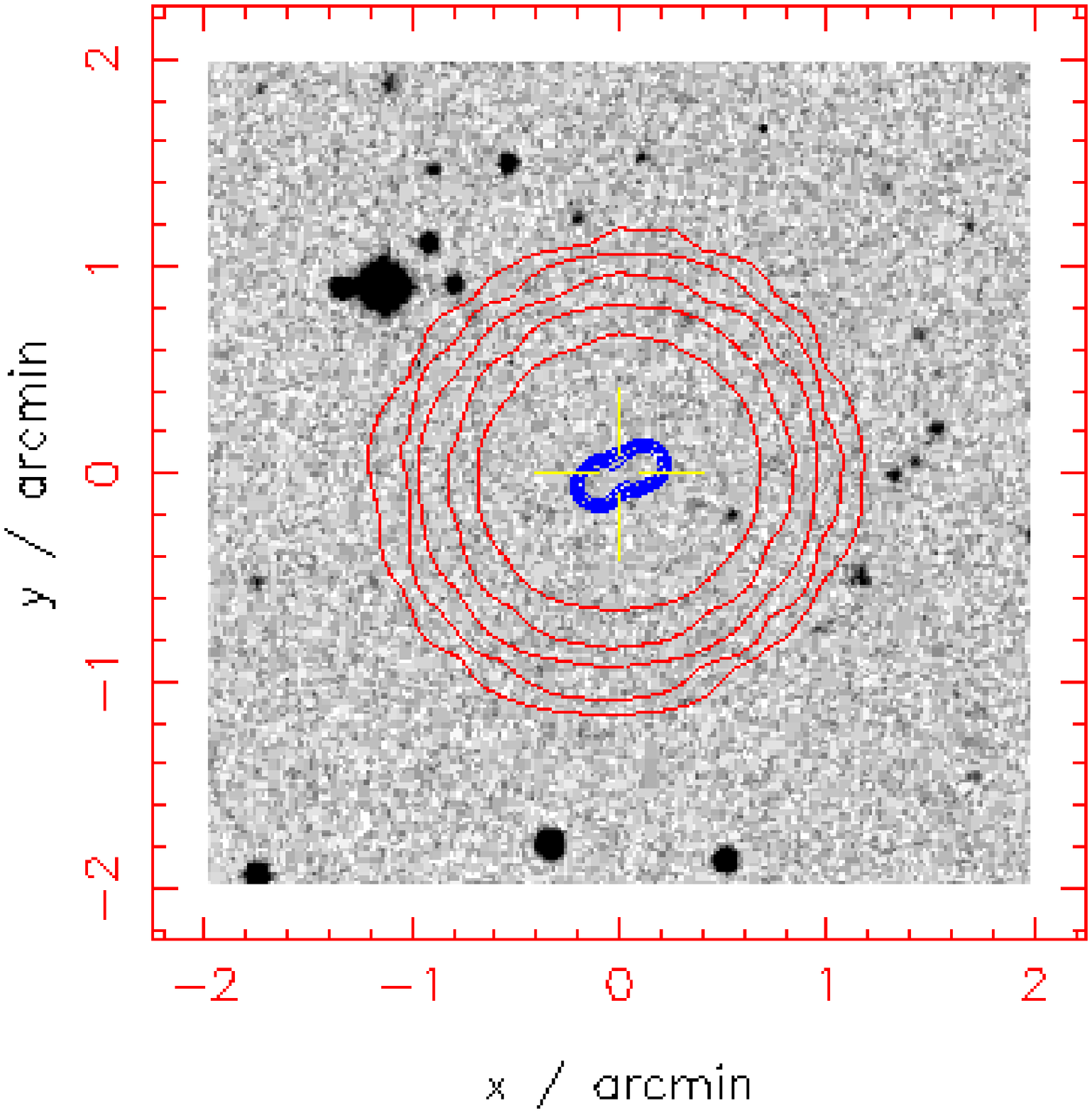}}
    \centerline{CoNFIG-009$^v$}
    \centerline{07 59 47.26 +37 38 50.20}
    \vfill
    \mbox{}
    \centerline{\includegraphics[scale=0.2,angle=270]{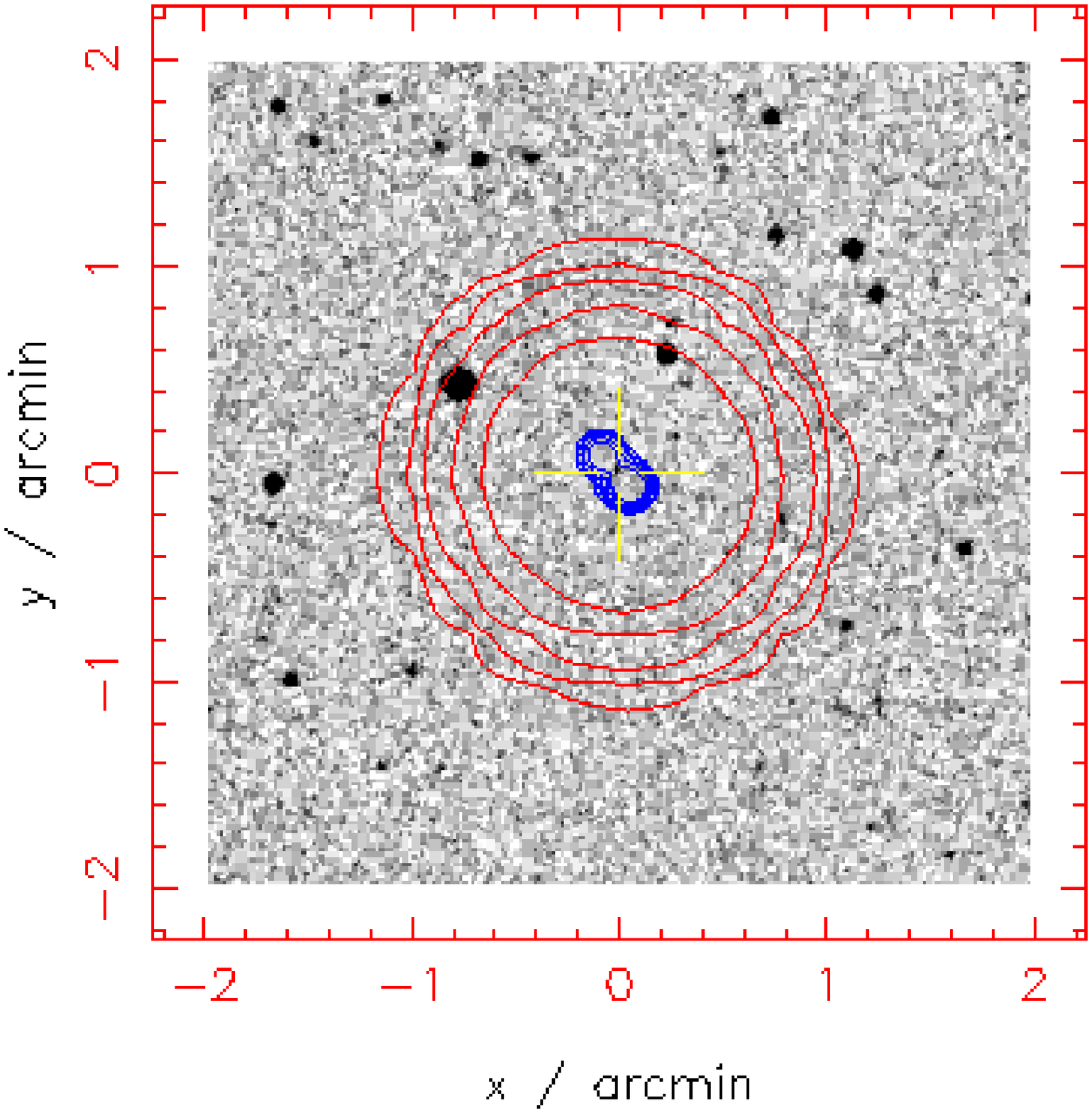}}
    \centerline{CoNFIG-010$^v$}
    \centerline{08 01 35.32 +50 09 43.00}
    \vfill
    \mbox{}
    \centerline{\includegraphics[scale=0.2,angle=270]{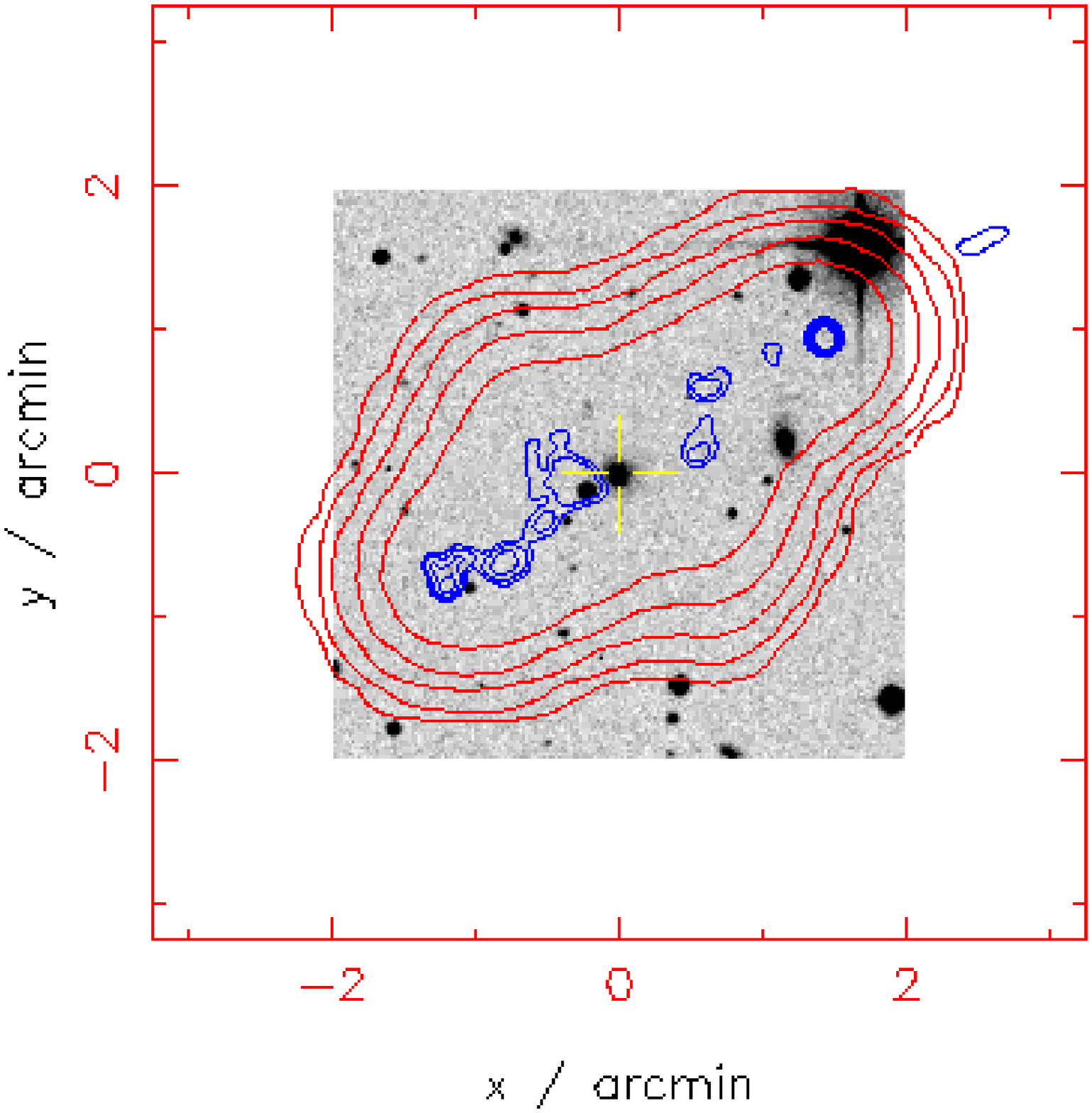}}
    \centerline{CoNFIG-011}
    \centerline{08 05 31.31 +24 10 21.30}
    \vfill
    \mbox{}
    \centerline{\includegraphics[scale=0.2,angle=270]{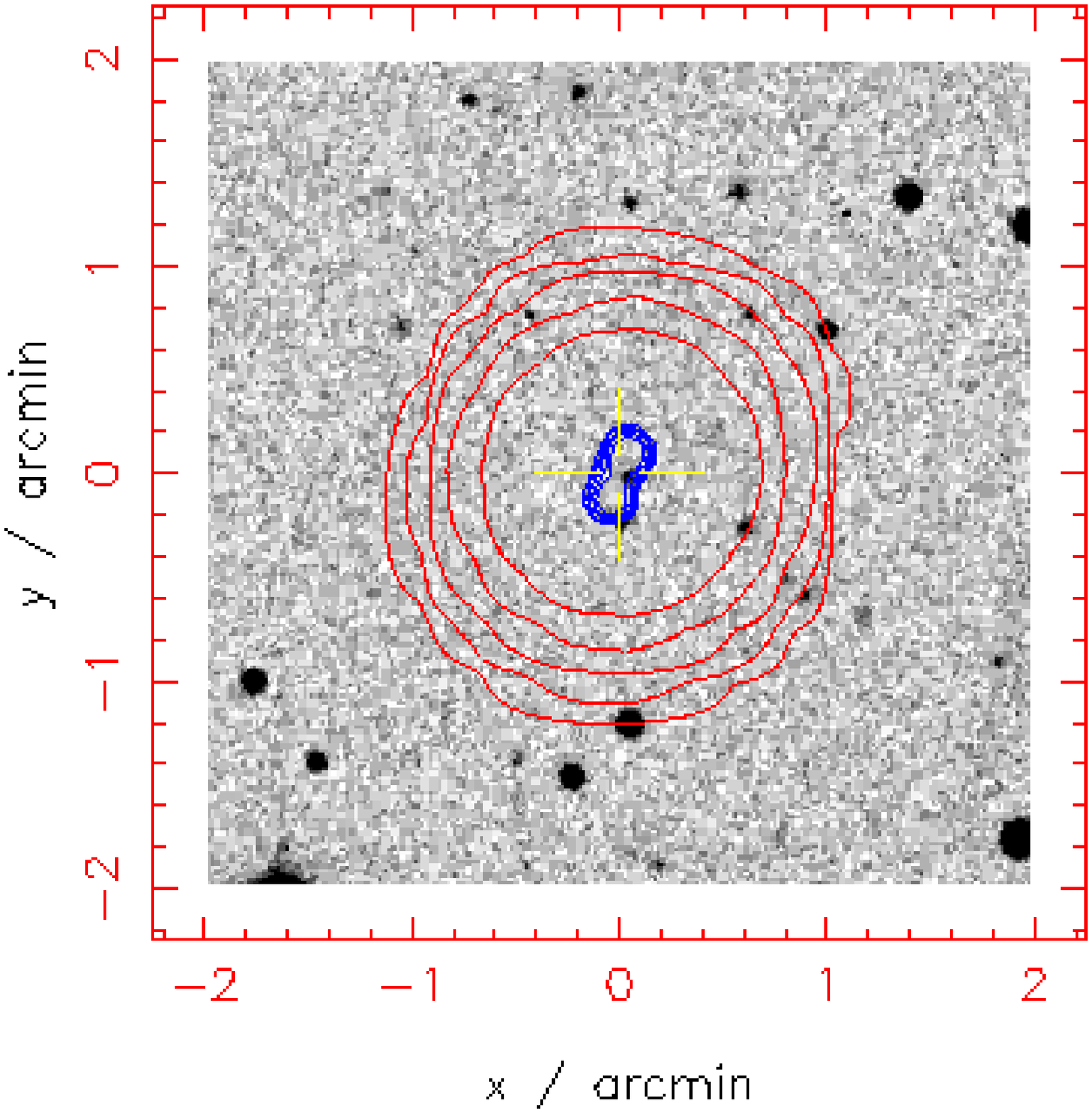}}
    \centerline{CoNFIG-012$^n$}
    \centerline{08 10  3.67 +42 28  4.00}
  \end{minipage}}
 \end{figure}

%%%% Images from VLA Obs

\begin{table}
  \section{VLA observations}\label{VLAobs}
  \subsection{Comments on sources with additional VLA observations}
  \caption[VLA observations details]{\label{VLAobstab}List of the sources with higher
  resolution contour plots. The sources identified by a star are part
  of the VLA 8 GHz A-configuration observation obtained on July 29,
  2007. For these sources, the exposure time and the predicted signal-to-noise
  are stated. Data were extracted from the VLA archives at other
  frequencies and configurations for some sources for which the 8 GHz
  A~-~configuration contours were not satisfactory.\\
  {\footnotesize $^1$ Because of the very low quality of the 8 GHz
data for this source, its morphology remains uncertain.\\
$^2$ The source actually consist of two separate
sources with \SoneG$<1.3Jy$ and was deleted from the sample.\\
$^3$ The source was classified as an FRII, but with
reserve.\\
$^4$ This source was originally classified as FRI
from the FIRST contours, but appeared to be an FRII when looking at
the 8 GHz contours.}
}
  \centerline{
    \begin{tabular}{|lllllcll|}
      \hline
      Source Name & RA & DEC & \SoneG  & Frequ. & Conf. &t$_{exp}$  & S/N \\
                  &    &     &    \phantom{0}(mJy) & & & (min) & \\
      \hline
      CoNFIG-009* & 07 59 47.26 & +37 38 50.20 & 1691.2 & \phantom{1}8 GHz & A&  \phantom{0}5.0 &  14.1\\
      CoNFIG-010* & 08 01 35.32 & +50 09 43.00 & 1471.7 & \phantom{1}8 GHz & A& 10.0 &  10.0\\
      CoNFIG-012 & 08 10 03.67 & +42 28 04.00 & 2056.6 & \phantom{1}8 GHz & A& &\\
      CoNFIG-015*$^2$ & 08 14 43.59 & +12 58 10.00 & 1603.3 & \phantom{1}8 GHz & A& 30.0 &   \phantom{0}5.7\\
      CoNFIG-016*$^3$ & 08 19 47.55 & +52 32 29.50 & 2104.2 & \phantom{1}5 GHz & B& 15.0 &   \phantom{0}5.7\\
      CoNFIG-017$^4$ & 08 21 33.77 & +47 02 35.70 & 1787.1 & \phantom{1}8 GHz & B& &\\
      CoNFIG-025* & 08 34 48.37 & +17 00 46.10 & 1882.8 & \phantom{1}8 GHz & A&  \phantom{0}5.0 &  14.1\\
      CoNFIG-034* & 08 54 39.35 & +14 05 52.10 & 2163.8 & \phantom{1}8 GHz & A&  \phantom{0}5.0 &  42.2\\
      CoNFIG-036* & 08 57 40.64 & +34 04 06.40 & 1798.4 & \phantom{1}8 GHz & A& 10.0 &   \phantom{0}8.9\\
      CoNFIG-037* & 08 58 10.07 & +27 50 50.80 & 1807.8 & \phantom{1}8 GHz & A&  \phantom{0}5.0 &  11.8\\
      CoNFIG-045* & 09 12 04.00 & +16 18 29.70 & 1374.6 & \phantom{1}8 GHz & A&  \phantom{0}5.0 &  21.7\\
      CoNFIG-048* & 09 22 49.93 & +53 02 21.20 & 1597.8 & \phantom{1}8 GHz & A&  \phantom{0}5.0 &  10.9\\
      CoNFIG-050 & 09 30 33.45 & +36 01 23.60 & 1875.1 & \phantom{1}5 GHz & B& &\\
      CoNFIG-054 & 09 42 15.35 & +13 45 49.60 & 3336.4 & \phantom{1}5 GHz & A& &\\
      CoNFIG-055* & 09 43 12.74 & +02 43 27.50 & 1331.5 & \phantom{1}8 GHz & A&  \phantom{0}5.0 &  32.8\\
      CoNFIG-060 & 09 51 58.83 &$-$00 01 26.80& 3152.1 & \phantom{1}5 GHz & A& &\\
      CoNFIG-068 & 10 11 00.36 & +06 24 40.20 & 2964.2 & \phantom{1}8 GHz & A& &\\
      CoNFIG-070 & 10 17 14.15 & +39 01 24.00 & 1392.2 & \phantom{1}8 GHz & A& &\\
      CoNFIG-080* & 10 41 39.01 & +02 42 33.00 & 2710.1 & \phantom{1}8 GHz & A&  \phantom{0}5.0 &  77.4\\
      CoNFIG-083 & 10 58 17.46 & +19 52 09.50 & 2143.0 & \phantom{1}5 GHz & A& &\\
      CoNFIG-084$^1$ & 10 58 29.62 & +01 33 58.20 & 3220.2 & \phantom{1}8 GHz & A& &\\
      CoNFIG-086* & 11 02 03.91 &$-$01 16 18.30& 2799.6 & \phantom{1}8 GHz & A&  \phantom{0}5.0 &  13.4\\
      CoNFIG-098 & 11 23 09.10 & +05 30 20.30 & 1721.1 & \phantom{1}5 GHz & A& &\\
      CoNFIG-101* & 11 34 38.46 & +43 28 00.50 & 1567.1 & \phantom{1}8 GHz & A&  \phantom{0}5.0 &  36.8\\
      CoNFIG-106* & 11 41 08.23 & +01 14 17.70 & 2690.8 & \phantom{1}8 GHz & A&  \phantom{0}5.0 &  94.1\\
      CoNFIG-114 & 11 53 24.51 & +49 31 09.50 & 1572.2 & \phantom{1}8 GHz & A& &\\
      CoNFIG-115* & 11 54 13.01 & +29 16 08.50 & 1620.3 & \phantom{1}8 GHz & A& 10.0 &   \phantom{0}6.8\\
      CoNFIG-122* & 12 04 02.13 &$-$04 22 43.90& 2141.3 & \phantom{1}8 GHz & A& 30.0 &   \phantom{0}4.9\\
      CoNFIG-128* & 12 15 29.80 & +53 35 54.10 & 2755.0 & \phantom{1}8 GHz & A&  \phantom{0}5.0 &  18.4\\
      CoNFIG-129 & 12 15 55.60 & +34 48 15.10 & 1506.8 & \phantom{1}8 GHz & A& &\\
      CoNFIG-134 & 12 24 54.62 & +21 22 47.20 & 2094.4 & \phantom{1}8 GHz & A& &\\
      CoNFIG-140 & 12 32 00.13 & $-$02 24 04.10 & 1646.7 & \phantom{1}8 GHz & A& &\\
      CoNFIG-148* & 12 53 03.55 & +02 38 22.30 & 1604.9 & \phantom{1}8 GHz & A& 30.0 &   \phantom{0}5.4\\
      CoNFIG-151 & 12 56 11.15 & $-$05 47 20.10 & 9711.2 & \phantom{1}8 GHz & A& &\\
      CoNFIG-154 & 13 05 36.05 & +08 55 15.90 & 1461.8 & \phantom{1}8 GHz & A& &\\
      CoNFIG-162* & 13 20 21.45 & +17 43 12.40 & 1573.2 & \phantom{1}8 GHz & A&  \phantom{0}5.0 &   \phantom{0}9.6\\
      CoNFIG-163 & 13 21 18.84 & +11 06 49.40 & 2238.0 & \phantom{1}8 GHz & A& &\\
      CoNFIG-171 & 13 38 08.07 & $-$06 27 11.20 & 2958.5 & \phantom{1}8 GHz & A& &\\
      CoNFIG-173 & 13 42 13.13 & +60 21 42.30 & 1493.3 & \phantom{1}8 GHz & A& &\\
      CoNFIG-189 & 14 17 23.95 & $-$04 00 46.60 & 1687.2 & \phantom{1}8 GHz & A& &\\
      CoNFIG-193 & 14 24 56.93 & +20 00 22.70 & 1808.5 & \phantom{1}8 GHz & A& &\\
      CoNFIG-194* & 14 25 50.67 & +24 04 06.70 & 1558.7 & \phantom{1}8 GHz & A& 30.0 &   \phantom{0}5.2\\
      CoNFIG-195 & 14 28 31.22 & $-$01 24 08.70 & 3157.4 & 15 GHz& C& &\\
      CoNFIG-197* & 14 36 57.07 & +03 24 12.30 & 2797.3 & \phantom{1}8 GHz & A&  \phantom{0}5.0 &  66.8\\
      CoNFIG-201*$^1$ & 14 48 39.98 & +00 18 17.90 & 1651.5 & \phantom{1}8 GHz & A&  \phantom{0}5.0 &  10.5\\
      CoNFIG-208* & 15 10 53.55 &$-$05 43 07.10& 3569.3 & \phantom{1}8 GHz & A&  \phantom{0}5.0 &  56.3\\
      CoNFIG-210*$^1$ & 15 12 25.35 & +01 21 08.70 & 2262.7 & \phantom{1}5 GHz & A&  \phantom{0}5.0 &  52.3\\
      CoNFIG-220* & 15 31 25.36 & +35 33 40.60 & 1820.7 & \phantom{1}8 GHz & A&  \phantom{0}5.0 & \phantom{0}9.1\\
      CoNFIG-224* & 15 37 32.39 & +13 44 47.70 & 1805.6 & \phantom{1}8 GHz & A&  \phantom{0}5.0 &  91.5\\
      CoNFIG-227* & 15 41 45.64 & +60 15 36.20 & 1337.4 & \phantom{1}8 GHz & A& 60.0 &   \phantom{0}5.0\\
      CoNFIG-234*$^1$ & 15 56 10.06 & +20 04 21.20 & 2313.7 & \phantom{1}8 GHz & A&  \phantom{0}5.0 &  91.3\\
      CoNFIG-244* & 16 17 15.75 & +21 07 29.40 & 1748.5 & \phantom{1}8 GHz & A& 10.0 &   \phantom{0}6.9\\
      CoNFIG-257* & 16 36 37.38 & +26 48 06.60 & 1336.1 & \phantom{1}8 GHz & A& 40.0 &   \phantom{0}5.8\\
      \hline
  \end{tabular}}
\end{table}

%-------------------------------------------------------------------------
% PAGE 1 (1-6)

\begin{figure}
\subsection{1.4 GHz and 8 GHz contour plots}
Contour plots from FIRST 1.4GHz (outermost contours) and VLA 8 GHz observations
(innermost contours). The contours are displayed in logarithmic
intervals from 5 mJy/beam to the peak flux per beam of the
source. The greyscale background corresponds to the SSS optical R-band
image. Stars show the FIRST positions, squares the NVSS position and
triangles the optical ID from SSS. The full version of these figures can
be found online on MNRAS.\\

  \hspace{3cm}
  \begin{minipage}[t]{2.5cm}
    \tiny
    \mbox{}
    \centerline{\includegraphics[scale=0.3,angle=270]{VLA009.ps}}
    \centerline{CoNFIG-009}
    \centerline{8 GHz, A configuration}
    \vfill
    \mbox{}
    \centerline{\includegraphics[scale=0.3,angle=270]{VLA010.ps}}
    \centerline{CoNFIG-010}
    \centerline{8 GHz, A configuration}
    \vfill
    \mbox{}
    \centerline{\includegraphics[scale=0.3,angle=270]{VLA012.ps}}
    \centerline{CoNFIG-012}
    \centerline{8 GHz, A configuration}
  \end{minipage}
  \hspace{6cm}
    \begin{minipage}[t]{2.5cm}
    \tiny
    \mbox{}
    \centerline{\includegraphics[scale=0.3,angle=270]{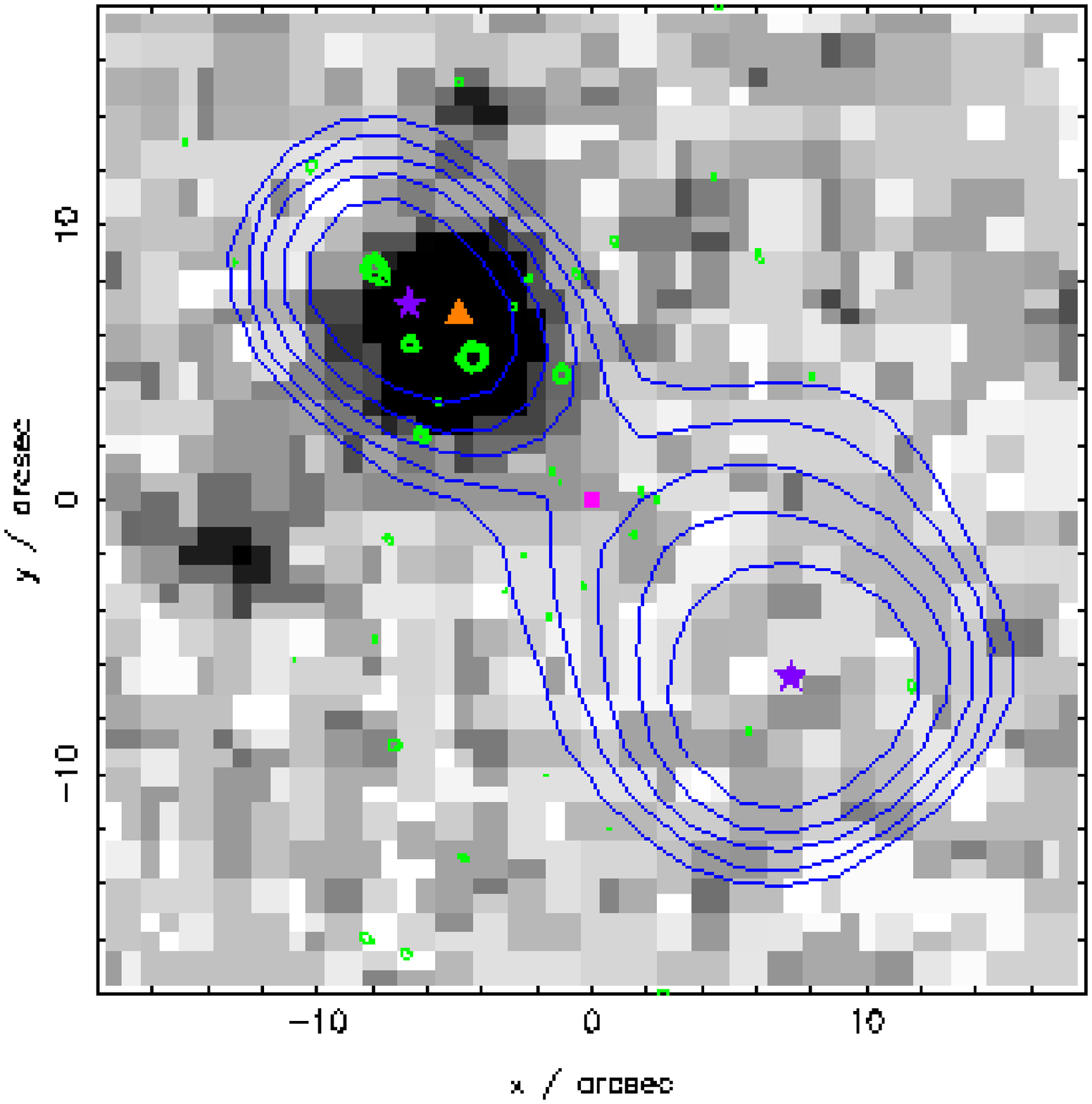}}
    \centerline{CoNFIG-015}
    \centerline{8 GHz, A configuration}
    \vfill
    \mbox{}
    \centerline{\includegraphics[scale=0.3,angle=270]{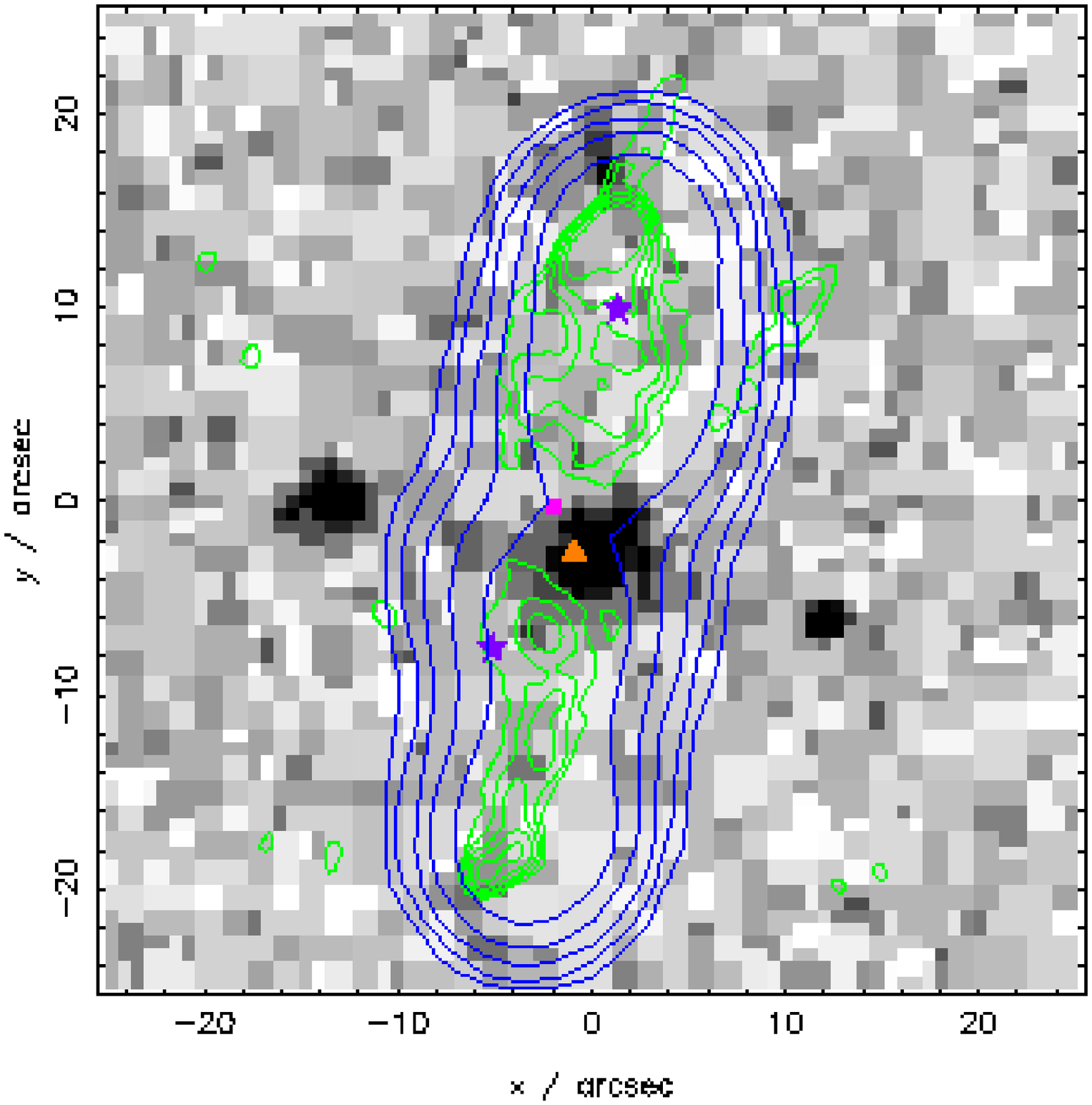}}
    \centerline{CoNFIG-016}
    \centerline{5 GHz, B configuration}
    \vfill
    \mbox{}
    \centerline{\includegraphics[scale=0.3,angle=270]{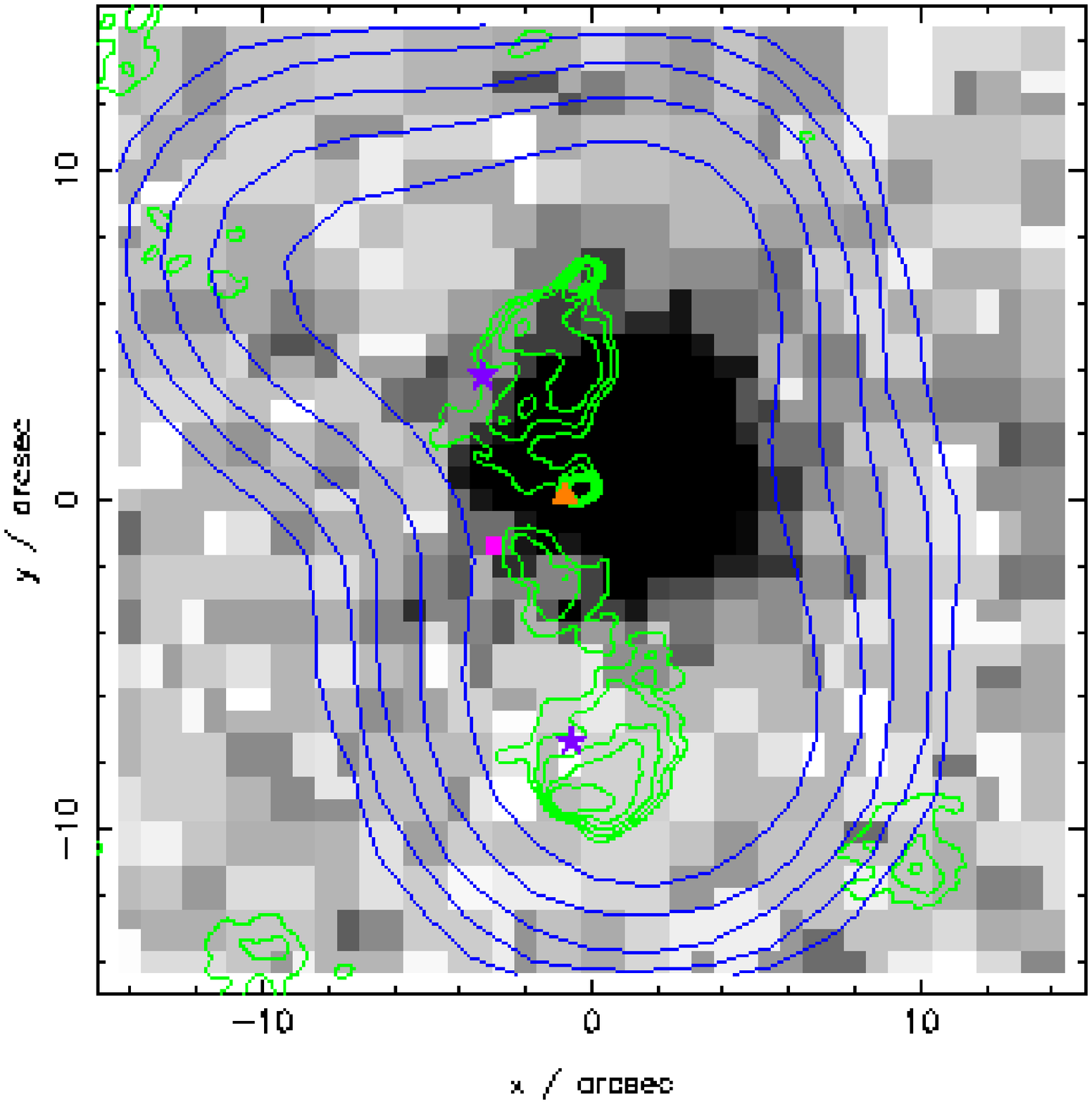}}
    \centerline{CoNFIG-017}
    \centerline{8 GHz, B configuration}
  \end{minipage}
 \end{figure}

\label{lastpage}

\end{document}